\documentclass[12pt,preprint]{aastex}
\newcommand{\be}{\begin{equation}}
\newcommand{\ee}{\end{equation}}
\newcommand{\ul}{\underline{\hspace{40pt}}}
\newcommand{\bea}{\begin{eqnarray}}
\newcommand{\eea}{\end{eqnarray}}
\newcommand{\tbnm}{\tablenotemark}

\newcommand{\bc}{\begin{center}}
\newcommand{\ec}{\end{center}}
\newcommand{\bml}{\begin{mathletters}}
\newcommand{\eml}{\end{mathletters}}

\newcommand{\cc}{\mbox{cm$^{-3}$}}
\newcommand{\kms}{\mbox{\rm km s$^{-1}$}}

\newcommand{\cm}{\mbox{\rm cm}}
\newcommand{\yr}{\mbox{\rm yr}}
\newcommand{\vdot}{\mbox{\boldmath{$\cdot$}}}

\newcommand{\simlt}{\lesssim}
\newcommand{\simgt}{\gtrsim}

\newcommand{\onefourth}{\mbox{$\case{1}{4}$}}
\newcommand{\xhat}{\mbox{\boldmath{$\hat{x}$}}}

\newcommand{\zhat}{\mbox{\boldmath{$\hat{z}$}}}

\newcommand{\vvec}{\mbox{\boldmath{$v$}}}
\newcommand{\yvec}{\mbox{\boldmath{$y$}}}

\newcommand{\Bvec}{\mbox{\boldmath{$B$}}}

\newcommand{\Evec}{\mbox{\boldmath{${\cal E}$}}}
\newcommand{\Fvec}{\mbox{\boldmath{${\cal F}$}}}
\newcommand{\Fvecn}{\mbox{$\Fvec_{\rm n}$}}
\newcommand{\Fveci}{\mbox{$\Fvec_{\rm i}$}}
\newcommand{\Fvecbeta}{\mbox{$\Fvec_{\beta}$}}

\newcommand{\Uvec}{\mbox{\boldmath{$U$}}}
\newcommand{\Uvecn}{\mbox{$\Uvec_{\rm n}$}}
\newcommand{\Uveci}{\mbox{$\Uvec_{\rm i}$}}
\newcommand{\Uveco}{\mbox{$\Uvec^{1}$}}
\newcommand{\Uvecno}{\mbox{$\Uvec^{1}_{\rm n}$}}
\newcommand{\Uvecio}{\mbox{$\Uvec^{1}_{\rm i}$}}
\newcommand{\Uvect}{\mbox{$\Uvec^{2}$}}
\newcommand{\Uvecnt}{\mbox{$\Uvec^{2}_{\rm n}$}}
\newcommand{\Uvecit}{\mbox{$\Uvec^{2}_{\rm i}$}}
\newcommand{\Uvecth}{\mbox{$\Uvec^{3}$}}
\newcommand{\Uvecnth}{\mbox{$\Uvec^{3}_{\rm n}$}}
\newcommand{\Uvecith}{\mbox{$\Uvec^{3}_{\rm i}$}}

\newcommand{\Uvecbetao}{\mbox{$\Uvec^{1}_{\beta}$}}
\newcommand{\Uvecbetat}{\mbox{$\Uvec^{2}_{\beta}$}}
\newcommand{\Uvecbar}{\mbox{$\overline{\Uvec}$}}
\newcommand{\UvecR}{\mbox{$\Uvec^{\rm R}_{\beta,j}$}}
\newcommand{\UvecL}{\mbox{$\Uvec^{\rm L}_{\beta,j}$}}
\newcommand{\UvecRbar}{\mbox{$\Uvecbar^{\rm R}_{\beta,j}$}}
\newcommand{\UvecRbarmo}{\mbox{$\Uvecbar^{\rm R}_{\beta,j-1}$}}

\newcommand{\UvecLbar}{\mbox{$\Uvecbar^{\rm L}_{\beta,j}$}}
\newcommand{\UvecLbarpo}{\mbox{$\Uvecbar^{\rm L}_{\beta,j+1}$}}
\newcommand{\Svec}{\mbox{\boldmath{${\cal S}$}}}
\newcommand{\Svecn}{\mbox{$\Svec_{\rm n}$}}
\newcommand{\Sveci}{\mbox{$\Svec_{\rm i}$}}
\newcommand{\kB}{\mbox{$k_{\rm B}$}}
\newcommand{\cri}{\mbox{$\zeta_{{}_{\rm{CR}}}$}}
\newcommand{\water}{\mbox{H$_2$O}}
\newcommand{\Htwo}{\mbox{H$_2$}}

\newcommand{\alphdr}{\mbox{$\alpha_{{}_{\rm DR}}$}}

\newcommand{\sigwin}{\mbox{$\left<\sigma w\right>_{\rm in}$}}
\newcommand{\siggeo}{\mbox{$\sigma_{\rm{geo}}$}}
\newcommand{\mn}{\mbox{$m_{\rm n}$}}
\newcommand{\mprot}{\mbox{$m_{\rm p}$}}

\newcommand{\nn}{\mbox{$n_{\rm n}$}}
\newcommand{\nno}{\mbox{$n_{\rm n0}$}}
\newcommand{\rhon}{\mbox{$\rho_{\rm n}$}}
\newcommand{\rhono}{\mbox{$\rho_{\rm n0}$}}
\newcommand{\Tn}{\mbox{$T_{\rm n}$}}
\newcommand{\Pn}{\mbox{$P_{\rm n}$}}
\newcommand{\Pno}{\mbox{$P_{\rm n0}$}}
\newcommand{\En}{\mbox{$E_{\rm n}$}}
\newcommand{\enint}{\mbox{$\epsilon_{\rm n}$}}
\newcommand{\vn}{\mbox{$v_{{\rm n}}$}}

\newcommand{\Gn}{\mbox{$G_{\rm n}$}}
\newcommand{\Ln}{\mbox{$\Lambda_{\rm n}$}}

\newcommand{\Fn}{\mbox{$F_{\rm n}$}}
\newcommand{\Fnel}{\mbox{$F_{\rm n,el}$}}
\newcommand{\Fnin}{\mbox{$F_{\rm n,inel}$}}

\newcommand{\sigin}{\mbox{$\langle \sigma w \rangle_{\rm in}$}}

\newcommand{\tni}{\mbox{$\tau_{{\rm ni}}$}}

\newcommand{\Sn}{\mbox{$S_{\rm n}$}}
\newcommand{\sn}{\mbox{$s_{\rm n}$}}
\newcommand{\mi}{\mbox{$m_{\rm i}$}}
\newcommand{\nni}{\mbox{$n_{\rm i}$}}
\newcommand{\rhoi}{\mbox{$\rho_{\rm i}$}}
\newcommand{\rhoio}{\mbox{$\rho_{\rm i0}$}}
\newcommand{\nnio}{\mbox{$n_{\rm i0}$}}
\newcommand{\Ti}{\mbox{$T_{\rm i}$}}
\newcommand{\Si}{\mbox{$S_{\rm i}$}}

\newcommand{\xxi}{\mbox{$x_{\rm i}$}}
\newcommand{\vi}{\mbox{$v_{{\rm i}}$}}

\newcommand{\tin}{\mbox{$\tau_{{\rm in}}$}}

\newcommand{\ve}{\mbox{$v_{{\rm e}}$}}
\newcommand{\me}{\mbox{$m_{\rm e}$}}
\newcommand{\nne}{\mbox{$n_{\rm e}$}}

\newcommand{\Te}{\mbox{$T_{\rm e}$}}

\newcommand{\Cs}{\mbox{$C_{\rm{n}}$}}
\newcommand{\vims}{\mbox{$V_{\rm ims}$}}
\newcommand{\vnms}{\mbox{$V_{\rm nms}$}}
\newcommand{\Alf}{\mbox{Alfv\'{e}n}}
\newcommand{\Cada}{\mbox{$\check{\rm C}$ada}}
\newcommand{\vnA}{\mbox{$V_{\rm{nA}}$}}
\newcommand{\viA}{\mbox{$V_{\rm{iA}}$}}
\newcommand{\Lims}{\mbox{$L_{\rm{ims}}$}}
\newcommand{\Lnms}{\mbox{$L_{\rm{nms}}$}}
\newcommand{\ampl}{\mbox{${\cal A}_{\rm p}$}}
\newcommand{\lgauss}{\mbox{$L_{\rm G}$}}
\newcommand{\Di}{\mbox{${\cal D}_{\rm ad}$}}
\newcommand{\Dth}{\mbox{${\cal D}_{\rm th}$}}
\newcommand{\Gonefunc}{\mbox{${\cal G}_{1}$}}
\newcommand{\Gtwofunc}{\mbox{${\cal G}_{2}$}}
\newcommand{\delx}{\mbox{$\Delta x$}}
\newcommand{\delt}{\mbox{$\Delta t$}}
\newcommand{\sourceopa}{\mbox{${\cal T}^{(\delt)}_{\cal S}$}}
\newcommand{\sourceopb}{\mbox{${\cal T}^{(\delt/2)}_{\cal S}$}}
\newcommand{\RGop}{\mbox{${\cal T}^{(\delt)}_{\rm RG}$}}
\newcommand{\xj}{\mbox{$x_{j}$}}
\newcommand{\xjmo}{\mbox{$x_{j-1}$}}
\newcommand{\xjpo}{\mbox{$x_{j+1}$}}
\newcommand{\xjphlf}{\mbox{$x_{j+\onehalf}$}}
\newcommand{\xjmhlf}{\mbox{$x_{j-\onehalf}$}}
\newcommand{\lamj}{\mbox{$\lambda_{{\rm max}j}$}}
\newcommand{\deltCFL}{\mbox{$\delt_{{\rm CFL}j}$}}
\newcommand{\deltsource}{\mbox{$\delt_{{\cal S}j}$}}

\newcommand{\qR}{\mbox{$q^{\rm R}_{j}$}}
\newcommand{\qL}{\mbox{$q^{\rm L}_{j}$}}
\newcommand{\Deltaj}{\mbox{$\overline{\Delta}_{j}$}}
\newcommand{\Deltajmhlf}{\mbox{$\Delta_{j-\onehalf}$}}
\newcommand{\Deltajphlf}{\mbox{$\Delta_{j+\onehalf}$}}

\newcommand{\ULveci}{\mbox{$\Uvec^{\rm L}_{\rm i}$}}
\newcommand{\URveci}{\mbox{$\Uvec^{\rm R}_{\rm i}$}}
\newcommand{\rhoiL}{\mbox{{$\rho^{\rm L}_{\rm i}$}}}
\newcommand{\rhoiR}{\mbox{{$\rho^{\rm R}_{\rm i}$}}}
\newcommand{\viL}{\mbox{{$v^{\rm L}_{\rm i}$}}}
\newcommand{\viR}{\mbox{{$v^{\rm R}_{\rm i}$}}}
\newcommand{\BL}{\mbox{{$B^{\rm L}$}}}
\newcommand{\BR}{\mbox{{$B^{\rm R}$}}}
\newcommand{\matA}{\mbox{\boldmath{${\sf A}$}}}
\newcommand{\lambdam}{\mbox{$\lambda_-$}}
\newcommand{\lambdaz}{\mbox{$\lambda_0$}}
\newcommand{\lambdap}{\mbox{$\lambda_+$}}
\newcommand{\vecR}{\mbox{\boldmath{$R$}}}
\newcommand{\vecL}{\mbox{\boldmath{$L$}}}
\newcommand{\Rm}{\mbox{\vecR $_{\rm -}$}}
\newcommand{\Rz}{\mbox{\vecR $_{\rm 0}$}}
\newcommand{\Rp}{\mbox{\vecR $_{\rm +}$}}
\newcommand{\Lm}{\mbox{\vecL $_{\rm -}$}}
\newcommand{\Lz}{\mbox{\vecL $_{\rm 0}$}}
\newcommand{\Lp}{\mbox{\vecL $_{\rm +}$}}
\newcommand{\rhoiSL}{\mbox{$\rho^{*{\rm L}}_{\rm i}$}}
\newcommand{\viSL}{\mbox{$v^{*{\rm L}}_{\rm i}$}}
\newcommand{\BSL}{\mbox{$B^{*{\rm L}}$}}
\newcommand{\rhoiSR}{\mbox{$\rho^{*{\rm R}}_{\rm i}$}}
\newcommand{\viSR}{\mbox{$v^{*{\rm R}}_{\rm i}$}}
\newcommand{\BSR}{\mbox{$B^{*{\rm R}}$}}
\newcommand{\viS}{\mbox{$v^{*}_{\rm i}$}}
\newcommand{\BS}{\mbox{$B^{*}$}}
\newcommand{\fLS}{\mbox{$f_{\rm LS}$}}
\newcommand{\PS}{\mbox{$P^{*}$}}
\newcommand{\PL}{\mbox{$P^{{\rm L}}$}}
\newcommand{\PR}{\mbox{$P^{{\rm R}}$}}
\newcommand{\fRS}{\mbox{$f_{\rm RS}$}}

\newcommand{\fLR}{\mbox{$f_{\rm LR}$}}
\newcommand{\viAL}{\mbox{$V^{{\rm L}}_{\rm iA}$}}
\newcommand{\fRR}{\mbox{$f_{\rm RR}$}}
\newcommand{\viAR}{\mbox{$V^{{\rm R}}_{\rm iA}$}}
\newcommand{\fL}{\mbox{$f_{\rm L}$}}
\newcommand{\fR}{\mbox{$f_{\rm R}$}}
\newcommand{\CiL}{\mbox{${\cal C}^{\rm L}_{\rm i}$}}
\newcommand{\CiR}{\mbox{${\cal C}^{\rm R}_{\rm i}$}}
\newcommand{\sL}{\mbox{$s^{\rm L}_{\rm i}$}}
\newcommand{\sR}{\mbox{$s^{\rm R}_{\rm i}$}}
\newcommand{\QL}{\mbox{$Q^{\rm L}_{\rm i}$}}
\newcommand{\QR}{\mbox{$Q^{\rm R}_{\rm i}$}}
\newcommand{\hL}{\mbox{$h^{\rm L}_{\rm i}$}}
\newcommand{\hR}{\mbox{$h^{\rm R}_{\rm i}$}}
\newcommand{\tL}{\mbox{$t^{\rm L}_{\rm i}$}}
\newcommand{\tR}{\mbox{$t^{\rm R}_{\rm i}$}}
\newcommand{\USLveci}{\mbox{$\Uvec^{* \rm L}_{\rm i}$}}
\newcommand{\USRveci}{\mbox{$\Uvec^{* \rm R}_{\rm i}$}}
\newcommand{\ULfanveci}{\mbox{$\Uvec^{\rm Lfan}_{\rm i}$}}
\newcommand{\URfanveci}{\mbox{$\Uvec^{\rm Rfan}_{\rm i}$}}
\newcommand{\viASL}{\mbox{$V^{* \rm L}_{\rm iA}$}}
\newcommand{\viASR}{\mbox{$V^{*\rm R}_{\rm iA}$}}
\newcommand{\rhoiLfan}{\mbox{{$\rho^{\rm Lfan}_{\rm i}$}}}
\newcommand{\rhoiRfan}{\mbox{{$\rho^{\rm Rfan}_{\rm i}$}}}
\newcommand{\viLfan}{\mbox{{$v^{\rm Lfan}_{\rm i}$}}}
\newcommand{\viRfan}{\mbox{{$v^{\rm Rfan}_{\rm i}$}}}
\newcommand{\BLfan}{\mbox{{$B^{\rm Lfan}$}}}
\newcommand{\BRfan}{\mbox{{$B^{\rm Rfan}$}}}
\newcommand{\delrhoi}{\mbox{$\delta\rho_{\rm i}$}}
\newcommand{\delrhon}{\mbox{$\delta\rho_{\rm n}$}}
\newcommand{\delvi}{\mbox{$\delta v_{\rm i}$}}
\newcommand{\delvn}{\mbox{$\delta v_{\rm n}$}}
\newcommand{\delB}{\mbox{$\delta B$}}
\newcommand{\fcarat}{\mbox{$\hat{f}$}}
\newcommand{\delvicarat}{\mbox{$\delta\hat{v}_{\rm i}$}}
\newcommand{\delvncarat}{\mbox{$\delta\hat{v}_{\rm n}$}}
\newcommand{\delrhoncarat}{\mbox{$\delta\hat{\rho}_{\rm n}$}}
\newcommand{\delrhoicarat}{\mbox{$\delta\hat{\rho}_{\rm i}$}}
\newcommand{\delBcarat}{\mbox{$\delta\hat{B}$}}
\newcommand{\ycarati}{\mbox{$\hat{\yvec}_{\rm i}$}}
\newcommand{\ycaratn}{\mbox{$\hat{\yvec}_{\rm n}$}}
\newcommand{\ycarattot}{\mbox{$\hat{\yvec}_{\rm tot}$}}
\newcommand{\Eplusi}{\mbox{$\Evec_{\rm i+}$}}
\newcommand{\Eplusn}{\mbox{$\Evec_{\rm n+}$}}
\newcommand{\Ezeron}{\mbox{$\Evec_{\rm n0}$}}
\newcommand{\Eminusi}{\mbox{$\Evec_{\rm i-}$}}
\newcommand{\Eminusn}{\mbox{$\Evec_{\rm n-}$}}
\newcommand{\Epmi}{\mbox{$\Evec_{\rm i\pm}$}}
\newcommand{\Epmn}{\mbox{$\Evec_{\rm n\pm}$}}
\newcommand{\aplusi}{\mbox{$a_{\rm i+}$}}
\newcommand{\aplusn}{\mbox{$a_{\rm n+}$}}
\newcommand{\aminusi}{\mbox{$a_{\rm i-}$}}
\newcommand{\aminusn}{\mbox{$a_{\rm n-}$}}
\newcommand{\azeron}{\mbox{$a_{\rm n0}$}}
\newcommand{\gplusi}{\mbox{$g_{\rm i+}$}}
\newcommand{\gplusn}{\mbox{$g_{\rm n+}$}}
\newcommand{\gzeron}{\mbox{$g_{\rm n0}$}}
\newcommand{\gminusi}{\mbox{$g_{\rm i-}$}}
\newcommand{\gminusn}{\mbox{$g_{\rm n-}$}}
\newcommand{\gpmi}{\mbox{$g_{\rm i\pm}$}}
\newcommand{\gpmn}{\mbox{$g_{\rm n\pm}$}}
\newcommand{\kims}{\mbox{$k_{\rm ims}$}}
\newcommand{\knms}{\mbox{$k_{\rm nms}$}}
\newcommand{\matM}{\mbox{{\boldmath{${\sf M}$}}}}
\newcommand{\matMi}{\mbox{$\matM_{\rm i}$}}
\newcommand{\matMn}{\mbox{$\matM_{\rm n}$}}
\newcommand{\matMtot}{\mbox{$\matM_{\rm tot}$}}
\newcommand{\Lpre}{\mbox{$L_{\rm pre}$}}
\newcommand{\Lone}{\mbox{$L_{1}(\vi)$}}
\newcommand{\lambdaeig}{\mbox{$\lambda_{\rm p}$}}
\newcommand{\keig}{\mbox{$k_{\rm p}$}}
\newcommand{\beig}{\mbox{$b_{\rm p}$}}
\newcommand{\phaseeig}{\mbox{$\chi_{\rm p}$}}

\slugcomment{To appear in {\sl ApJ}.}
\shortauthors{Ciolek \& Roberge}
\shorttitle{Line Emission from Multifluid Shocks}
\begin{document}
\title{Molecular Line Emission from Multifluid Shock Waves. I. Numerical Methods and
Benchmark Tests}
\author{Glenn E.\ Ciolek and Wayne G.\ Roberge}
\affil{New York Center for Astrobiology
\\ and \\
Department of Physics, Applied Physics and Astronomy \\
Rensselaer Polytechnic Institute,  110 8th Street, Troy, NY 12180}
\email{cioleg@rpi.edu, roberw@rpi.edu}
\begin{abstract}
We describe a numerical scheme for studying time-dependent, multifluid, 
magnetohydrodynamic shock waves in weakly ionized interstellar clouds and cores.
Shocks are modeled as propagating perpendicular to the magnetic field and 
consist of a neutral molecular fluid plus a fluid of ions and electrons. The 
scheme is based on operator splitting, wherein time integration of the governing 
equations is split into separate parts. In one part independent homogeneous Riemann 
problems for the two fluids are solved using Godunov's method. In the other equations 
containing the source terms for transfer of mass, momentum, and energy between the 
fluids are integrated using standard numerical techniques. We show that, for the 
frequent case where the thermal pressures of the ions and electrons are $\ll$ magnetic 
pressure, the Riemann problems for the neutral and ion-electron fluids have a similar 
mathematical structure which facilitates numerical coding. Implementation of the scheme 
is discussed and several benchmark tests confirming its accuracy are presented, 
including (i) MHD wave packets ranging over orders of magnitude in length and time 
scales; (ii) early evolution of mulitfluid shocks caused by two colliding clouds; 
and (iii) a multifluid shock with mass transfer between the fluids by cosmic-ray 
ionization and ion-electron recombination, demonstrating the effect of ion mass 
loading on magnetic precursors of MHD shocks. An exact solution to a MHD Riemann 
problem forming the basis for an approximate numerical solver used in the 
homogeneous part of our scheme is presented, along with derivations of the analytic 
benchmark solutions and tests showing the convergence of the numerical algorithm.
\end{abstract}
\section{Introduction and Motivation}
\label{sec-intro}
Shock waves in molecular clouds and star-forming regions are energetic events which can process 
the interstellar gas and dust they entrain. 
Shock-excited molecular {\water} has been
detected by {\it Herschel} (van Dishoeck et al.\ 2011) and shocked species such as {\water} and
CO are among the targets of opportunity for {\it SOFIA} (Eisl\"{o}ffel et al.\ 2012). 
Observations of the kinematics and proper motions of protostellar outflows and their 
associated shocks yield ages as young as $\sim 10^{2}$--$10^{4}~\yr$ (e.g., Gueth et 
al. 1998; Hartigan et al.\ 2001).
Because the time scale for these shocks to develop steady flow is typically $\simgt 10^4$\,yr
(see \S\ref{sec-models}),
time-dependent solutions are generally required to model observations of
the shock-excited emission.
For instance Gusdorf et al.\ (2011) applied approximate time-dependent
models to {\Htwo}, ${\rm SiO}$, and {\water} lines in the BHR71 
bipolar outflow shocks and concluded that the best fit to the
observations occurred in models having ages $ < 2000~\yr$.
However there was a fairly large degeneracy in the parameter space which 
allowed reasonable fits. 

Modeling shocks in weakly-ionized interstellar clouds is a complex task due largely
to the complex nature of the clouds themselves: molecular clouds are threaded 
by interstellar magnetic fields of order $10$--$10^{3}~\mu{\rm G}$ 
(Crutcher 2004; Rao et al.\ 2009; Tang et al.\ 2009) and are only weakly
ionized, with ion fractional abundances $\simlt 10^{-5}$ (Caselli 2002; Miettinen et al.\ 2011).
Because the bulk of the matter is neutral, clouds have a finite electrical conductivity;
consequently the neutral gas and charged particles collectively constitute a nonideal
magnetohydrodynamic (MHD) system.
Moreover the neutral particles respond indirectly to magnetic forces via momentum
exchange with the (rare) ions and electrons, so that a multifluid treatment of the
dynamics is required.
Early work on multifluid MHD shocks
established that the ions can be accelerated ahead of the neutral shock front 
in a ``magnetic precursor" (Mullan 1971; Draine 1980).
Ion collisional drag in the precursor accelerates, compresses, and heats the neutrals, thereby 
passing information about the disturbance on to the neutral gas upstream
from the shock.
Depending on the shock speed, the time scale for 
collisions between the ions and neutrals, and the rate at which the neutral gas can 
cool, the resulting MHD shock can be a discontinuous flow (J-shock) with a supersonic 
to subsonic transition at a shock front, a continuous flow (C-shock) which
is supersonic everywhere (Draine 1980; Chernoff 1987; Draine \& McKee 1993), or
a continuous flow with a supersonic to subsonic transition (C*-shock, Roberge \& Draine 
1990). 

The majority of theoretical work has been done on {\it steady}\/ multifluid MHD shock waves.
The chemical abundances and excitation of certain molecular species 
in shock waves propagating perpendicular to the ambient magnetic field (``perpendicular shocks")
were studied by Draine et al.\ (1983).
Steady perpendicular shock models with increasingly 
elaborate physics (including the effects of water emission, extensive chemical networks, etc.)
were subsequently presented by
Flower et al.\ (1985, 1986), Pilipp et al.\ (1990), Tielens et al.\ (1994), 
Kaufman \& Neufeld (1996a,b), Flower et al.\ (1996), Schilke et al.\ (1997), and 
Guillet et al.\ (2009). 
Steady MHD shocks propagating obliquely to the magnetic field have also been studied 
(e.g., Wardle \& Draine 1987; Caselli et al.\ 1997).
However the applicability of steady models is limited when it comes 
to interpreting observations of shocks in star-forming regions for reasons
noted above.

A study of time-dependent, multidimensional C~shocks using a two-fluid
finite-difference formulation, and including the inertia of the ions, was presented by 
T\'{o}th (1994). 
Time-dependent simulations of the formation of C-shocks in models which 
neglected the inertia of the ion-electron fluid were produced by Smith \& Mac Low (1997).
MacLow \& Smith (1997) investigated the nonlinear development of Wardle instabilities in 
three-dimensional C-shocks (Wardle 1990) using a two-fluid, time-dependent,
finite-difference MHD code, including the inertia of the ions. 
Ciolek \& Roberge (2002) and Ciolek et al.\ (2004) 
simulated the formation and evolution of one-dimensional perpendicular shocks, accounting 
for the inertial effect of charged dust grain fluids (ion inertia was
ignored, however).
Time-dependent, one-dimensional, perpendicular shock models including the 
inertia of the ions, heating and cooling, and an extensive network of reactions for 33 
different chemical species, were described by LeSaffre et al.\ (2004a,b). 
Multidimensional and multifluid (including charged dust grain species, but neglecting ion inertia) 
evolutionary MHD shock models were discussed by Falle (2003) and Van Loo et al.\ (2009); the
same numerical code was also used to investigate the effect of upstream density 
perturbations on dusty C-shocks in Ashmore et al.\ (2010). Finally, a one-dimensional study 
of the redistribution of mass and magnetic flux and potential protostellar core formation 
due to ambipolar diffusion (the drift of the charged particles with respect to the neutrals) in 
transient C-shocks arising from colliding flows was presented by Chen \& Ostriker (2012); 
in their calculation the inertia of the ions (and self-gravity of the system) was not 
included.

A pseudo-time-dependent (quasi-Lagrangian), one-dimensional MHD shock model has been
employed in several studies of multifluid shocks (e.g., Chieze et al.\ 1998;
Gusdorf et al.\ 2008; Flower \& Pineau des For\'{e}ts 2010, 2012). In these models, 
time dependence is mimicked by setting the partial derivative $\partial/\partial t$ equal 
to zero in the governing equations for conservation of mass, momentum, energy, and magnetic
flux (see eqs.\  [\ref{neutmassconteq}] - [\ref{inducteq}]
below), and then replacing the spatial derivatives $\partial/\partial x$ in those equations
with a ``flow derivative" $(1/v)d/dt$, where $v$ is the fluid velocity of either the neutrals 
or the ions. The pseudo-time-dependent approach has grown to include increasingly more
detailed molecular and chemical networks. 

To study MHD turbulence in weakly ionized clouds, some have advocated a multifluid MHD method referred 
to as the ``heavy ion approximation" (Li et al.\ 2006; Oishi \& MacLow 2006; Li 
et al.\  2008). This approximation was introduced to overcome numerical difficulties
associated with large values of the ion {\Alf} speed, which are typical in molecular 
clouds (see eq. [\ref{viAdef}] below) and limit the size of the time steps a 
numerical code can take to produce accurate, stable simulations (via the CFL 
condition, see eq.\ [\ref{deltCFLeq}]). 
In the heavy ion approximation the masses of the ions are artificially raised to 
increase the mass density of the ion fluid so that smaller, more tractable ion {\Alf} 
speeds (and thus, CFL-limited time steps) are attained. To keep the collisional drag 
terms between the neutral and ion fluids unchanged, the coupling constants which appear in
the ion-neutral frictional forces (e.g., see eqs.\ [\ref{Fneq}]-[\ref{tnieq}]) are
also adjusted so as to offset the effect the inflation of the ion masses has on the forces.
However, Tilley \& Balsara (2010) showed that, while the collisional
force terms remain unchanged in the heavy ion approximation, characteristic length
scales (which involve different combinations of the ion masses and coupling coefficients)
associated with the dissipation range of MHD turbulence are not calculated
correctly using this method. 

Roberge \& Ciolek (2007, hereafter RC07) examined the initial phases of MHD shock
formation in weakly ionized clouds, including the effects of ion inertia.
They noted that for sufficiently small times the disturbance in the ion-electron
fluid is linear, and derived
explicit analytic solutions for the time evolution of a perpendicular shock.
RC07 found that at very early times the inertia of the ions determines
the formation and propagation of the magnetic precursor to a neutral shock. 
At later times the ions' inertia becomes negligible, their motion is force free,
and the evolution of the precursor is nearly self-similar.

The essential mathematical difficulty in modeling time-dependent, multifluid shock waves
is caused by the transfer of mass, momentum, and energy between the fluids
by elastic ion-neutral scattering, ionization/recombination, and numerous other atomic
and molecular processes.
In the absence of coupling between the fluids, the multifluid shock problem
would reduce to independent solutions of Euler's equations for the neutral gas
and the equations of ideal MHD for the ion-electron fluid.
In both cases the governing equations are hyperbolic partial differential equations (PDEs)
which can be solved using well-known techniques.
Many of the algorithms for Euler's equations rely on one's ability to solve the ``Riemann problem''
(e.g., Courant \& Friedrichs 1948),
wherein the gas initially is in two uniform but different states separated by a discontinuity.
In particular,
Godunov (1959) realized that the evolution of gas in two adjacent cells of a finite difference grid is
indeed a Riemann problem and exploited this fact to produce an algorithm, Godunov's method, whose
descendants account for a large subset of all algorithms for gas dynamics (e.g., see Toro 2009 and Pirozzoli 2010).
Riemann solutions are combinations of the characteristic waves of a fluid--- sound waves, rarefactions,
and contact discontinuities in the case of a gas--- which are orchestrated to satisfy certain matching conditions
where the waves intersect (see App.~\ref{sec-exactmhdriemann} for an example).
Riemann-Godunov algorithms also exist for ideal MHD but are complicated by the fact that seven characteristic
waves exist when the flow propagates at an arbitrary angle with respect to the magnetic field
(see Fig.~2 of Dai \& Woodward [1994] or eq.~[7] of Torrilhon [2003]).

When interfluid coupling is included, the only changes to the hyperbolic PDEs for each fluid are
the addition of certain ``source terms'' which describe mass, momentum, and energy transfer
between the fluids (see eq.~[\ref{neutmassconteq}]--[(\ref{inducteq}]).
Toro (2009) noted that problems of this nature could be solved by a technique
called operator splitting and gave a simple, nonhydrodynamical example.
Operator splitting combines separate solutions of the ``homogeneous problem'' (source
terms set to zero) obtained, e.g., with a Riemann-Godunov alogrithm, and the ``inhomogeneous problem'' (evolution
due only to the source terms) obtained with standard methods for ordinary differential equations (ODEs).
Tilley et al.\ (2012) have recently presented an operator-splitting scheme for 
multifluid MHD which treats general time-dependent 
flows for a two-fluid system of neutrals and ions traveling at any orientation with respect to the magnetic field.
Their algorithm retains the inertia of both the neutrals and the ions.
Collisional drag between the two fluids (with a velocity-independent
collision rate) is included along with an energy equation with source terms for each fluid.
The algorithm of Tilley et al.\ (2012) does not account for mass exchange between
the neutral and ion-electron fluids, presumably because this is not important for the
applications of interest to them. 
Tilley et al.\ (2012) presented various benchmarks tests, including the development of a Wardle instability,
and found that the scheme performed well.

In this paper, the first in a series on shocks in molecular clouds, we present a 
split-operator method similar to the algorithm of Tilley et al. (2012), but tailored
to the one-dimensional geometry appropriate for shock waves. This allows us to exploit 
the unique property of perpendicular shocks in weakly ionized plasmas: there exists an 
exact solution to the MHD Riemann problem, which is fundamental to our algorithm. While
the assumption of perpendicular shocks is obviously a special case, it does treat the 
fundamental mathematical problem of coupled hyperbolic PDEs including mass, momentum, 
and energy transfer. Deferring the case of oblique shocks to future work also makes 
strategic sense because modeling perpendicular shocks has a particular advantage:
for likely cloud and shock parameters, the thermal pressure of the ion-electron fluid 
can be neglected compared to magnetic pressure (\S~\ref{sec-governeq}). In this regime 
the charged fluid acoustic modes play no part in the MHD Riemann problem and the number
of characteristic waves in the ion-electron fluid reduces from seven to just three --- the
same number of waves that occur in the gas dynamic Riemann problem. As a result, it turns 
out that the Riemann problem for the charged fluid can be solved exactly 
(App.~\ref{sec-exactmhdriemann}). From this exact solution an approximate but accurate 
solver can be constructed for use in the numerical solution of the MHD Riemann problem in 
the split-operator method. In our algorithm we also include the effect of mass transfer 
between the neutral and charged fluids, which can have profound effects on multifluid 
shocks (Flower et al.\ 1985). 

The plan of our paper is as follows: in \S~2 we present a formulation of the model, including the 
governing equations and assumptions, the respective Riemann 
problems for the neutral gas and ion-electron fluid, and a description of how the 
split-operator scheme is implemented to evolve model shocks in time.
We assess the accuracy of our numerical solutions with various benchmark tests in \S~3.
Our results are summarized in \S~4. 
In the appendices we present the exact solution to the MHD Riemann problem for 
perpendicular flows (App.~\ref{sec-exactmhdriemann}), an approximate solver based on this
solution (App.~\ref{sec-approxmhdriemann}), a derivation of the analytic solutions used as 
benchmark tests (App.~\ref{sec-benchmarks}), and tests confirming the order of spatial and
temporal convergence of our numerical algorithm (App.~\ref{sec-convergence}).

\section{Formulation}
\label{sec-formulate}

Our area of interest is shocks and related flows in interstellar molecular clouds. These clouds
contain neutral particles with number density $\nn$ and particle mass $\mn$ plus a weakly 
ionized plasma of singly-charged ions and electrons having number densities $\nni$ and 
$\nne$, and particle masses $\mi$ and $\me$, respectively. 
Since we are not concerned with 
the problem of gravitational collapse and deal with systems typically having length scales
much smaller than the Jeans length, self-gravity of the gas is ignored.
We adopt a cartesian 
coordinate system ($x$,$y$,$z$) and restrict our attention to models in which all of the 
physical variables are functions of time and the $x$-coordinate only. 
We assume that there is a magnetic field $\Bvec = B(x,t)\zhat$, and that each of the fluids has a 
velocity $\vvec_{\alpha} = v_{\alpha}(x,t)\xhat$, with $\alpha = {\rm n, i, e}$. 
Thus we consider perpendicular shocks only; however we note that
the split-operator method described in this paper can be extended to other geometries. 

In the molecular cloud and core environments we study, the magnetic field strengths and
neutral gas densities are such that the ion and electron fluids each have Hall
parameters (= charged particle gyrofrequency times the mean collision time with the
neutral gas) $\gg 1$. This means that the ions and electrons gyrate about a
magnetic field line many times before suffering a collision with a neutral particle, and
can therefore be considered to be attached to the field. The magnetic field is thus
``frozen into" the charged fluid, and the electrons and ions move together with $\ve \simeq \vi$.

Finally, we ignore the effects of interstellar dust grains. Grains can become charged and
numerical simulations have shown that under certain conditions they can affect the evolution
of MHD shocks (Ciolek \& Roberge 2002; Ciolek et al.\ 2004; Van Loo et al.\ 2009);
the influence of dust grains in MHD flows using a split-operator method will be 
considered at a later time.

\subsection{Governing Equations and Assumptions}
\label{sec-governeq}

Comprehensive derivations of the multifluid MHD equations for astrophysical
flows are presented in Draine (1986) and Mouschovias (1987). For the geometry adopted
here they are
\bea
\label{neutmassconteq}
\frac{\partial \rhon}{\partial t} + \frac{\partial (\rhon \vn)}{\partial x} &=& \Sn \\
\label{neutmtmeq}
\frac{\partial (\rhon \vn)}{\partial t}  + 
\frac{\partial}{\partial x}\left(\rhon \vn^{2} + \Pn \right)
&=& \Fn \\
\label{neutenergyeq}
\frac{\partial \En}{\partial t} + \frac{\partial}{\partial x}\left([\En + \Pn]\vn\right) 
&=& \Fn\vn - \frac{1}{2}\Sn\vn^{2} + \Gn - \Ln \\
\label{ionmassconteq}
\frac{\partial \rhoi}{\partial t}  + \frac{\partial(\rhoi \vi)}{\partial x} &=& -\Sn \\
\label{ionmtmeq}
\frac{\partial(\rhoi \vi)}{\partial t}  
+ \frac{\partial}{\partial x}\left(\rhoi \vi^{2} + \frac{B^{2}}{8 \pi}\right) 
&=&  -\Fn \\
\label{inducteq}
\frac{\partial B}{\partial t} + \frac{\partial(B\vi)}{\partial x} &=& 0~~~.
\eea
Equations (\ref{neutmassconteq})--(\ref{neutenergyeq}) express the conservation of mass, momentum,
and energy for the neutral gas.
Equations (\ref{ionmassconteq})--(\ref{ionmtmeq}) express mass and momentum conservation for the
ion-electron fluid and (\ref{inducteq}) is the induction equation.
We also impose macroscopic charge neutrality,
\be
\label{chargeneutralityeq}
e(\nni - \nne) = 0~~.
\ee

To the set above one should generally add separate energy equations for the ions and electrons,
which are needed to calculate the ion and electron temperatures, \Ti\ and \Te.
However if elastic collisions dominate the transfer of energy between the ions and neutrals,
as is usually the case, then
\be
\label{Tieq}
\Ti \approx \Tn + \frac{\mn}{3 \kB} \left(\vi - \vn\right)^{2}
\ee
(Chernoff 1987), an approximation we adopt.
There is no analogous approximation for \Te\ because energy exchange between the electrons
and neutrals is dominated by complex inelastic processes such as electron impact excitation and ionization.
We temporarily omit the electron energy equation and estimate \Te\ when it is needed
in one of our benchmark calculations (see \S\ref{sec-masstransfer}).

The quantities $\rhon$ ($=\mn \nn$) and $\rhoi$ ($=\mi \nni$) are the mass densities of 
the neutral gas and ion-electron fluid (we neglect the electrons' contribution to \rhoi).
The neutral gas has thermal pressure
\be
\label{idealgaslaw}
\Pn = \frac{\rhon \kB \Tn}{\mn}~,
\ee
where $\Tn$ is the neutral temperature and $\kB$ is the Boltzmann constant.
Its energy density is
\be
\label{energydenseq}
\En = \frac{1}{2}\rhon \vn^{2} + \rhon \enint~~,
\ee
where $\enint$ is the internal energy per unit mass.
In general it is necessary to find \enint\ by a kinetic calculation of
the level populations of rotationally and vibrationally excited \Htwo, CO, \water, etc.\ 
(e.g., Flower et al.\ 2003).
Here we temporarily forego this complication and use the internal
energy for an ideal gas in thermodynamic equilibrium,
\be
\label{eninteq}
\enint = \frac{\Pn}{\rhon(\gamma - 1)}~~,
\ee
with $\gamma=5/3$ in the calculations presented below.
However it is important to note that the split-operator method does not preclude
a kinetic calculation of \enint\ or nonideal equations of state for \enint.

In the ion-electron momentum equation (\ref{ionmtmeq}) we have dropped the thermal pressure force 
because it is typically much smaller than the magnetic pressure force.
The ratio of the ion+electron to magnetic pressure is
\be
\label{plasratio}
\frac{\nni \kB (\Ti + \Te)}{(B^{2}/8\pi)} = 1.39 \times 10^{-6} \left(\frac{50~\mu{\rm G}}{B}\right)^{2}
\left(\frac{\nn}{10^{4}~\cc}\right)\left(\frac{\xxi}{10^{-7}}\right)\left(\frac{\Ti + \Te}{10^{3}~{\rm K}}\right),
\ee
where $\xxi = \nni/\nn$ is the fractional ionization 
and we have normalized quantities to representative values for a shock wave.
We conclude that thermal pressure of the ions and electrons is indeed 
negligible for the conditions of interest here. 

The source terms \Sn, \Fn, and $\Gn-\Ln$ are the net rates per unit volume
at which mass, momentum, and thermal energy are added to the neutral
gas,
where \Ln\ is rate of radiative cooling and \Gn\ is the net rate of heating
by all other processes.
In \S\ref{sec-models} we describe benchmark calculations with various assumptions
about the source terms.
In all cases momentum exchange is dominated by elastic ion-neutral scattering with
\be
\label{Fneq}
\Fn \approx \Fnel = -\frac{\rhon}{\tni}(\vn - \vi) ~~. 
\ee
The time scale for drag to accelerate the ion-electron fluid is
\be
\label{tineq}
\tin = \frac{(\mi + \mn)}{\rhon \sigwin}\left(1 + \left[\frac{\siggeo|\vn - \vi|}{\sigwin}\right]^{2}\right)^{-1/2}~~,
\ee
where $\sigwin$ is the Langevin collision rate (Giousmousis \& Stevenson 1958;
Flower 2000) and 
$\siggeo$ is the geometric cross section.
It follows from Newton's 3rd Law that the time to accelerate the {\em neutral}\/ gas is
\be
\label{tnieq}
\tni = (\rhon/\rhoi)\tin~~ \gg \tin.
\ee
The ion-neutral and neutral-ion drag times, \tin\ and \tni\ respectively, are fundamental time
scales for the flow.

\subsection{The Spit Operator Method}
\label{sec-splitop}

The system of governing PDEs (\ref{neutmassconteq})-(\ref{inducteq}) has the 
conservative form
\be
\label{conssystemeq}
\frac{\partial \Uvec}{\partial t} + \frac{\partial \Fvec}{\partial x} = \Svec~~,
\ee
where $\Uvec \equiv \left[\Uvecn~,~\Uveci\right]^{\rm T}$ is the array of conserved 
dependent variables, $\Fvec \equiv \left[\Fvecn~,~\Fveci\right]^{\rm T}$ is the 
corresponding array of fluxes, and 
$\Svec \equiv \left[\Svecn~,~\Sveci\right]^{\rm T}$ is the array of source terms, with
\bea
\label{Uvecndef}
\Uvecn &\equiv& \left[\rhon~,~\rhon\vn~,~\En\right]^{\rm T}~~, \\
\label{Uvecidef}
\Uveci &\equiv& \left[\rhoi~,~\rhoi\vi,~B\right]^{\rm T}~~, \\
\label{Fvecndef}
\Fvecn &\equiv& \left[\rhon\vn~,~\rhon\vn^{2} + \Pn~,~(\En + \Pn)\vn\right]^{\rm T}~~, \\
\label{Fvecidef}
\Fveci &\equiv& \left[\rhoi\vi~,~\rhoi\vi^{2} + B^{2}/8\pi,~B\vi\right]^{\rm T}~~, \\
\label{Svecndef}
\Svecn &\equiv& \left[\Sn~,~\Fn~,~\Fn\vn - \frac{1}{2}\Sn\vn^{2}+\Gn-\Ln\right]^{\rm T}~~,\\
\label{Svecidef}
\Sveci &\equiv& \left[-\Sn~,~-\Fn~,~0\right]^{\rm T}~~.
\eea
There are several different ways to solve system (\ref{conssystemeq}). 
In this paper we apply a scheme known as {\em operator splitting},
an authoritative discussion of which can be 
found in Toro (2009; see Ch.~15).
The basic idea is to split the full problem into two parts,
called the homogeneous and inhomogeneous sub-problems.
In the former one solves the hyperbolic system
\be
\label{homogeneousprob}
\frac{\partial \Uvec}{\partial t} + \frac{\partial \Fvec}{\partial x} = 0~~
\ee
on a discretized computational domain 
using, say, a Riemann-Godunov (RG) solver (see \S\ref{sec-algorithm}).
In the latter one solves
\be
\label{sourceprob}
\frac{d \Uvec}{d t} = \Svec
\ee
using, say, a Runge-Kutta algorithm for ODEs.

The accuracy of the resulting solution depends on the manner in which the
sub-problems are coordinated.
For example, $\Uvec(x,t)$ can be advanced from time $t$ to $t+\delt$ with first-order
accuracy in \delt\ by setting
\be
\label{splitopeqa}
\Uvec(x,t + \delt) = \sourceopa \RGop \left[\Uvec(x,t)\right]~~,
\ee
where the operator $\RGop$ advances the solution from time $t$ to $t+\delt$
by solving (\ref{homogeneousprob}).
The result is then fed to $\sourceopa$, which advances 
the solution from $t$ to $t+\delt$ by solving (\ref{sourceprob}).
Second-order accuracy can be attained by using what is sometimes referred to as 
``Strang splitting" (Strang 1968; Toro 2009),
\be
\label{splitopeqb}
\Uvec(x,t + \delt) = \sourceopb \RGop \sourceopb \left[\Uvec(x,t)\right]~,
\ee
where the inhomogeneous sub-problem is advanced by a ``predictor'' half step $\delt/2$,
the homogeneous sub-problem is advanced by a full step $\delt$, and the
inhomogeneous sub-problem is integrated again over another half step.
We use Strang splitting in our algorithm (see \S\ref{sec-algorithm}).

In the operator-splitting method, the homogeneous subproblem 
(\ref{homogeneousprob}) can be further separated into 
independent Riemann-Godunov problems for each fluid:
\bea
\label{hydroRGprob}
\frac{\partial \Uvecn}{\partial t} + \frac{\partial \Fvecn(\Uvecn)}{\partial x} &=&0~~,\\
\label{mhdRGprob}
\frac{\partial \Uveci}{\partial t} + \frac{\partial \Fveci(\Uveci)}{\partial x} &=&0~~.
\eea
This means that the neutral gas and ion-electron fluid do not directly influence 
one another in either of their separate RG problems. The two fluids are therefore
{\em dynamically uncoupled} during this stage of the calculation. 
Operationally, 
this has the advantage that existing well-developed numerical techniques (such as 
approximate Riemann solvers for gas dynamics) can be applied readily to 
RG problem (\ref{hydroRGprob}) for the neutral gas. 
In Appendix \ref{sec-approxmhdriemann} we describe an accurate MHD
Riemann solver for RG problem (\ref{mhdRGprob}) for the ion-electron fluid.
Thus there is an overall symmetry to the 
solution of the two separate Riemann-Godunov problems when the operator 
splitting scheme is used. Exploiting this
symmetry greatly enhances the efficiency, and simplifies the writing, of a numerical code 
designed to model multifluid shock waves. Another advantage of our scheme is that by having RG
problem (\ref{mhdRGprob}) as one of the governing equations, the inherent MHD
hyperbolic (i.e., wave) structure is unambiguously built into the time evolution.
That is, the inertia of the ion-electron fluid is always included.
Of course the neutral gas and ion-electron fluid do influence one another. 
{\em Dynamical recoupling} of the fluids occurs during the source integration stage 
(\ref{sourceprob}) of the split-operator method. As we show in our benchmark tests (\S~3), 
the interaction between the fluids is indeed accurately accounted for during this
stage of the calculation.

\subsection{Outline of the Algorithm}
\label{sec-algorithm}

Models are calculated on a spatial domain consisting of a set of $N$ fixed 
mesh points $\{\xj, j = 1,\cdots,N\}$ with uniform spacing $\delx = \xj - \xjmo$.
Variables are calculated at $\xj$. Located midway between cell $j$ (centered about $\xj$) 
and cell $j+1$ (centered about $\xjpo$) is the cell face $\xjphlf = \xj + \delx/2$.

To advance a model in time, a stable numerical time step $\delt$ has to be chosen.
One possibility is to impose the Courant-Friedrichs-Lewy (CFL, e.g., Toro 2009) 
condition at each mesh point $j$,
\be
\label{deltCFLeq}
\deltCFL = \frac{\nu \delx}{\lamj}~,
\ee 
where $\lamj$ is the maximum (for all wave modes including both fluids) 
wave speed at grid point $\xj$ and $\nu$ is a number such that $0 < \nu < 1$.  
In our application it is typically the case 
that the MHD wave speeds far exceed those of the neutral gas.
Therefore we set
\be
\label{lamjeq}
\lamj = (|\vi| + \vims)_{j}~~,
\ee
where $\vims$ is the ion magnetosound speed.
When ion-electron pressure is neglected, $\vims = \viA$, where the ion {\Alf} speed
\be
\label{viAdef}
\viA \equiv \frac{B}{\sqrt{4\pi \rhoi}}
= 6.90 \times 10^{2} \left(\frac{B}{50~\mu{\rm G}}\right)\left(\frac{25~\mprot}{\mi}\right)^{1/2}
\left(\frac{10^{-7}}{\xxi}\right)^{1/2}\left(\frac{10^{4}~\cc}{\nn}\right)^{1/2}~\kms
\ee
($\mprot$ is the proton mass).
However the source terms $\Svecn$ and $\Sveci$ also contain important time scales, 
including the collisional drag times, $\tin$ and $\tni$, as well as scales
related, e.g., to the heating and cooling of the gas. Let $\deltsource$ be
some fraction of the smallest time scale associated with the source terms at
$\xj$.
To have a stable numerical time integration throughout the entire computational 
mesh, it is then necessary that the time step be such that
\be
\label{comptimestep}
\delt \leq \min[\deltsource, \deltCFL]~.
\ee
Once a stable step size has been determined, integration proceeds as given by
equation (\ref{splitopeqb}) in the fashion described in the subsections below. 

As discussed early on by Paleologou \& Mouschovias (1983, see their App. A), 
large disparities exist between the magnitudes of flow time scales and collisional 
time scales in the multifluid MHD equations rendering the governing equations
mathematically ``stiff". This means that $\delt$ will often have to be less than
the smallest physical time scale in a system containing other time scales which
are much greater in size. In this circumstance a large number of small time steps
will then be required to follow a state that is evolving on a much longer 
natural time scale. An extreme example occurs in the third benchmark test 
presented in section \S~\ref{sec-thirdwavemodel} which is followed in its 
development until several neutral-ion collision times $\tni$ have elapsed, up
to a time $\approx 2.6 \times 10^{5}~\yr$ ($\sim 10$ to 100 times greater
than the ages of shocks in star-forming regions that we intend to study --- see 
\S~\ref{sec-intro}). To stably integrate that model to that time using an explicit method,
$\delt$ had to be kept below the ion-neutral time scale 
$\tin \approx 1.2 \times 10^{-2}~\yr$; $\delt = 0.4\tin$ was actually used for that model.
Thus, $\sim 8 \times 10^{7}$ time steps were taken to reach completion. While this is indeed
a large number of computational steps, it is not prohibitively so with modern computational 
methods. 

\subsubsection{Step 1: First Source Integration}
\label{sec-firstsourcestep}

Given some initial values $\Uvec(x,t)$, system (\ref{sourceprob}) is advanced
in time by $\delt/2$. 
This can be carried out using well-known and tested numerical integration methods such as 
Runge-Kutta schemes, or explicit multistep schemes, etc.\ (e.g., see Ch.\ 6 of Atkinson 
1989; or Ch.\ 16 of Press et al.\ 1996). For the benchmark
calculations presented in \S~3, we used a second-order Runge-Kutta integrator. 
The result of Step~1 is an updated state vector $\Uveco = [\Uvecno~,~\Uvecio]^{\rm T}$
which becomes the input for Step~2. 

\subsubsection{Step 2: Riemann-Godunov Integration}
\label{sec-RGstep}

In Step~2 the dynamically decoupled RG problems (\ref{hydroRGprob}) and (\ref{mhdRGprob})
are advanced a full time step. 
The solution for the conserved dependent variables at the end of this integration 
step, $\Uvect = [\Uvecnt~,~\Uvecit]^{\rm T}$, is given by
\be
\label{RGnumeric}
\Uvecbetat(\xj) = \Uvecbetao(\xj) - \frac{\delt}{\delx}\left(\Fvecbeta(\xjphlf) - 
\Fvecbeta(\xjmhlf)\right)~,
\ee
with $\beta = {\rm n, i}$. 

The variables are known at the grid points, \xj, but the fluxes in equation (\ref{RGnumeric})
are required at the cell faces.
Second-order
accuracy in space and time is attained by first ``reconstructing" the data and then interpolating
the variables from the grid points to the cell faces using total variation diminishing (TVD) slope limiter methods
such as those employed in the  MUSCL-Hancock scheme (Toro 2009). 
For any dependent variable $q$ in the cell centered about 
$\xj$, reconstruction and interpolation to the cell faces are carried out by setting
\be
\label{qinterp}
\qR = q(\xj) + \frac{1}{2} \Deltaj~,~\qL = q(\xj) - \frac{1}{2} \Deltaj~~,
\ee
where R refers to the right face of the cell $j$ (i.e., $\xjphlf - \delta$, 
$\delta \rightarrow 0$) and L refers to its left face (i.e., $\xjmhlf+\delta$, 
$\delta \rightarrow 0$). $\Deltaj$ is the ``limited average slope value'' of $q$. 
There are many different kinds of slope limiter.
For example, the MINBEE slope limiter is
\bea
\label{MINBEEslope}
\Deltaj = \left\{\begin{array}{ll} \max\left[0,\min(\Deltajmhlf,\Deltajphlf)\right] ~~{\rm if}~~ \Deltajphlf > 0, \\
\min\left[0,\max(\Deltajmhlf,\Deltajphlf)\right]~~{\rm if}~~ \Deltajphlf < 0,
\\
\end{array} 
\right. \\
\Deltajphlf \equiv q(\xjpo)-q(\xj)~,~ \Deltajmhlf \equiv q(\xj)-q(\xjmo)
\eea
(Toro 2009), and a generalized version of the monotonic slope limiter of van Leer (1979) is
\bea
\label{minmodslope}
\Deltaj 
&=& \sigma_{j} \min\left[\Theta|\Deltajphlf|~,~\frac{1}{2}|\Deltajphlf+\Deltajmhlf|~,~\Theta|\Deltajmhlf|\right], 
~~1 \leq \Theta \leq 2~,\\
\label{sigmaeq}
\sigma_{j} 
&=&\left\{\begin{array}{ll} \hspace{0.8em} 1 ~~{\rm if}~~\Deltajmhlf > 0~,~\Deltajphlf > 0~, \\
-1 ~~{\rm if}~~\Deltajmhlf < 0~,~\Deltajphlf < 0~, \\
\hspace{0.8em} 0 ~~{\rm otherwise.}
\end{array}
\right.
\eea
We normally use the van Leer limiter but both are easy to implement and yield 
good results. We also impose the positivity conditions of Waagan (2009) to the slope limiting and 
interpolation procedure for the neutral gas, which ensures that $\rhon$, $\Pn$, $\Tn$ and 
$\enint$ are always $> 0$ at all $\xj$, even for very hypersonic flows.

Another TVD-limiter which can be used for the data reconstruction of the charged fluid and 
magnetic field variables is the one derived by {\Cada} \& Torrilhon (2009)
It has the form:
\be
\label{Cadareconstruction}
\qR = q(x) + \frac{1}{2}\phi({\cal R}_{j})\Deltajphlf~~,~~
\qL = q(x) - \frac{1}{2}\phi({\cal R}_{j}^{-1})\Deltajmhlf~,
\ee
where
\be
\label{phidef}
{\cal R}_{j} \equiv \frac{\Deltajmhlf}{\Deltajphlf}
\ee
and
\be
\label{Cadalimiter}
\phi({\cal R}) =
\max\left[0,\min\left(\frac{2+{\cal R}}{3},\max\left[-0.5{\cal R},\min\left(2{\cal R},
\frac{2+{\cal R}}{3},1.6\right)\right]\right)\right]~;
\ee
for flows with smooth minima and maxima, we have found that this reconstruction 
method and limiter function produces smaller relative errors, including the regions
about the extrema, when compared to that which results from (\ref{qinterp})
with the limiters (\ref{MINBEEslope}) and (\ref{minmodslope}). The {\Cada}
\& Torrilhon (2009) scheme is used in the eigenmode convergence test models
presented in Appendix \ref{sec-convergence}.

Integrating equations (\ref{hydroRGprob}) and (\ref{mhdRGprob}) over a virtual (predictor)
time increment $\delt/2$ further improves the left and right cell face values in each 
cell:
\bml
\bea
\label{barrRpredictor}
\UvecRbar = \UvecR - \frac{\delt}{2\delx}\left(\Fvecbeta(\UvecR)-\Fvecbeta(\UvecL)\right), \\
\UvecLbar = \UvecL - \frac{\delt}{2\delx}\left(\Fvecbeta(\UvecR)-\Fvecbeta(\UvecL)\right) 
\eea
\eml
(Toro 2009). 

Data reconstruction and interpolation yields variable pairs \{$\UvecRbar$, $\UvecLbarpo$\} at 
$\xjphlf$ and \{$\UvecRbarmo$, $\UvecLbar$\} at $\xjmhlf$, which can be used to calculate the 
fluxes $\Fvecbeta(\xjphlf)$ and $\Fvecbeta(\xjmhlf)$ needed in equation (\ref{RGnumeric}). 
Calculation of the fluxes is accomplished by solving the Riemann problem at the cell
faces using the method of Godunov (1959; for an extremely comprehensive discussion 
see Toro 2009). Solution of the RG problems at each cell interface is carried out using
approximate Riemann solvers, one for the neutral gas and another for the ion-electron fluid.
The approximate MHD Riemann solver used for the charged fluid is derived in 
Appendix~\ref{sec-approxmhdriemann}. For the neutral gas we use the 
approximate gas dynamic Riemann solver described in \S~5 of Almgren et al.\ 
(2010; reportedly based on unpublished work 
by P. Colella, 1997). The latter solver is similar to the approximate
Riemann solver presented in Toro (2009, Ch.\ 9), but has been extended to RG problems
that also include nonideal gases (e.g., Colella \& Glaz 1985). Using the
solver of Almgren et al. allows the modeling of fluids having nonideal 
equations of state, or those that are not in LTE.
Our approximate gas dynamic Riemann solver is very similar in structure
to our approximate MHD Riemann solver. (Hence the general symmetry
of the problem when using the split-operator method, as noted above.)

The approximate gas dynamic and MHD Riemann solvers are used to obtain the (Godunov) fluxes 
at the faces of each cell, $\Fvecbeta(\xjphlf)$ and $\Fvecbeta(\xjmhlf)$.
The details of how this is done for the ion-electron fluid MHD problem are 
presented in Appendix~\ref{sec-approxmhdriemann}, and the details for the neutral gas
problem follow by analogy (or, see \S~5 of Almgren et al.\ 2010). These values are
then inserted into equation (\ref{RGnumeric}), thereby giving $\Uvecbetat$ and
$\Uvect$. The updated variables $\Uvecbetat$ are the input for Step~3.

\subsubsection{Step 3: Final Source Integration} 
\label{sec-finalsourcestep}

\begin{figure}
\epsscale{0.77}
\plotone{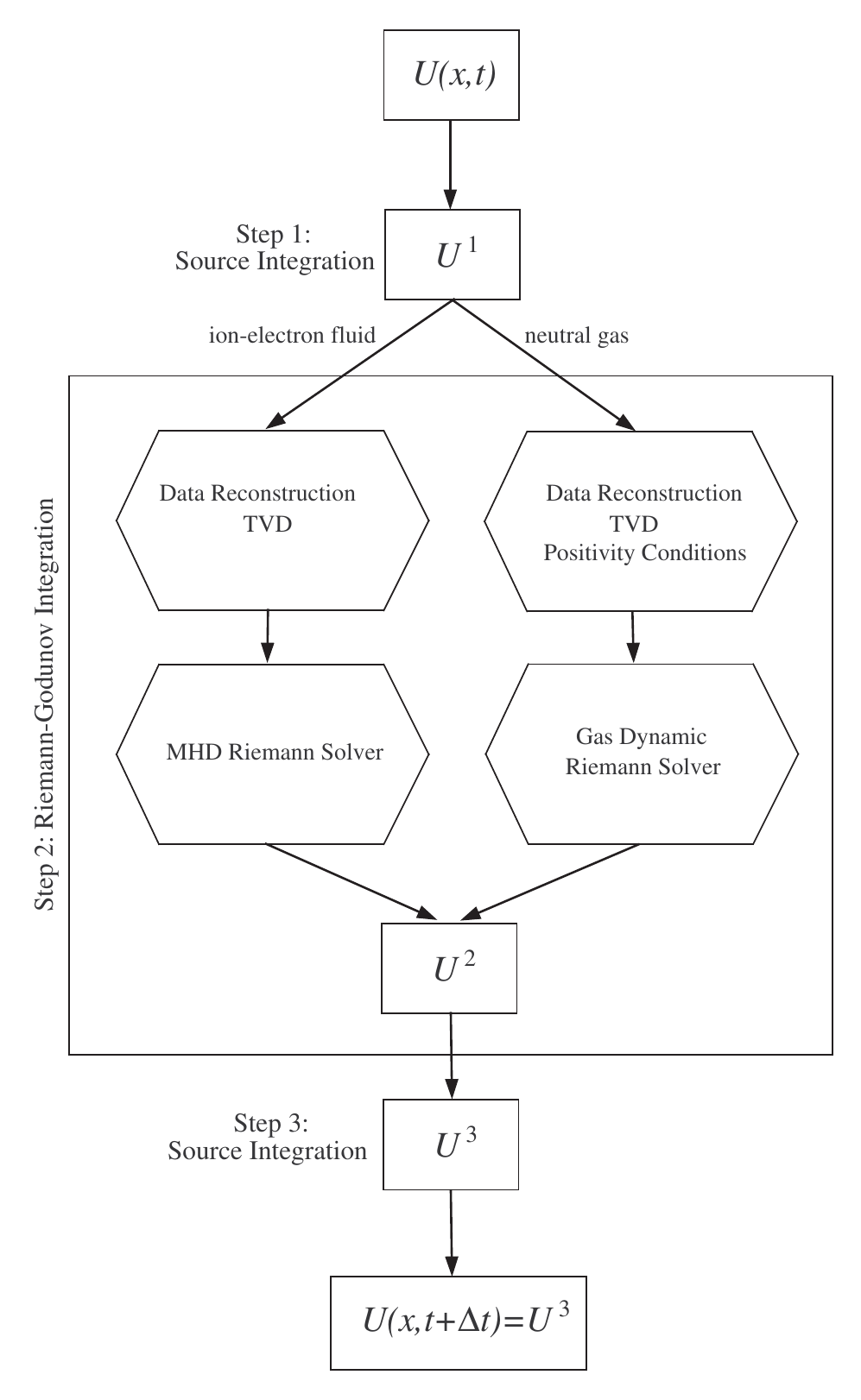}
\caption{Schematic diagram of our split-operator algorithm to simulate multifluid
MHD shocks and related flows in interstellar clouds and cores.
\label{fig-schematic}
}
\end{figure}

Another source integration of size $\delt/2$ is performed as in Step 1. 
This yields the last update of the physical variables, $\Uvecth = [\Uvecnth~,~\Uvecith]^{\rm T}$. 
The split-operator integration scheme is thus completed with the second-order accurate (in 
space and time) final result $\Uvec(x,t+\delt) \doteq \Uvecth$.
A schematic of our algorithm is presented in Figure~\ref{fig-schematic}.  
Because of the modular and general nature of the algorithm we have 
described, altering or tailoring a multifluid MHD split-operator code to one's specific 
needs, if necessary, would not be difficult. For instance, if one wishes to use a 
different method for interpolation of the variables, such as a piecewise parabolic method 
(PPM; Colella \& Woodward 1984), or a weighted essentially nonoscillatory method (WENO; 
Liu, Osher, \& Chan 1994), instead of the MUSCL-Hancock TVD scheme we describe, the 
alternate method can be substituted in the indicated data reconstruction substeps for both
the neutral gas and ion-electron fluids.


\section{Benchmark Tests}
\label{sec-models}

In this section we present some numerical calculations which establish the validity of 
the split-operator method for multifluid MHD shocks and other flows in weakly ionized
interstellar clouds. The accuracy of the algorithm is tested by comparing 
our numerical results to accurate analytic solutions. Data regarding the convergence and 
second-order scaling of our algorithm are presented in Appendix \ref{sec-convergence}. 
In all of the test models the neutrals are assumed to be an ideal gas of $\Htwo$ 
molecules having a ratio of specific heats $\gamma = 5/3$, and the ion mass $\mi$
is set to a generic value of 25$\mprot$, which could represent ions such as $\rm{HCO}^+$
and $\rm{Na}^+$. For momentum transfer by elastic ion-neutral collisions we use a Langevin collision rate 
$\sigin = 1.7 \times 10^{-9}~{\rm cm^{3} s^{-1}}$ (McDaniel \& Mason 1973)
and geometric cross section
$\siggeo = \pi(r_{\rm H_{2}} + r_{\rm i})^{2} = 2.86 \times 10^{-15}~{\rm cm^{2}}$
($r_{\rm H_{2}}$ and $r_{\rm i}$ are the molecular and ionic radii, respectively).

\subsection{Linear Wave Packets}
\label{sec-linwaves}

In this test we set the source terms $\Sn$, $\Gn$, and $\Ln$ to zero in equations
(\ref{neutmassconteq}) - (\ref{inducteq}) and assume that momentum transfer
is entirely due to elastic ion-neutral scattering.
We initiate small-amplitude disturbances in the plasma and follow the subsequent evolution of the 
charged and neutral fluids. As discussed in Ciolek \& Roberge (2002, \S~2.3.2) and 
Mouschovias et al. (2011, \S~3.2), there exist two important length scales 
relevant to MHD waves propagating perpendicular to the magnetic field. 
The first length scale is the ion magnetosound wave upper cutoff 
$\Lims = 4 \pi \vims \tin$. Ion magnetosound waves with wave speed $\vims = \viA$ can
propagate when their wavelength is less than $\Lims$. The second length scale is 
the neutral magnetosound wave lower cutoff length scale $\Lnms = \pi \vnA^{2} \tni/\vnms$.
Magnetosound waves in the neutral fluid travel at the neutral magnetosound speed,
\be
\label{vnmsdef}
\vnms = \left( \Cs^{2} + \vnA^{2} \right)^{1/2}~,
\ee
where
\be
\label{Csdef}
\Cs \equiv \left(\frac{\partial \Pn}{\partial \rhon}\right)^{1/2}_{\rm \sn}
= \left(\frac{\gamma \Pn}{\rhon}\right)^{1/2}
= 0.262 \left(\frac{\gamma}{5/3}\right)^{1/2} 
\left(\frac{T}{10~{\rm K}}\right)^{1/2}
\left(\frac{2~\mprot}{\mn}\right)^{1/2}~\kms 
\ee
is the adiabatic 
sound speed (the derivative is taken at constant neutral entropy $\sn$), and
\be
\label{vnAdef}
\vnA \equiv \frac{B}{\sqrt{4 \pi \rhon}} = 0.771 \left(\frac{B}{50~\mu{\rm G}}\right)
\left(\frac{2~\mprot}{\mn}\right)^{1/2} \left(\frac{10^{4}~\cc}{\nn}\right)^{1/2}~\kms
\ee
is the neutral {\Alf} speed.
Neutral magnetosound waves can propagate for wavelengths greater 
than $\Lnms$. At wavelengths between $\Lims$ and $\Lnms$ there is no MHD
wave propagation; any mode excited on these scales produces 
{\em ambipolar diffusion} of the ions, electrons, and magnetic field through the neutral gas.

To selectively excite different wave modes we impose Gaussian initial perturbations in the 
ion density and magnetic field of the form
\bea
\label{Bperturb}
B(x,0)&=&B_{0}\left[1 + \ampl \exp\left(\frac{-x^{2}}{\lgauss^{2}}\right)\right] \\
\label{iperturb}
\nni(x,0)&=&\nnio \frac{B(x,0)}{B_{0}} ~~.
\eea
The charged fluid is initially taken to be stationary with $\vi(x,0) = 0$. The 
neutral fluid is initially uniform and at rest, with $\nn(x,0)=\nno$, 
$\Pn(x,0)=\Pno$, and $\vn(x,0)=0$. Quantities with a ``0" subscript are constant. 
$\ampl$ is the dimensionless amplitude of the perturbation, and $\lgauss$ is 
its width; wave modes excited by this perturbation will have wavelengths 
clustered about $\lgauss$. 

Below we present three benchmark tests having 
different packet widths, $\lgauss$, but the same perturbation amplitude $\ampl = 0.01$.
All three tests have $\nno = 2 \times 10^{4}~\cc$ and 
$x_{\rm i0} = \nnio/\nno = 3.16 \times 10^{-8}$.
All particle species are assigned the same reasonable but somewhat
arbitrary temperature: $T_{\rm n0} = T_{\rm i0} = T_{\rm e0} = 10~{\rm K}$. 
The unperturbed magnetic field strength is $B_{0} = 50~\mu{\rm G}$, yielding an 
ion magnetosound speed $\vims = \viA = 868~\kms$, and a neutral {\Alf} speed 
$\vnA = 0.545~\kms$. The sound speed $\Cs = 0.262~\kms$ and the neutral 
magnetosound speed $\vnms = 0.605~\kms$. 
It follows that $\tin = 1.26 \times 10^{-2}~\yr$, 
$\tni = 3.19 \times 10^{4}~\yr$, $\Lims = 4.33 \times 10^{14}~\cm$, and 
$\Lnms = 1.55 \times 10^{17}~\cm$. 

\subsubsection{Wave Packet 1: $\lgauss = 9.35 \times 10^{11}~\cm$}  
\label{sec-firstwavemodel}

Because this wave packet has $\lgauss \ll \Lims$, we expect that
the initial perturbation will generate two traveling wave packets in the ions
and the magnetic field, one traveling leftward from the origin and the other rightward. 
Although acoustic waves in the neutrals can exist on these scales (see, e.g., Fig. 1 of Ciolek 
et al.\ 2004 or Fig.\ 6 of Mouschovias et al.\ 2011), they will not be excited by this 
perturbation: the neutrals are initially uniform and motionless and the neutral-ion
drag time, {\tni}, is much longer than the decay time scale $\tau_{\rm dec}$ for the waves 
(see below). We expect that in this model the ion and magnetic field wave packets will 
propagate through a stationary neutral fluid.
   
\begin{figure}
\epsscale{0.90}
\plotone{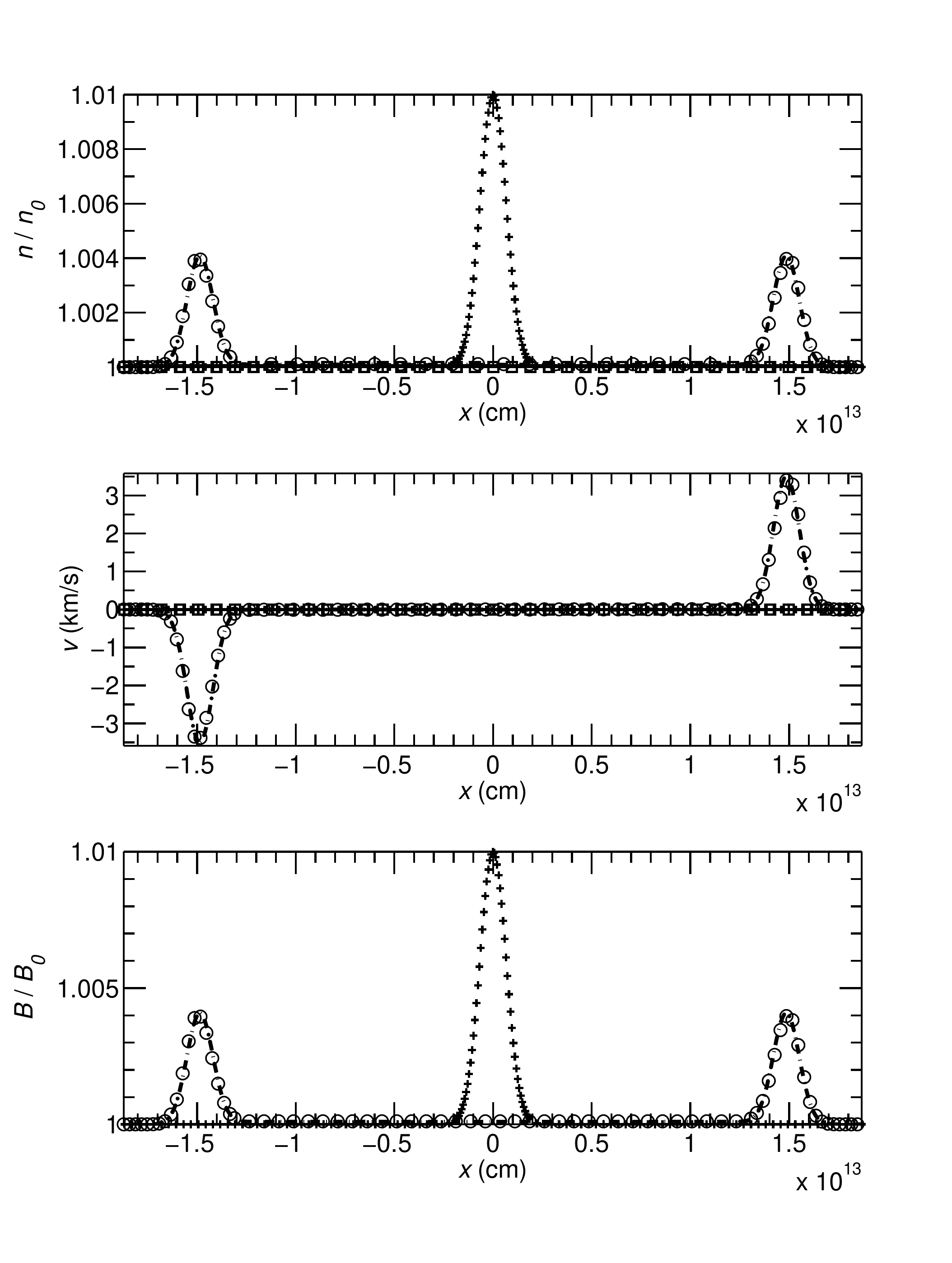}
\vspace{-3.2ex}
\caption{Model results for the initial perturbation (\ref{Bperturb}), (\ref{iperturb}),
with $\lgauss = 9.35 \times 10^{11}~\cm$ and $\ampl = 0.01$, shown at time 
$t = 5.42 \times 10^{-3}~\yr$. Squares indicate initial values for the neutral gas in each
panel. Initial ion and magnetic field curves are crosses. Neutrals at time $t$ are the solid 
lines, and the ions are circles. For clarity some data points have been omitted. Also shown 
(dash-dot lines) are the analytic wave solutions (\ref{Bwave})-(\ref{viwave}). {\it Top}:
number density of the neutrals and ions, normalized to their values in the unperturbed 
state. {\it Middle}: fluid velocities. {\it Bottom}: magnetic field, 
normalized to its unperturbed value.
\label{fig-firstwave}
}
\end{figure}

Given the assumption of stationary neutrals, analytic expressions for the small-amplitude
wave packets in the ion-electron fluid, which are expected to occur in this model, 
can be derived from the governing equations (\ref{neutmassconteq})-(\ref{inducteq}) in the limit where
the wavelengths ($\sim \lgauss$) of the Fourier components comprising the packets are 
$\ll \Lims$.   For the initial conditions (\ref{Bperturb}), 
(\ref{iperturb}), and $\vi(x,0)=0$, the solutions are 
\bml
\bea
\label{Bwave}
B_{\rm ims}(x,t)&=&B_{0}\left[ 1 + \frac{\ampl}{2}\exp\left(-\frac{t}{2\tin}\right)\left\{\exp\left(\frac{-(x-\vims t)^{2}}{\lgauss^{2}}\right)
+ \exp\left(\frac{-(x+\vims t)^{2}}{\lgauss^{2}}\right)\right\} \right. \nonumber \\ 
& & \hspace{1.5em} 
\left. + \frac{\sqrt{\pi}\ampl \lgauss}{8\vims\tin}\exp\left(-\frac{t}{2\tin}\right)
\left\{{\rm erf}\left(\frac{x+\vims t}{\lgauss}\right)
- {\rm erf}\left(\frac{x-\vims t}{\lgauss}\right)\right\} \right]~, \hspace{2em} \\
\label{iwave}
\nni_{\rm ,ims}(x,t) &=&\nnio \frac{B_{\rm ims}(x,t)}{B_{0}}~, \\
\label{viwave}
\vi_{\rm ,ims}(x,t)&=&\frac{\ampl}{2} \vims \exp\left(-\frac{t}{2\tin}\right)\left[\exp\left(\frac{-(x-\vims t)^{2}}{\lgauss^{2}}\right)
- \exp\left(\frac{-(x+\vims t)^{2}}{\lgauss^{2}}\right)\right]~~
\eea
\eml
(see Appendix \ref{sec-imssolns}).
As expected, the solution consists of two traveling Gaussian 
packets, with propagation velocities $\pm \vims$ and amplitudes $\ampl/2$.
Momentum exchange (i.e., friction) with the fixed neutrals causes the pulses to decay exponentially
on a  characteristic time scale $\tau_{\rm dec} = 2 \tin$. The decay time exceeds \tin\ by
a factor of 2 because of the equipartition of magnetic and kinetic energy in 
the ion magnetosound waves which make up the wave packets. 
Elastic collisions with the neutrals directly reduce the kinetic energy of the ion-electron fluid;
the energy stored in the magnetic field is reduced indirectly as it is 
gradually converted into ion-electron kinetic energy.  

Figure \ref{fig-firstwave} shows the results for this model 
at $t = 5.42 \times 10^{-3}~\yr$ using our fully nonlinear split-operator method 
code on a mesh with 5000 grid points. Also shown in each panel are the analytic
wave solutions (\ref{Bwave})-(\ref{viwave}) evaluated at the same time. The
split-operator code results are in excellent agreement with the analytic solutions, 
with the relative error in the ion density having a maximum value of 
$2.5 \times 10^{-4}$. The split-operator scheme we have employed has
accurately reproduced all of the fundamental aspects of the physical evolution for
this model. This includes the propagation of the magnetosound pulses, the 
dependence of their speed on the magnetic field strength and inertia of the ions
(inherent to the solution from the RG step of the integration), as well as 
the decay of the two pulse peaks caused by ion-neutral friction (which can only result from the 
source integration steps).
Although the ions and neutrals are dynamically decoupled during the RG stage of the integration, 
recoupling and realistic interaction of the two fluids are faithfully 
reproduced during the source integration stages of the algorithm.

\subsubsection{Wave Packet 2: $\lgauss = 3.74 \times 10^{15}~\cm$}
\label{sec-secondwavemodel}

This wave packet has $\Lims \ll \lgauss \ll \Lnms$. As a result, there will be no propagating 
MHD waves in this case. Instead, the initial perturbation will give 
rise to an ambipolar diffusion mode, in which the ions, electrons, and magnetic field 
diffuse outward from the origin through a neutral fluid which is still effectively 
stationary on the length and time scales characterizing this particular model. The 
motion of the ions is essentially inertialess or force-free; that is, the 
driving magnetic pressure force is almost exactly balanced by the retarding
ion-neutral friction (\ref{ionmtmeq}).
The linearized ambipolar diffusion mode is found to have the analytic solutions
\bml
\bea
\label{Bdiff}
B_{\rm ad}(x,t) &=& B_{0} \left[1 + \frac{\ampl}{(1 + 4 \Di t/\lgauss^{2})^{1/2}}
\exp\left(\frac{-x^{2}/\lgauss^{2}}{1 + 4 \Di t/\lgauss^{2}}\right)\right] \nonumber \\
& &  + \frac{2 B_{0} \ampl \Di \tin/\lgauss^{2}}{(1+4\Di t/\lgauss^{2})^{3/2}}
\left[1 - \frac{2 x^{2}/\lgauss^{2}}{1 + 4 \Di t/\lgauss^{2}}\right]
\exp\left(\frac{-x^{2}/\lgauss^{2}}{1+4\Di t/\lgauss^{2}}\right) \nonumber \\
& & - \frac{2 B_{0} \ampl \Di \tin/\lgauss^{2}}{(1-4\Di t/\lgauss^{2})^{3/2}}
\left[1 - \frac{2 x^{2}/\lgauss^{2}}{1 - 4 \Di t/\lgauss^{2}}\right]
\exp\left(\frac{-t}{\tin}\right)
\exp\left(\frac{-x^{2}/\lgauss^{2}}{1-4\Di t/\lgauss^{2}}\right)~, \hspace{2.3em} \\
\label{idiff}
\nni_{\rm ,ad}(x,t) &=& \nnio \frac{B_{\rm ad}(x,t)}{B_{\rm 0}}~~, \\
\label{vidiff}
\vi_{\rm ,ad}(x,t) &=& \frac{2 \Di \ampl x}{\lgauss^{2}(1 + 4 \Di t/\lgauss^{2})^{3/2}}~~ 
\exp\left(\frac{-x^{2}/\lgauss^{2}}{1 + 4 \Di t/\lgauss^{2}}\right) \nonumber \\
& & - \frac{2 \ampl \Di x}{\lgauss^{2}(1 - 4\Di t/\lgauss^{2})^{3/2}}
\exp\left(\frac{-t}{\tin}\right)
\exp\left(\frac{-x^{2}/\lgauss^{2}}{1-4\Di t/\lgauss^{2}}\right)~~
\eea
(a derivation is given in Appendix \ref{sec-diffsoln}).
The quantity
\be
\label{Dicoef}
\Di \equiv  \vims^{2}\tin
\ee
\eml
is the diffusion coefficient of the ions and magnetic field through the neutrals. The
pulse diffuses on the ambipolar diffusion time scale 
$\tau_{\rm dec} = \tau_{\rm ad} = \lgauss^{2}/4\Di = 37.0~\yr$

\begin{figure}
\epsscale{0.90}
\plotone{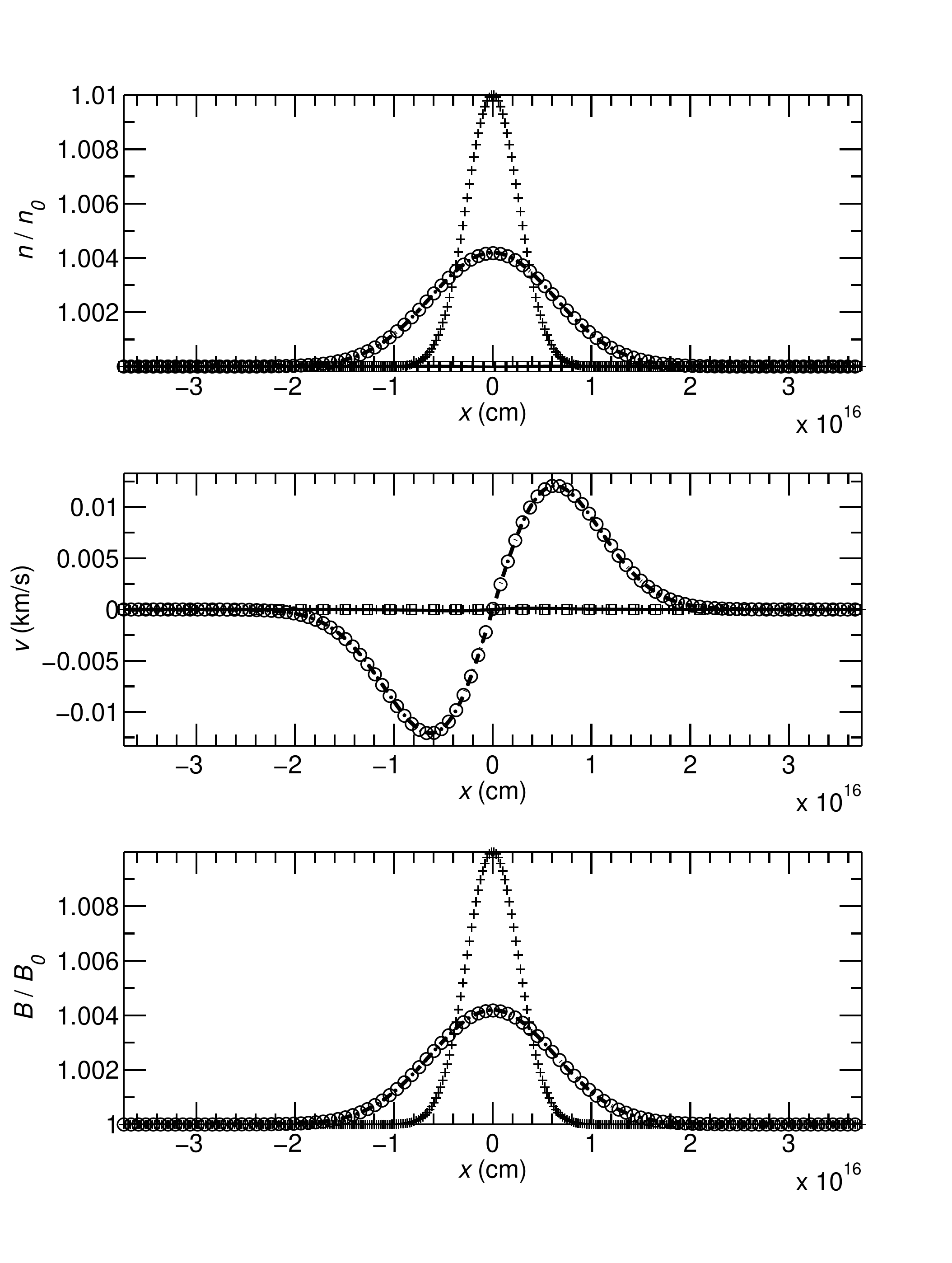}
\caption{As in Fig.\ \ref{fig-firstwave} but for
$\lgauss=3.74 \times 10^{15}~\cm$, and time $t = 174~\yr$. Also shown
in each panel (dash-dot curves) are the analytic ambipolar diffusion 
solutions (\ref{Bdiff})-(\ref{vidiff}) at time $t$.
\label{fig-secondwave}
}
\end{figure}

Figure \ref{fig-secondwave} shows the results of the split-operator method calculation,
in the same format as in Figure \ref{fig-firstwave}. The output time
is $t= 174~\yr$. Also shown in each panel are the analytic ambipolar 
diffusion solutions (\ref{Bdiff})-(\ref{vidiff}) at that same time (dash-dot curves). 
The split-operator numerical results are seen to be in superb agreement with the 
analytic solutions, having a maximum relative error in the ion density 
equal to $1.7 \times 10^{-5}$. The split-operator algorithm has a step (the
RG step) in which both the neutral and ion-electron fluids are described by hyperbolic
PDEs, i.e., PDEs which describe wave propagation. In contrast, the solution
in Fig.~\ref{fig-secondwave} exhibits diffusion, which is described by 
{\em parabolic}\/ PDEs (Courant \& Hilbert 1953; Dennery \& Kryzwicki 1996).
The transition from hyperbolic to parabolic behavior as the packet width increases is made possible
by the source term for momentum exchange between the ions and neutrals, which comes into play during
the source integration steps (Fig. \ref{fig-schematic}).
The combination of source- and RG- integration steps preserves the underlying
physics, which dictates that no propagating MHD waves should exist 
for wave packets with dimensions $\sim \lgauss$ if $\Lims \ll \lgauss \ll \Lnms$.

\subsubsection{Wave Packet 3: $\lgauss = 2.24 \times 10^{17}~\cm$}
\label{sec-thirdwavemodel}

This wave packet has $\lgauss > \Lnms$, which means it will excite 
low-frequency ($< 1/\tni$) neutral magnetosound waves over sufficiently large length and 
time scales. In these waves, magnetic forces on the ions are transmitted to the neutrals 
by collisional drag, allowing the neutrals, ions, and magnetic field to participate in
collective wave behavior. Because $\rhon \gg \rhoi$, the overall inertia of
the collective motion is due to the neutral fluid. The wave speed is therefore the 
neutral magnetosound speed $\vnms$ (eq. [\ref{vnmsdef}]), with the bulk neutral fluid
acting as if it is responding directly to magnetic forces on these scales. 

For wavelengths $\geq \Lnms$ the ions can still be described as being in force-free
motion because $\Lnms \gg \Lims$. Using this fact, we find that the 
Fourier modes for perturbations with $\lgauss \gg \Lnms$ yield the analytic solutions
\bml
\bea
\label{Bnmseq}
\hspace{-2em} B_{\rm nms}(x,t) &=& B_{0}\left[1
+ \frac{\ampl\Gonefunc}{2}\left\{\left(1 - \frac{2(x+\vnms t)\Dth/\lgauss^{2}}{\vnms[1+(2\Di t/\lgauss^{2})]}\right)
\exp\left(\frac{-(x+\vnms t)^{2}}{\lgauss^{2}[1+(2\Di t/\lgauss^{2})]}\right)
\right. \right. \nonumber \\
& & \hspace{7em} \left. +\left(1 + \frac{2(x-\vnms t)\Dth/\lgauss^{2}}{\vnms[1+(2\Di t/\lgauss^{2})]}\right)
\exp\left(\frac{-(x-\vnms t)^{2}}{\lgauss^{2}[1+(2\Di t/\lgauss^{2})]}\right)\right\}
\nonumber \\
& & \hspace{3em}
\left. 
+ \ampl\Gtwofunc\left(\frac{\Cs}{\vnms}\right)^{2}\exp\left(\frac{-x^{2}}{\lgauss^{2}[1 + (4\Dth t/\lgauss^{2})]}\right)
\right] ~~,\hspace{-3em}\\
\label{nninmseq}
\hspace{-4em}\nni_{\rm ,nms}(x,t) &=& \nnio \frac{B_{\rm nms}(x,t)}{B_{0}}~, \\
\label{vnnmseq}
\hspace{-4em} \vn_{\rm ,nms}(x,t) &=&
\frac{\ampl \vnms\Gonefunc}{2}
\left[ 
\left(\frac{2(x+\vnms t)\Dth/\lgauss^{2}}{\vnms[1+(2\Di t/\lgauss^{2})]}-1\right)
\exp\left(\frac{-(x+\vnms t)^{2}}{\lgauss^{2}[1+(2\Di t/\lgauss^{2})]}\right) \right. \nonumber \\
& & \hspace{6em}
\left. +\left(1 + \frac{2(x-\vnms t)\Dth/\lgauss^{2}}{\vnms[1+(2\Di t/\lgauss^{2})]}\right)
\exp\left(\frac{-(x-\vnms t)^{2}}{\lgauss^{2}[1+(2\Di t/\lgauss^{2})]}\right) \right] \nonumber \\
& & - \frac{2 \ampl \Gtwofunc \Dth x}{\lgauss^{2}[1 + (4\Dth t/\lgauss^{2})]}
\exp\left(\frac{-x^{2}}{\lgauss^{2}[1+(4\Dth t/\lgauss^{2})]}\right)~,
\hspace{-3em} \\
\label{nnnmseq}
\hspace{-4em} \nn_{\rm ,nms}(x,t) &=& \nno\left[1 
+ \frac{\ampl\Gonefunc}{2} \left\{
\left(1 - \frac{2(x+\vnms t)\Dth/\lgauss^{2}}{\vnms[1 + (2\Di t/\lgauss^{2})]}\right)
\exp\left(\frac{-(x+\vnms t)^{2}}{\lgauss^{2}[1+(2\Di t/\lgauss^{2})]}\right)\nonumber \right. \right.\\
& & \hspace{8em} 
\left.
+ \left(1 + \frac{2(x-\vnms t)\Dth/\lgauss^{2}}{\vnms[1 + (2\Di t/\lgauss^{2})]}\right)
\exp\left(\frac{-(x-\vnms t)^{2}}{\lgauss^{2}[1+(2\Di t/\lgauss^{2})]}\right)\nonumber 
\right\} \\
& & \hspace{3em}
\left. - \ampl\Gtwofunc
\exp\left(\frac{-x^{2}}{\lgauss^{2}[1+(4\Dth t/\lgauss^{2})]}\right)\right]~, \\
\label{vinmseq}
\hspace{-4em} \vi_{\rm ,nms}(x,t) &=& \vn_{\rm, nms}(x,t) \nonumber \\
& & \hspace{-8em} + \frac{\ampl\Di\Gonefunc}{\lgauss^{2}[1 + (2\Di t/\lgauss^{2})]}
\left[\left(1 + \frac{2(x+\vnms t)\Dth/\lgauss^{2}}{\vnms[1+(2\Di t/\lgauss^{2})]}\right)
(x+\vnms t) -\frac{\Dth}{\vnms}\right]\exp\left(\frac{-(x+\vnms t)^{2}}{\lgauss^2[1+(2\Di t/\lgauss^{2})]}\right) \nonumber \\
& & \hspace{-8em}
+\frac{\ampl \Gonefunc \Di}{\lgauss^{2}[1+(2\Di t/\lgauss^{2})]}
\left[\frac{\Dth}{\vnms} + \left(1 - \frac{2(x-\vnms t)\Dth/\lgauss^{2}}{\vnms[1+(2\Di t/\lgauss^{2})]}\right)(x-\vnms t)\right]
\exp\left(\frac{-(x-\vnms t)^{2}}{\lgauss^{2}[1+(2\Di t/\lgauss^{2})]}\right) \nonumber \\
& & 
+ \frac{2 \ampl \Gtwofunc\Dth x}{\lgauss^{2}[1+(4\Dth t/\lgauss^{2})]}
\exp\left(\frac{-x^{2}}{\lgauss^{2}[1+(4\Dth t/\lgauss^{2})]}\right)~,
\eea
\eml
where
\bea
\label{Gonefuncdef}
\Gonefunc &\equiv& \left[1 + \left(\Cs/\vnms\right)^{2}\right]^{-1}
\left[1 + \left(2 \Di t/\lgauss^{2}\right)\right]^{-1/2}~, \\
\label{Gtwofuncdef}
\Gtwofunc &\equiv& \left[1 + \left(\Cs/\vnms\right)^{2}\right]^{-1}
\left[1 + \left(4 \Dth t/\lgauss^{2}\right)\right]^{-1/2}
\eea
(for a derivation, see Appendix \ref{sec-nmssolns}). Appearing in these expressions is the 
neutral thermal diffusion coefficient
\be
\label{Dthcoef}
\Dth \equiv \left(\frac{\Cs}{\vnms}\right)^{2} \Di = \frac{\Cs^{2} \tni}{1+(\Cs/\vnA)^{2}}.
\ee
The last equality follows from equations (\ref{Dicoef}), (\ref{tnieq}), (\ref{viAdef}), 
(\ref{vnAdef}), and (\ref{vnmsdef}). The analytic solutions (\ref{Bnmseq})--(\ref{nnnmseq})
show that there will be Gaussian wave packets propagating with velocities $\pm \vnms$, 
verifying that it is the inertia of the neutral fluid which determines the rate at which 
the packets propagate. The solutions also reveal that the neutral 
magnetosound pulses decay by ambipolar diffusion on a time scale 
$\tau_{\rm dec} = 2 \tau_{\rm ad} = \lgauss^{2}/2 \Di = 2.66 \times 10^{5}~\yr$; the
decay time is a factor of 2 longer than the diffusion time scale because, once again,
there is equipartition of magnetic and kinetic energy.

\begin{figure}
\epsscale{0.90}
\plotone{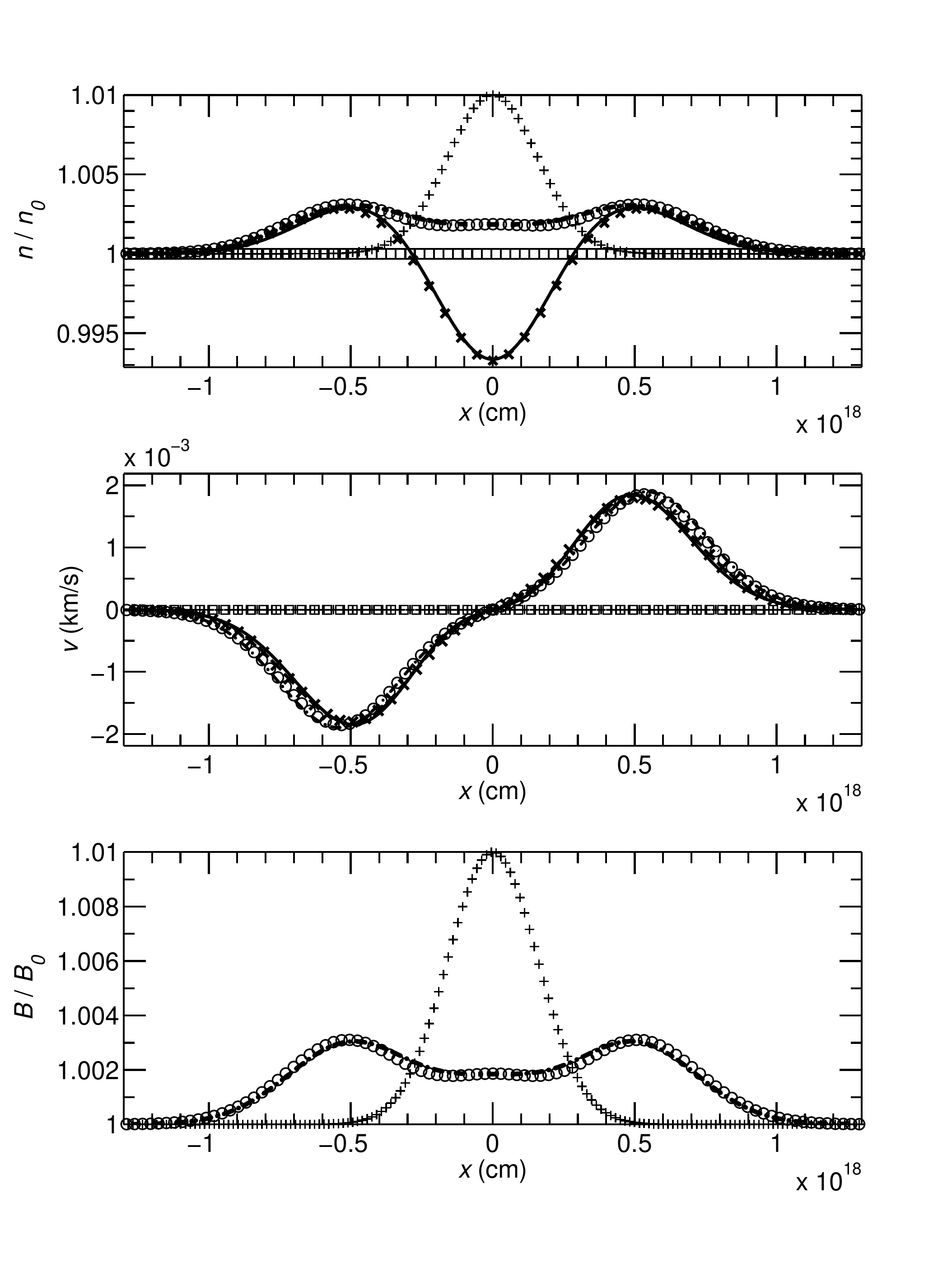}
\caption{As in Fig. \ref{fig-firstwave} but for $\lgauss=2.24 \times 10^{17}~\cm$
and time $t = 2.55 \times 10^{5}~\yr$. {\it Top}: neutral and ion densities. Also
shown are the analytic solutions (\ref{nninmseq}) and (\ref{nnnmseq}) for the ions 
(dash-dot) and the neutrals ($\times$). {\it Middle}: neutral and ion velocities. Included
are the solutions (\ref{vnnmseq}) and (\ref{vinmseq}) for the neutrals ($\times$) and 
the ions (dash-dot). {\it Bottom}: magnetic field. The solution (\ref{Bnmseq}) is also 
displayed (dash-dot). 
\label{fig-thirdwave}
}
\end{figure}

Figure \ref{fig-thirdwave} shows the results of the split-operator algorithm
at time $t = 2.55 \times 10^{5}~\yr$ calculated on a mesh with 2000 points.
The figure shows that
there are indeed two oppositely directed wave packets emerging from the initial disturbance 
about the origin. There is a decrease in the neutral gas density from the central 
region as the matter once located there is set into motion and becomes ``scooped up" by 
collisions with the ions, which are being driven outward by magnetic field pressure 
gradients. The transported neutral gas piles up just outside the 
depleted central region and increases the density in the two adjacent pulses moving away 
from the origin. Note that the ion fluid velocity slightly leads that of the neutral fluid. 
This effect is caused by the inertia of the neutrals, which delays the 
acceleration of the neutrals from rest up to the same velocity as the ions.
Figure \ref{fig-thirdwave} also shows the analytic solutions (\ref{Bnmseq})--(\ref{nnnmseq})
at the same output time. The numerical results are in very good agreement with the 
analytic solutions, with a relative neutral density difference $\leq 3.1 \times 10^{-4}$ 
and magnetic field relative difference $\leq 2.2 \times 10^{-4}$.

Because the initial state of the neutral fluid is static and completely lacks 
density or thermal pressure gradients, the RG integration step for the neutrals 
does not initiate the wave motion in the neutrals in this example. 
Rather it is the collisional drag of the magnetically-driven 
ions during the source integration steps which starts the neutrals moving after a 
time $\simgt \tni$ has passed for regions near the origin. Once this happens, as we 
noted above, gradients in the neutral density and pressure form, and thermal pressure 
gradient forces start to act on the neutrals during the RG integration step for that 
fluid. From that point on, there is a combination of wave-driving thermal pressure and 
magnetic forces (through ion collisions) on the neutrals during all integration stages of 
the split-operator method scheme. (It is this combination of thermal pressure and magnetic
forces acting on the neutrals which explains the appearance of both the sound and {\Alf} 
speeds in the expression for the neutral magnetosound speed $\vnms$, eq. [\ref{vnmsdef}].)

To summarize: the benchmark tests for wave packets show that our split-operator method successfully 
incorporates the relevant physics and accurately follows the evolution of both propagating
and diffusive flows traveling perpendicular to the magnetic field, 
over many orders of magnitude in both the temporal and spatial scales. 

\subsection{Early Evolution of Shocks Caused by a Cloud-Cloud Collision}
\label{sec-RCmodels}

We now consider a more dynamic test which shows that the split-operator method is also suitable for 
multifluid shock waves.
RC07 considered the collision of two identical clouds and found
analytic solutions (using Green functions) which describe the early-time behavior of
the ensuing disturbances in the ion-electron fluid.
For the parameters RC07 considered, the collision produced forward- and reverse J-shocks
in the neutral gas.
They referred to the disturbances in the ion-electron fluid as ``driven waves'' because 
they are driven by frictional coupling to the neutral flow.
Their results describe initial stages in the formation of magnetic precursors on 
the J-shocks. 

The flow geometry in RC07 is the same as the geometry in this paper, with fluid velocities
perpendicular to the magnetic field. The clouds are
identical and semi-infinite; the collision takes place when their free surfaces coincide
at the plane $x=0$ at $t=0$. Prior to the collision the charged and neutral fluids in each cloud move 
together with $\vn(x,0)=\vi(x,0)= v(x)$, where
\bea
\label{RCinitialvelocities}
v(x) = \left\{ \begin{array}{ll} -10~\kms ~{\rm for}~x > 0~, \\
+ 10~\kms ~{\rm for}~x < 0~.
\end{array}
\right.
\eea
Each cloud has an initial neutral density $\nn(x,0) = \nno = 2 \times 10^{4}~\cc$, ion 
density $\nnio(x,0)= \nnio = 5.72 \times 10^{-4}~\cc$, fractional ionization 
$\xxi(x,0)=\nnio/\nno = 2.86 \times 10^{-8}$, temperatures 
$\Tn(x,0) = \Ti(x,0)= 10~{\rm K}$, and magnetic field $B(x,0)=B_{0} = 50~\mu{\rm G}$. 
The ion magnetosound speed is $\vims = \viA = 912~\kms$ in the undisturbed charged fluid. 

We adopt the same source terms as RC07, who set $\Sn=\Gn=\Ln=0$ and assumed that momentum transfer is
due solely to elastic ion-neutral scattering.
RC07 assumed further that $\tin$ is independent of the relative velocity 
between the ions and neutrals. To make a legitimate comparison with the RC07 solution,
we drop the term $\propto |\vn - \vi|^{2}$ on the right-hand side of equation (\ref{tineq}) 
when calculating \Fn. The ion-neutral mean collision time in the 
clouds is then $\tin = 1.26 \times 10^{-2}~\yr$, and the neutral-ion drag time is
$\tni = 3.52 \times 10^{4}~\yr$. Because \tni\ is much longer than the 
times considered by RC07, the motion of the neutral gas is unaffected by
coupling to the ions and magnetic field. 

\begin{figure}
\epsscale{0.90}
\plotone{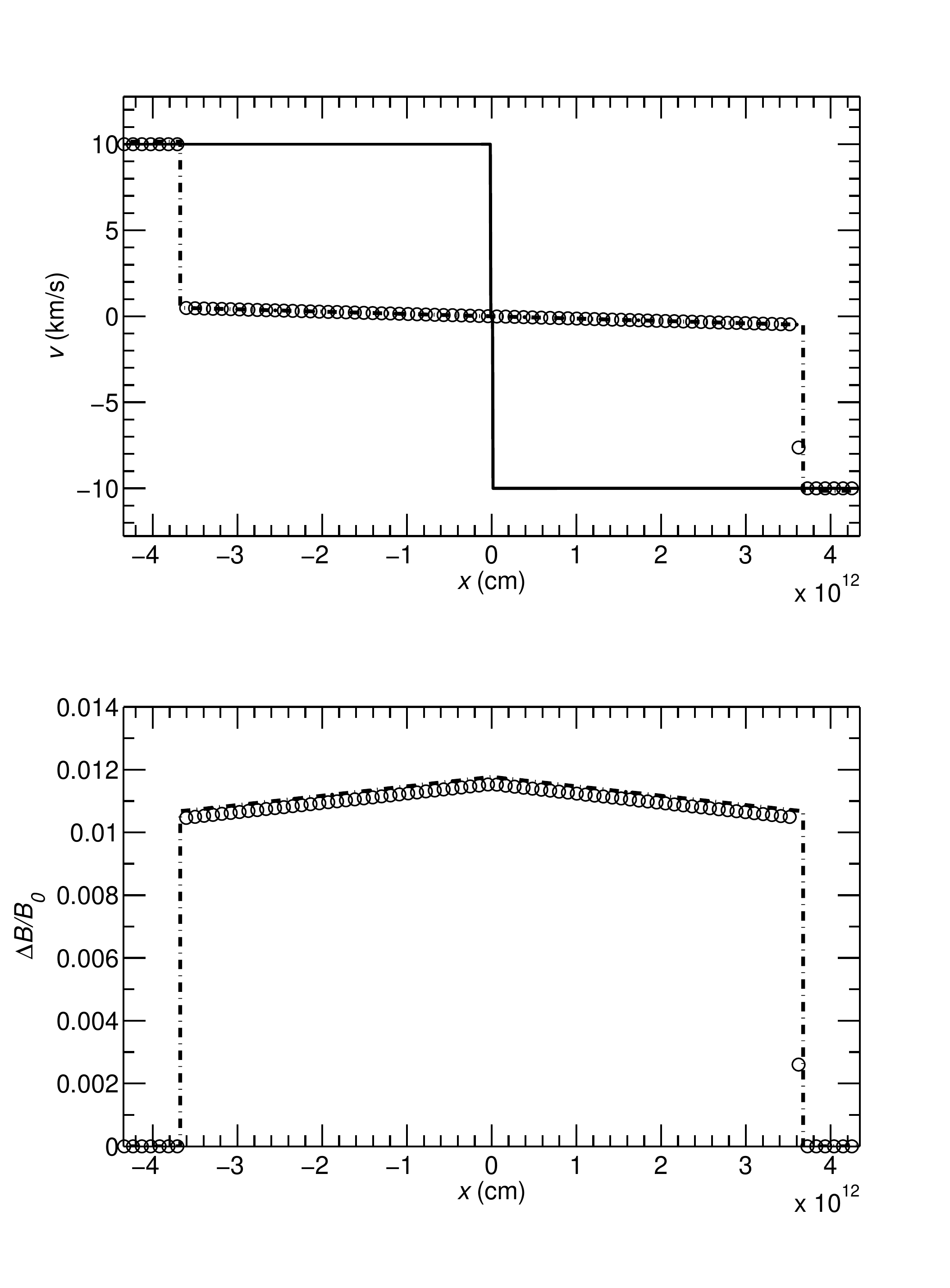}
\caption{Colliding clouds at time $t=0.1 \tin = 1.26 \times 10^{-3}~\yr$.
{\it Top panel:} Velocities of the neutrals (solid line) and ions (circles). Also shown
is the approximate analytic RC07 solution (dash-dot line) for the ion fluid velocity. {\it Bottom panel:}
Magnetic field increase above the initial value in each cloud $\Delta B = B(x,t)-B_{0}$, 
relative to the initial field strength (circles). The dash-dot line shows the RC07 solution for the magnetic field.
\label{fig-firstRCtime}
}
\end{figure}

Figure \ref{fig-firstRCtime} shows the multifluid MHD split-operator code results for the
colliding clouds test at $t=0.1 \tin = 1.26 \times 10^{-3}~\yr$. 
At this very early time ($\ll \tin$), the ions and magnetic field have 
yet to be affected by friction from the neutrals. Two discontinuities 
in the magnetic field and charged fluid, symmetric about the origin, travel 
outward at the ion magnetosound speed $\vims$, with their fronts located at
$|x_{\rm idisc}| = 3.63 \times 10^{12}~\cm$. There are also two symmetric neutral shocks 
propagating away from $x=0$ at a speed $v_{\rm nshk} = 3.34~\kms$. The neutral shock 
fronts are located at $|x_{\rm nshk}| = 1.33 \times 10^{10}~\cm$ so they are not 
visible on the scale of Figure \ref{fig-firstRCtime}. Also 
shown (dash-dot curves) in each panel of the figure are the analytic solutions of RC07.
The split-operator results agree very well with the RC07 solution,
with a relative difference in the magnetic field that is $< 0.015$ behind the 
discontinuities.

\begin{figure}
\plotone{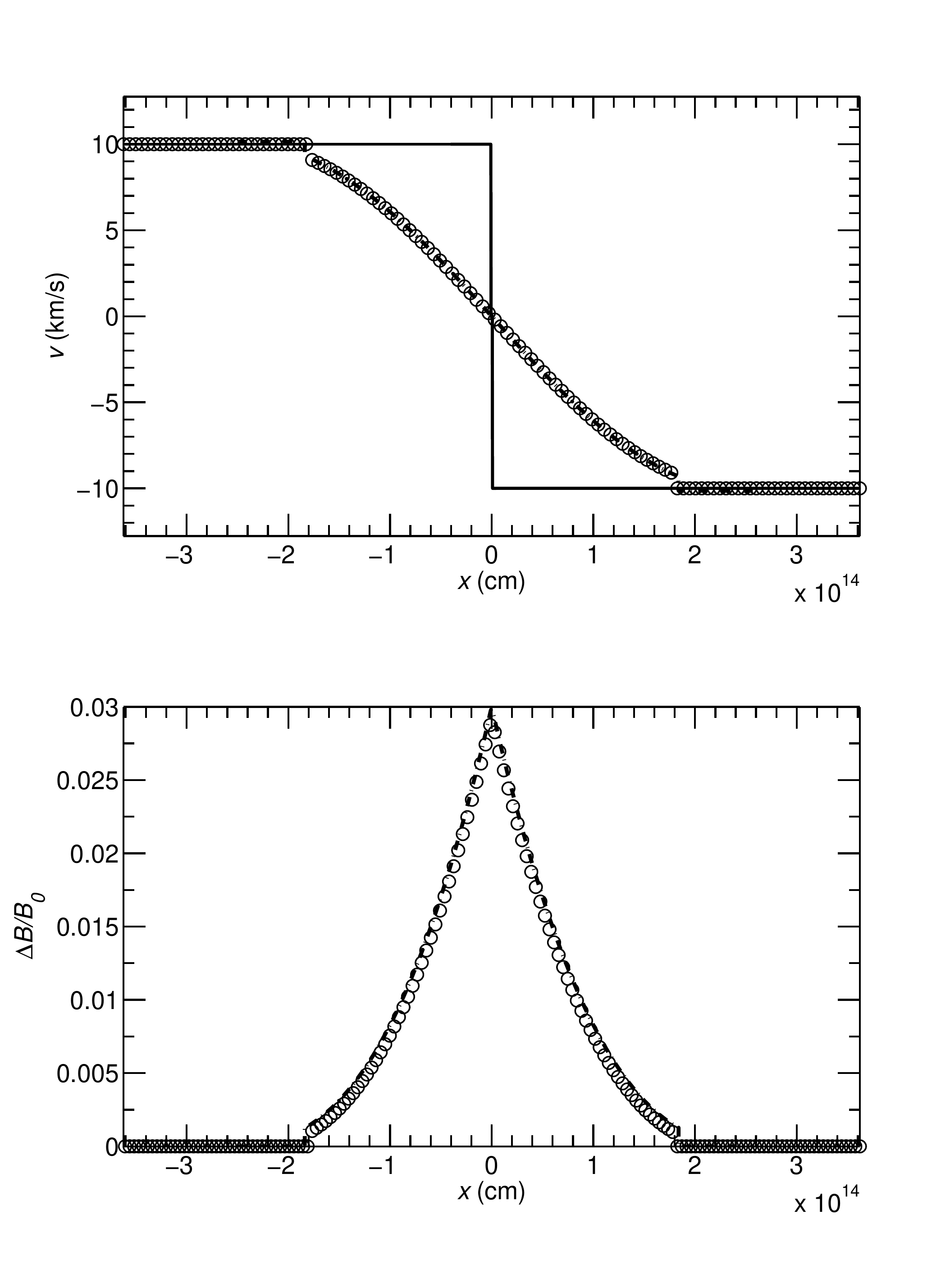}
\caption{Colliding model clouds at time $t=5\tin=6.29 \times 10^{-2}~\yr$. All 
normalizations and symbols have the same meaning as in Fig. \ref{fig-firstRCtime}.
\label{fig-secondRCtime}
}
\end{figure}

Results for the colliding clouds test at
$t = 5 \tin = 6.29 \times 10^{-2}~\yr$ are presented in Figure \ref{fig-secondRCtime}. 
The shock fronts in the neutral fluid continue to travel away from the origin at $3.34~\kms$, 
and are located at $|x_{\rm nshk}| = 6.63 \times 10^{11}~\cm$. At this stage of the 
evolution, pronounced precursors in the ion velocity and the magnetic field extend
to much greater distances beyond the neutral shock fronts. At the time shown ($> \tin$), 
collisional drag from the inflowing neutrals is having a pronounced effect on the 
precursors. Notably, they continue to be led by discontinuities in the velocity and 
magnetic field heading outward at the ion magnetosound speed $\vims$, but the jumps at the
discontinuities are significantly diminished in magnitude and strength compared to the 
earlier time (Fig. \ref{fig-firstRCtime}). The RC07 solutions are also plotted in 
Figure \ref{fig-secondRCtime}. Once again the split-operator results 
agree extremely well with the analytic solutions.

\begin{figure}
\plotone{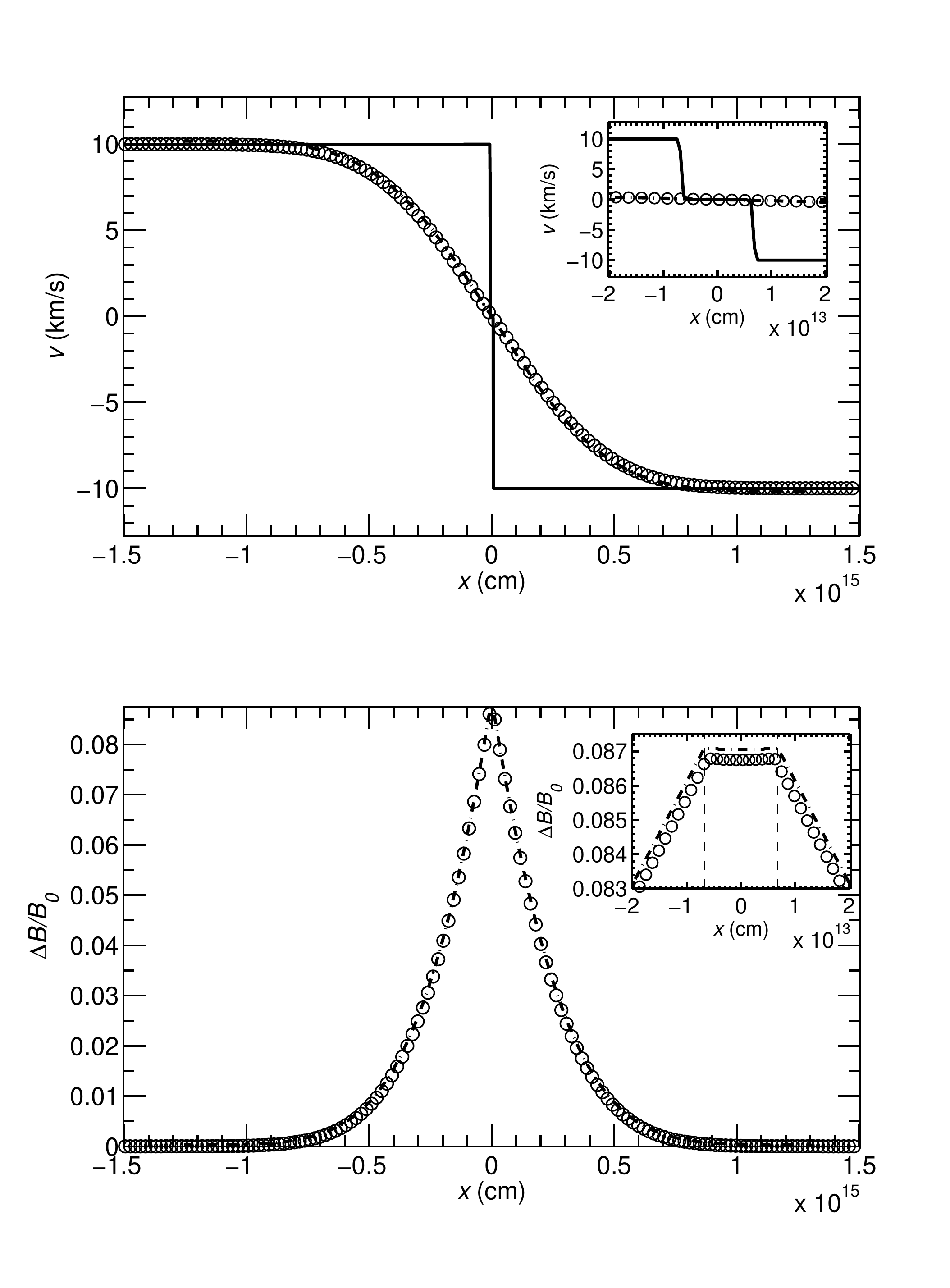}
\caption{Colliding model clouds at time $t = 50 \tin = 0.629~\yr$. All normalizations
and symbols have the same meaning as in Fig. \ref{fig-firstRCtime}. The insets show values
about the collision point at $x = 0$, including the region between the leftward and 
rightward traveling neutral shock fronts, each located by the vertical dashed lines.
\label{fig-thirdRCtime}
}
\end{figure}

The solutions at time $t = 50 \tin = 0.629~\yr$ are shown in Figure~\ref{fig-thirdRCtime}
along with the RC07 solutions. The agreement between the 
split-operator results and the analytic solutions is again very good. At this time 
($\gg \tin$) the ion and magnetic field discontinuities have virtually 
vanished, with the precursors ahead of the neutral shocks having profiles that are smooth 
and continuous. Inside the precursors the motion of the ions is dictated by a balance between
the collisional drag from the neutrals flowing toward the origin and the outwardly 
directed magnetic pressure gradient. This is consistent with the RC07 solution, which 
found that on these time scales the solution for the charged fluid motion tended 
toward a force-free diffusion mode.

Figure \ref{fig-thirdRCtime} also contains insets showing the fluid velocities and 
magnetic field on an expanded scale which resolves the region between the neutral J-shocks. The 
vertical dashed lines in the insets mark the location of the forward- and reverse 
shocks at that time. The RC07 solutions agree well with 
the split-operator numerical simulation, even on these greatly magnified scales,
with the numerical solution for the magnetic field differing from the RC07 solution
by less than 0.3\% within the inset region. This is an especially interesting result:
to be able to use the method of Fourier transforms to obtain their analytic solutions, 
RC07 were forced to assume that the neutral gas density is uniform--- even in 
the region between the two neutral shock fronts--- despite the fact that for strong (i.e.,
large Mach-number) adiabatic shocks the density behind a shock front 
increases by a factor of 4 (for $\gamma=5/3$, see, e.g., Zeldovich \& Raizer 1966 or Spitzer 1978). 
RC07 argued on physical grounds that their analytic solution would nevertheless be valid.
The excellent agreement between the analytic solutions and numerical results between the
two shock fronts (where the density of the neutrals does indeed have a fourfold increase)
confirms that their reasoning was correct. 

The colliding clouds test results and their agreement with the analytic solutions of
RC07 show that our operator-splitting scheme faithfully reproduces the
physics of dynamic MHD multifluid astrophysical flows.
The split-operator code successfully models the 
evolution of shocks and discontinuities in both the neutral gas and the charged fluid, having no 
difficulty in handling the interaction between the fluids which leads to the formation of a 
magnetic precursor.

\subsection{Shocks with Mass Transfer}
\label{sec-masstransfer}

As a final test we present a model which allows for the transfer of mass between 
the neutral and charged fluids in shocks.
We introduce a nonzero source term for the conversion of ion-electron mass to neutral mass, with
\be
\label{Sndefeq}
\Sn = \mi \left(\alphdr\nni^{2} - \cri \nn\right)~,
\ee
where $\cri$ is the cosmic-ray ionization rate and $\alphdr$ is the rate coefficient
for dissociative 
recombination of molecular $\rm{HCO}^{+}$ with electrons; note that in this expression we 
have used $\nne = \nni$, which follows from (\ref{chargeneutralityeq}). In this test we 
use the representative ionization rate $\cri = 5 \times 10^{-17}~{\rm s}^{-1}$ (Dalgarno 
2006) and take \alphdr\ from the UMIST Astrochemistry Database ({\tt http://www.udfa.net}; Millar, 
Farquhar, \& Willacy 1997),
\be
\label{alphdreq}
\alphdr = 2.4 \times 10^{-7} \left(\frac{300~{\rm K}}{\Te}\right)^{0.69}~{\rm cm}^{3}~{\rm s}^{-1}~.
\ee
Mass exchange also adds to the momentum source term, so that
\be
\Fn = \Fnel + \Fnin,
\ee
where
\be
\label{Fnindefeq}
\Fnin  =  \mi\left(\alphdr \nni^{2}\vi - \cri \nn \vn \right).
\ee
In this test we again take $\Gn = \Ln = 0$. Neglecting radiative cooling means 
that the temperatures in this test will be much larger than a realistic model with
cooling included. 
However it allows us to isolate the effects of mass-transfer between 
the neutrals and the ions as a test of the split-operator code. 
The rate coefficient for dissociative recombination depends on the
electron temperature, which is not calculated in the current version of our code.
For expediency we set
\be
\label{Teeq}
\Te = \max(\Tn,~0.15 \Ti)~,
\ee  
in rough agreement with realistic simulations of steady multifluid shocks (e.g., see Figs.~1--3
of Draine et al.\ 1983).

\begin{deluxetable}{ccccccc}
\tablecaption{Initial Conditions for the Mass Transfer Test \label{table-masstransfer}}
\tablewidth{0pt}
\tablehead{
\colhead{} &
\colhead{$~\vn$ ($=\vi$)} &
\colhead{$\nn$} &
\colhead{$\nni$} &
\colhead{$\Tn$ ($=\Ti=\Te$)} &
\colhead{$B$} &
\colhead{$\vims$} 
}
\startdata
$x < x_{0}:$ & $20~\kms$ & $2.5 \times 10^{4}~\cc$ & $8.12 \times 10^{-4}~\cc$ &
15 K & $50~\mu{\rm G}$ & $765~\kms$ \\ \\
$x > x_{0}:$ & $0$ & $2 \times 10^{4}~\cc$ & $6.31 \times 10^{-4}~\cc$ &
10 K & $25~\mu{\rm G}$ & $434~\kms$ \\ \\
\enddata
\end{deluxetable}

Initially this model has a discontinuity at $x_{0} = 1.12 \times 10^{16}~\rm{cm}$. For 
$x < x_{0}$ the neutral gas and ions are uniform and flowing in the $+x$-direction
with a common supersonic velocity, while for $x > x_{0}$ the matter is uniform and 
stationary.
The other initial conditions are listed in Table \ref{table-masstransfer}.
The initial ion density was calculated from the expression 
\be
\label{massequil}
\nni = \left(\frac{\cri \nn}{\alphdr}\right)^{1/2}~~,
\ee
which assumes that the creation and destruction rates of ions 
in eq.~(\ref{Sndefeq}) are equal.

\begin{figure}
\epsscale{0.85}
\plotone{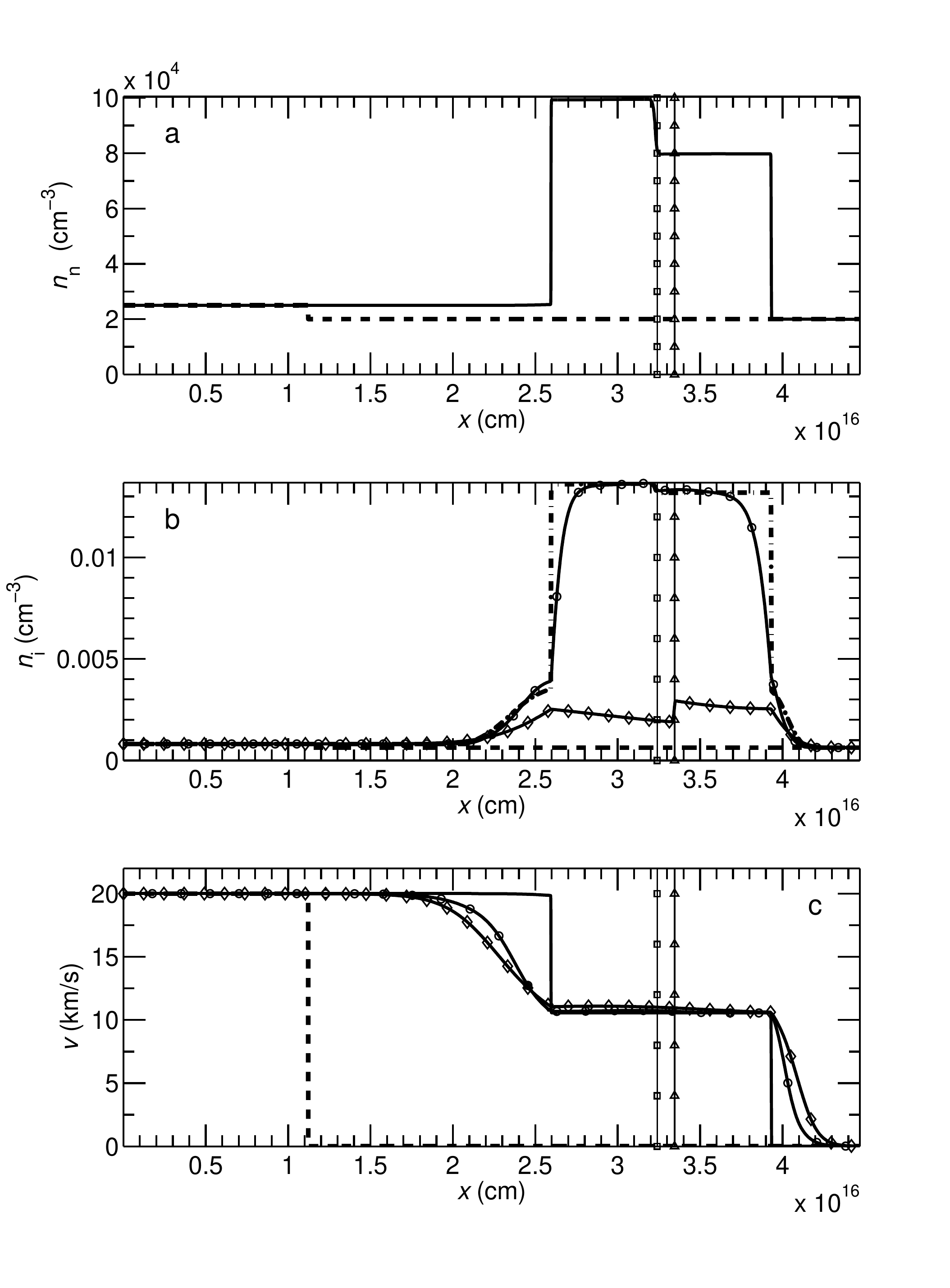}
\vspace{-3ex}
\caption{Mass transfer test at $t=632~\yr$. In each panel the initial state is 
shown as black dashed lines. Also displayed are the locations of a neutral contact 
discontinuity (vertical line with squares) in the MT model and an ion contact discontinuity
(vertical line with triangles) in the NMT model. ({\it a}) Neutral density.
({\it b}) Ion density (curve with circles) in the MT model and NMT model (curve with 
diamonds). The quasi-mass-equilibrium relation (\ref{massequil}) is 
the dash-dot curve. ({\it c}) Velocities of the neutrals (solid curve) and ions (curve with
circles) in the MT model, and the ions in the NMT model (curve with diamonds). 
\label{fig-masstransfer}
}
\end{figure}

\addtocounter{figure}{-1}
\begin{figure}
\epsscale{0.85}
\plotone{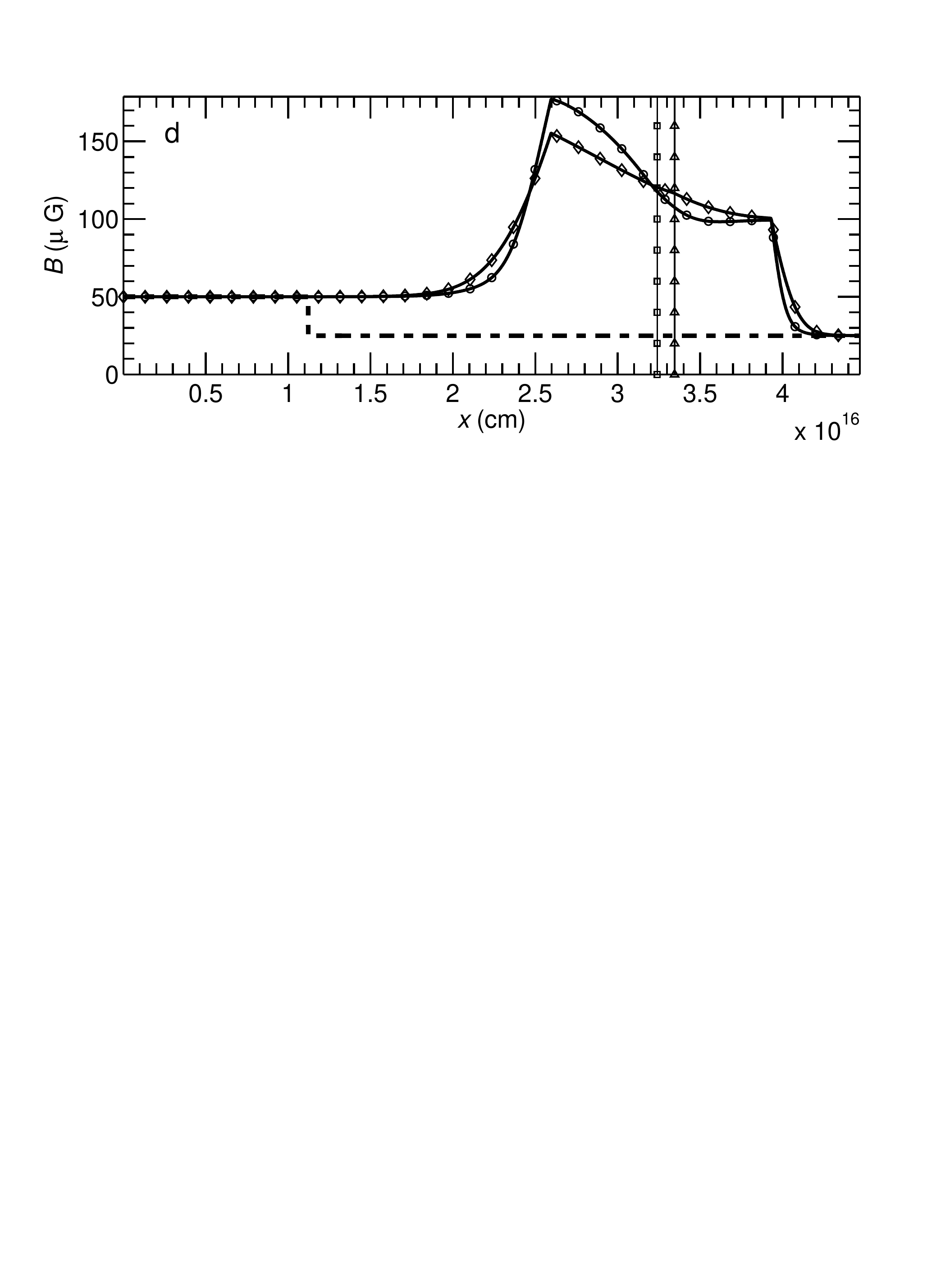}
\caption{{\it Cont.} ({\it d}) Magnetic field in the MT  model (curve with circles) and 
the NMT model (curve with diamonds).}
\end{figure}

Figure \ref{fig-masstransfer} presents the mass transfer (MT) test solution
at $t = 632~\yr$. 
For contrast we also show results for a model with no mass transfer (NMT)
which is identical otherwise.
At the relatively early time shown in the figure 
($\ll \tni \sim 3 \times 10^{4}~\yr$) the neutral gas is virtually unaffected by ion-electron 
drag. The effect of mass transfer on the neutral gas is also imperceptible in
the MT model for the following two reasons: (i) even if all of the ions and electrons were
to recombine, the added mass to the neutrals would be negligible because 
$\rhoi \simlt 10^{-6}\rhon$; (ii) the rate of change of the neutral density caused by cosmic-ray 
ionization is orders of magnitude smaller than the density change caused by advection.

In Figure~$\ref{fig-masstransfer}a$ it is evident that the collision of the inflowing and 
stationary material at $x_{0}$ has resulted in two neutral J-shocks: a left shock located
at $x_{\rm L,shk} = 2.60 \times 10^{16}~{\rm cm}$ traveling with a 
velocity $v_{\rm L,shk} = +7.41~\kms$, plus a right shock at 
$x_{\rm R,shk}=3.93 \times 10^{16}~{\rm cm}$ with velocity $v_{\rm R,shk} = +14.1~\kms$. 
The figure also reveals a contact discontinuity in the neutral gas at 
$x_{\rm n,con}= 3.22 \times 10^{16}~{\rm cm}$. The neutral density profile is plotted
for the MT model; the neutral density in the NMT is indistinguishable from that of the MT
model. 

The effect of mass transfer on the ion fluid in the MT model
can be seen in Figure $\ref{fig-masstransfer}b$. Between the shock fronts the neutral 
gas is heated, with a corresponding increase in the temperatures of the ions and electrons.
This results in a decrease in the number of ion-electron recombinations there (see eq. 
[\ref{alphdreq}]). At the same time, shock compression of the neutral gas leads to a 
greater number of cosmic-ray created ions (see eq. [\ref{Sndefeq}]). The interplay of
these two effects leads to ion densities between the shock fronts which are a factor $\approx 5$
greater compared to the NMT model. Not only does the NMT model have a 
significantly smaller ion density in the shocked region, it also has an ion contact 
discontinuity located at $x_{\rm i,con} = 3.32 \times 10^{16}~{\rm cm}$.
There is no ion contact discontinuity in the MT model because the source integration
steps in the algorithm completely overwhelm the RG step (Fig.~\ref{fig-schematic}).
That is, the ion density profile is determined almost entirely by 
chemistry rather than advection.
There is a small jump in the ion density at the {\em neutral}\/ contact discontinuity
caused by the jump in the neutral density, which is proportional to the rate of ionizations per
unit volume.
Increased $\Ti$ and  $\Te$ (due to significant drift speeds between the ions and the neutrals
[see eqs.~[\ref{Tieq}] and [\ref{Teeq}]) inside the magnetic precursors upstream from the shock fronts 
reduces the rate of ion-electron recombinations there.
This raises the ion density inside the precursors to values noticeably larger than those in the NMT model.

The width of the magnetic precursors depends on the amount of ion mass loaded onto
the magnetic field lines there, as a comparison of the MT and NMT models shows (Fig.~$\ref{fig-masstransfer}c$).
The width of a magnetic precursor scales as $\Lpre \sim \vims^{2} \tin/v_{\rm shk}$, where
$v_{\rm shk}$ is the shock speed; this result can be derived by balancing magnetic pressure
and ion-neutral drag in the precursor (Draine 1980) or by setting the 
precursor crossing time $t_{\rm pre} = \Lpre/v_{\rm shk}$ equal to the ion-magnetic field 
diffusion time ahead of the front $t_{\rm diff} = \Lpre^{2}/\Di$ (see eq. [\ref{Dicoef}]).
Because $\vims = \viA$ in our models, it follows from equation 
(\ref{viAdef}) that $\Lpre \propto \nni^{-1}$. Hence, the model with the greater ion density 
$\nni$ will have the smaller values of $\Lpre$. 
This explains why the precursors extend further in the NMT model than in the MT model.
 
Figure~$\ref{fig-masstransfer}d$ reveals a consequence of the dependence of precursor 
width on ion density: the distribution of magnetic flux 
is noticeably different in the MT and NMT models. 
Compression of the magnetic field by a shock is made less abrupt by having a precursor of greater 
width. This accounts for the smaller magnetic field increase at the left shock in the NMT 
model than in the MT model. (Note, however, that magnetic flux is conserved in both models:
the total magnetic flux contained in the region from $1.5 \times 10^{16}~{\rm cm}$ to 
$4.5 \times 10^{16}~{\rm cm}$ [= area under the curve] is 
the same for both.) A by-product of the mass transfer test, then, is that we have 
demonstrated how loading of mass onto magnetic field lines affects the structure of a magnetic
precursor and the magnetic field profile elsewhere in a shock. These effects feed back into the
dynamics, because the magnetic pressure gradient is one of the main driving forces in the
ion-electron momentum transport equation (\ref{ionmtmeq}). 

The fidelity of the mass-transfer test can be quantified by noting that
from the left-hand side of the ion mass equation (\ref{ionmassconteq}) we can 
define an ion advection time 
\bml
\be
\label{tiadv}
\tau_{\rm iadv} \equiv \frac{L}{\vi} = 317 \left(\frac{L}{10^{16}~{\rm cm}}\right)
\left(\frac{10~\kms}{\vi}\right)~\yr 
\ee
where $L$ is a characteristic length scale.
From the right-hand side of the same 
equation we can also define a recombination time scale 
\be
\label{tirec}
\tau_{\rm irec} \equiv \frac{\rhoi}{\mi\alphdr\nni^{2}} = 66.0 
\left(\frac{2 \times 10^{-3} \cc}{\nni}\right)
\left(\frac{\Te}{300~{\rm K}}\right)^{0.69}~\yr~, 
\ee
and a cosmic-ray ionization time 
\be
\label{tiCR}
\tau_{\rm iCR} \equiv \frac{\rhoi}{\mi\cri\nn} = 63.4 
\left(\frac{\nni}{2 \times 10^{-3}~\cc}\right) 
\left(\frac{2 \times 10^{4}~\cc}{\nn}\right)
\left(\frac{5 \times 10^{-17}~{\rm s}^{-1}}{\cri}\right)~\yr~ 
\ee
\eml
(see eq. [\ref{Sndefeq}]). For the physical conditions in the MT model, 
$\tau_{\rm iadv}  \simgt \tau_{\rm irec},\tau_{\rm iCR}$ on length scales 
$L_{\rm ieq} \simgt 5 \times 10^{15}~{\rm cm}$. For model ages greater than $\tau_{\rm iref}$
and $\tau_{\rm iCR}$ and length scales greater than $L_{\rm ieq}$, ion advection can be 
ignored, and there should be approximate equality between the rates of mass creation and 
destruction. That is, quasi-mass-equilibrium with $\Sn \approx 0$ should occur with $\nni$
given by equation (\ref{massequil}). It is therefore expected that the 
ion density in the MT run for the time displayed should be set by the condition of 
quasi-mass-equilibrium in regions having length scales $\simgt L_{\rm ieq}$. 

To test this hypothesis, the values of $\nni$ predicted by the quasi-equilibrium equation 
(\ref{massequil}) are also plotted in Figure $\ref{fig-masstransfer}b$ (dash-dot curve). It
is seen that the predicted values do match the actual MT model results well in most places,
except at the shock fronts where the assumption of large length scales becomes invalid.
Away from the shock fronts the agreement between the MT model and equation 
(\ref{massequil}) is generally quite good. For instance in the region between the shocks,
including the region about the neutral contact discontinuity, the relative differences of 
the quasi-equilibrium and model values for $\nni$ range from $3 \times 10^{-3}$ to $0.01$.
The very good agreement between the values predicted by equation (\ref{massequil}) for the 
ion density with the actual results of the fully dynamical MT model (in the regions 
where the mass-quasi-equilibrium approximation is valid) illustrates the accuracy 
of the split-operator scheme for multifluid shocks with mass transfer
by ionization and recombination.

\section{Summary}
\label{sec-summary}

Because many protostellar outflow shocks are much younger than the time scale
\tni\ to accelerate the neutral gas, it is likely they are not steady flows.
A time-dependent treatment is therefore generally required to model outflow shocks and their emission.
In this paper we have presented a method for modeling perpendicular, time-dependent, multifluid MHD shocks using 
operator splitting. This scheme splits the time integration of the evolutionary equations 
into separate steps, one dealing with the solution of independent homogeneous Riemann 
problems for the neutral and charged fluids using Godunov's method, plus other steps where 
only the equations describing interactions between the fluids are integrated.
Our method exploits the fact that the thermal pressure of the charged fluid is usually
negligible compared to its magnetic pressure. 
Neglecting the thermal pressure reduces the number of characteristic
waves in the ion-electron fluid to three, the same number as in one-dimensional 
gas dynamics (Toro 2009). We have shown that under these circumstances the MHD Riemann 
problem can be solved exactly. Using this exact solution, we have constructed
an approximate MHD solver which is used in the Riemann-Godunov integration of the charged fluid in
our split-operator scheme. The similarity in structure of the Riemann problems for both
fluids in our split-operator scheme makes it straightforward to adapt well-established 
gas-dynamic numerical techniques (TVD slope limiters, other data reconstruction methods such
as WENO, etc.) for use in the RG integration of the charged fluid. The symmetry in the 
RG problems for both the neutral and charged fluids also greatly simplifies the numerical 
coding. (In fact, following our scheme it should not be too difficult to modify an already 
existing non-magnetic single-fluid RG hydrodynamics code to one for use in multifluid MHD.) 

Since the inertia of a fluid and the wave modes it supports are fundamental to the solution of
the Riemann problem, the multifluid split-operator method outlined here has no difficulty
dealing with flows involving MHD waves or transients in the charged fluid. Several tests 
spanning a wide variety of time- and length scales were performed to demonstrate the 
versatility and accuracy of our algorithm. The numerical results are in very good
agreement with analytic solutions in all of our benchmark tests.
The latter include a model for the formation of 
MHD shocks resulting from the collision of two identical clouds, a problem solved 
analytically by RC07; the split-operator code successfully captures the
ion density and magnetic field transients which propagate away from the collision surface
at early times.
Our code also correctly reproduces the multifluid shocks with magnetic precursors which
develop at much later times, when the charged fluid is force free and evolves diffusively.  

Another benchmark test involved MHD shocks with transfer of mass between the neutral and charged 
fluids. A numerical model with mass transfer has significantly greater ion densities
behind the shocks and in the magnetic precursor compared to a model 
with no mass transfer. The enhanced loading of ion mass onto magnetic field lines
in the precursors of the mass transfer model results in precursors of smaller width, and 
there is a corresponding difference in the distribution of magnetic flux between 
models with and without mass transfer.
Analysis of the characteristic time scales for cosmic-ray ionization and 
ion-electron dissociative recombination in the mass transfer model 
suggested that for sufficiently large length scales the ion density should be
close to the value that would be predicted when ionization and recombination
balance one another.
The ion density profile in the mass transfer
model is indeed well fit by the quasi-equilibrium relation, except in
the vicinity of the shock fronts where the criterion of large length
scales (or, equivalently, negligible ion mass advection) is violated. 

\acknowledgements{This work was supported by the New York Center for Astrobiology, a member
of the NASA Astrobiology Institute, under grant \#NNA09DA80A. Code development and model 
runs were performed on the facilities of the Computational Center for Nanotechnology 
Innovations (CCNI), which is partnered with Rensselaer Polytechnic Institute.}

\appendix
\section{The Exact MHD Riemann Solution}
\label{sec-exactmhdriemann}
\begin{figure}
\plotone{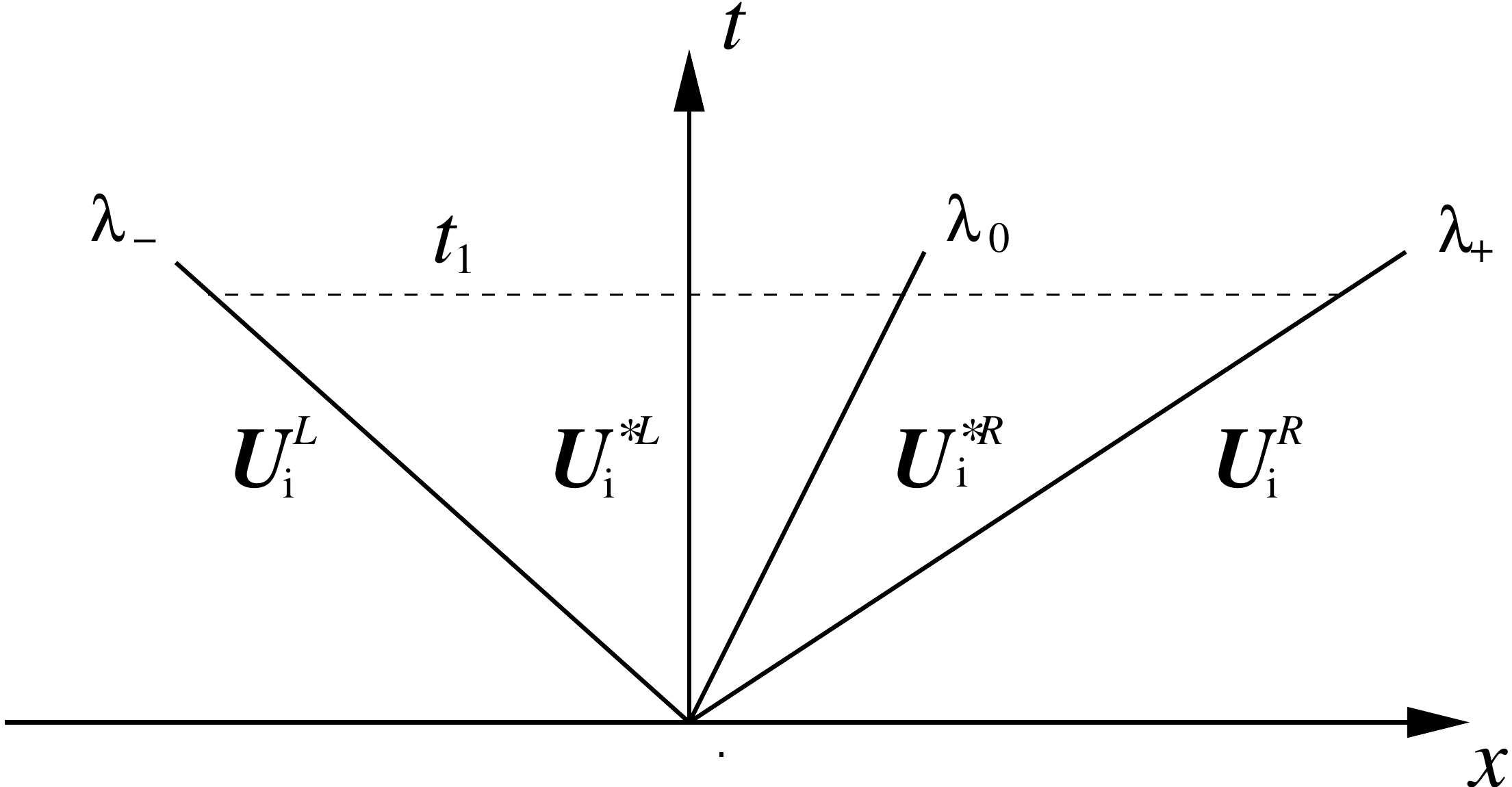}
\caption{The three characteristics emanating from the initial discontinuity.
\label{fig-characteristics}
}
\end{figure}
The three-dimensional MHD Riemann problem for an arbitrary flow geometry and 
a plasma with nonzero thermal pressure is probably too complex to be solved exactly (Brio \& Wu 1988; Torrilhon 2003). 
Because of this, several approximate MHD Riemann solvers have been developed (e.g., Dai \& Woodward 1994, 1997;
Ryu \& Jones 1995; Balsara 1998; Li 2005; Mignone 2007). 
However for the special case studied here--- one-dimensional flow, perpendicular geometry, and no thermal pressure--- 
the MHD Riemann problem for the charged fluid has a simple, exact solution.
We give it here.

\subsection{Characteristics}
\label{sec-characteristics}

The Riemann problem for the charged fluid is to solve
equation (\ref{mhdRGprob}),
\be
\frac{\partial \Uveci}{\partial t} + \frac{\partial \Fveci(\Uveci)}{\partial x} =0 ~~,
\label{eq-MHDRiemann}
\ee
subject to Riemann initial conditions,
\be
\label{riemannconditions}
\Uveci(x,0) = \left\{
\begin{array}{ll}
\ULveci\ \equiv \left[ \rhoiL, ~\rhoiL\viL, ~\BL \right]^{\rm T} & ~~{\rm if}~x<0 \\
                                                                           &                \\
\URveci\ \equiv \left[ \rhoiR, ~\rhoiR\viR, ~\BR \right]^{\rm T} & ~~{\rm if}~x>0 \\
\end{array}
\right. ,
\ee
where the constant vectors \URveci\ and \ULveci\ are the initial states on the right and 
left half planes.
The first step is to rewrite eq.~(\ref{eq-MHDRiemann}) in the equivalent form
\be
\frac{\partial{\Uveci}}{\partial t} 
+
\matA\,\vdot\,
\frac{\partial{\Uveci}}{\partial x} 
= 0
\label{eq-motionjac}
\ee
and examine the eigenvalues and eigenvectors of the Jacobian matrix,
\be
{\sf A}_{k\ell}
\equiv
\frac{\partial \Fveci_{k}}{\partial \Uveci_{\ell}}
=
\left[
\begin{array}{ccc}
       0     &   1       &  0      \\
    -v_i^2   &  2v_i     &  B/4\pi \\
-Bv_i/\rho_i &  B/\rho_i & v_i     \\
\end{array}
\right].
\label{eq-matA-comp}
\ee
The eigenvalues are
\be
\label{ieigenvals}
\left\{\lambdam,  ~\lambdaz, ~\lambdap \right\}
=
\left\{\vi-\viA, ~\vi, ~\vi+\viA \right\},
\ee
where $\viA$
is the ion {\Alf} speed (eq. [\ref{viAdef}]). 
The corresponding right (column) eigenvectors are
\bml
\be
\Rm = \left[\rhoi, ~\rhoi\left(\vi-\viA\right), ~B \right]^{\rm T},
\label{eq-Rmcomponents}
\ee
\be
\Rz =  \left[\rhoi, ~\rhoi\vi, ~0 \right]^{\rm T},
\label{eq-Rzcomponents}
\ee
and
\be
\Rp =  \left[\rhoi, ~\rhoi\left(\vi+\viA\right),  ~B \right]^{\rm T}.
\label{eq-Rpcomponents}
\ee
\eml
The solution also depends on the left (row) eigenvectors, which are
\bml
\be
\Lm = \frac{1}{2v_{\rm iA}}
\left[
\frac{v_{\rm i}}{\rho_{\rm i}}, ~-\frac{1}{\rho_{\rm i}}, ~\frac{V_{\rm iA}}{B}
\right],
\label{eq-Lmcomponents}
\ee
\be
\Lz = 
\left[
\frac{1}{\rho_{\rm i}}, ~0, ~-\frac{1}{B}
\right],
\label{eq-Lzcomponents}
\ee
and
\be
\Lp = \frac{1}{2v_{\rm iA}}
\left[ 
-\frac{v_{\rm i}}{\rho_{\rm i}}, ~\frac{1}{\rho_{\rm i}}, ~\frac{V_{\rm iA}}{B}
\right].
\label{eq-Lpcomponents}
\ee
\eml
The left- and right eigenvectors are orthonormal.

Introduce the characteristics curves, $x_m(t)$, defined by
\be
\label{eq-characcurves}
\frac{dx_m}{dt} =
\lambda_{m},  ~~~m=-,0,+.
\label{eq-char-def}
\ee
For Riemann initial conditions, the characteristics emanating from the initial discontinuity
at $x=0$ are straight lines (Fig.~\ref{fig-characteristics}). 
Each represents a 
discontinuity in the solution which may be a shock wave, rarefaction, or contact 
discontinuity. Since discontinuities propagate along characteristics, and the initial data
are uniform to the left and right of the origin, Figure~\ref{fig-characteristics} shows 
immediately that $\Uveci=\ULveci$ to the left of \lambdam\ and $\Uveci=\URveci$ to the right
of \lambdap. The problem therefore reduces to finding \Uveci\ in the ``star region'' between
\lambdam\ and \lambdap. Let \rhoiSL, \viSL, and \BSL\ denote the 3 unknowns in the star region
between \lambdam\ and \lambdaz\ and \rhoiSR, \viSR, and \BSR\ be the 3 unknowns between 
\lambdaz\ and \lambdap. These are found by (i) classifying each characteristic as a shock, 
rarefaction, or contact discontinuity; and (ii) joining the solutions in such a way that 
the Rankine-Hugoniot (RH) jump conditions are satisfied across shocks and the generalized 
Riemann invariants are conserved across rarefactions and contact discontinuities.
For a detailed discussion of the underlying theory, see the monograph by Toro (2009).

\subsection{The \lambdaz\ Characteristic}
\label{sec-lambdaz}

The classification of \lambdaz\ depends on the ``eigenvalue gradient,''
\be
\mbox{\boldmath$\nabla$}\lambdaz
\equiv
\left[
\frac{\partial\lambdaz}{\partial U_{\rm i1}},
~\frac{\partial\lambdaz}{\partial U_{\rm i2}},
~\frac{\partial\lambdaz}{\partial U_{\rm i3}}
\right].
\ee
It is easy to show that
\be
\mbox{\boldmath$\nabla$}\lambdaz
~=~
\frac{1}{\rho_i}
\left[-v_i,   ~1,  ~0\right]
\ee
and hence that 
\be
\mbox{\boldmath$\nabla$}\lambdaz\vdot\Rz=0.
\label{eq-lambdazcrit}
\ee
According to the theory of hyperbolic\footnote{If $B\ne 0$ the eigenvalues of \matA\ are
real and nondegenerate.  This guarantees that eqs.~(\ref{eq-motionjac}) are strictly
hyperbolic (e.g., Jeffrey 1976).} PDEs, equation (\ref{eq-lambdazcrit})
establishes that \lambdaz\ is always a contact discontinuity. Conservation of the Riemann
invariants across the $m$th characteristic implies that
\be
\frac{dU_1}{R_{m1}} ~=~ \frac{dU_2}{R_{m2}} ~=~ \frac{dU_3}{R_{m3}}
\ee
(see Toro 2009, \S2.4.4) and writing this out for $m=0$ gives
\be
{d\rho_i \over \rho_i}
=
{d\left(\rho_i v_i\right) \over \rho_i v_i}
=
{dB \over 0}.
\label{eq-RiemRz}
\ee
Integrating equation~(\ref{eq-RiemRz}) across the contact discontinuity yields
\be
\label{lamioeq}
\viSL = \viSR \equiv \viS
\ee
and
\be
\BSL = \BSR \equiv \BS.
\ee
Thus the ion density may undergo a jump across the contact discontinuity but the other 
fluid variables are continuous, in precise analogy with contact discontinuities in 
gas dynamics. Notice that the number of unknowns in the star region has now been 
reduced to four:
\viS, \BS, \rhoiSL, and \rhoiSR.

\subsection{The $\lambdap$ and $\lambdam$ Characteristics}
\label{sec-lambdapm}

The eigenvalue gradients for the other characteristics are
\be
\mbox{\boldmath$\nabla$}\lambda_{\pm}
=
\frac{1}{\rho_i}
\left[
\begin{array}{ccc}
-\frac{1}{\rhoi}\left(\vi \mp \viA/2\right), &
\frac{1}{\rhoi}, &
\mp \frac{1}{\sqrt{4 \pi \rhoi}}
\end{array}
\right],
\ee
from which it follows that
\be
\mbox{\boldmath$\nabla$}\lambda_{\pm}
\cdot
\vecR_{\pm}
=
\mp\frac{3}{2}\viA.
\label{eq-lambdapmcrit}
\ee
It seems reasonable to assume on physical grounds that $B \ne 0$ everywhere for $t>0$ (no 
vacuum) so that the RHS of expression (\ref{eq-lambdapmcrit}) is always nonzero. Then the 
theory of hyperbolic PDEs establishes that \lambdam\ and \lambdap\ are never contact 
discontinuities. The \lambdam\ wave is a shock wave if the pressure in the star region 
exceeds the pressure in the left region ($\BS>\BL$) and a rarefaction wave otherwise.
Similarly, \lambdap\ is a shock if and only if $\BS>\BR$. These conclusions follow from the
fact that shocks are compressive and rarefactions are not, plus our assumption that the 
pressure of the plasma is entirely magnetic. For a given set of initial conditions, only 
one combination of shocks and/or rarefactions at \lambdam\ and \lambdap\ gives a solution which
satisfies the matching conditions across all three discontinuities. In 
\S\ref{sec-flowconfig} we give a simple criterion for identifying the correct combination.

The matching conditions for shock waves are the RH jump conditions,
which require \Fveci\ to be conserved across the shock front in a frame comoving with the shock.
We omit details of the derivation and simply give the results.
For a shock at \lambdam\ (a ``left shock''), we find that the ion velocities
on opposite sides of the shock front are related by
\be
\viS = \viL + \fLS\left(\BS\right),
\ee
where
\be
\label{eq-fLS-def}
\fLS\left(\BS\right)
\equiv
-\left[\,
\frac{ \left(\PS-\PL\right) }{ \rhoiL }\,
\left(1-\frac{\BL}{\BS}\right)
\right]^{1/2}
\ee
and $P(B)\equiv B^2/8\pi$ is the magnetic pressure. Similarly, for a right shock we find
\be
\viS = \viR + \fRS\left(\BS\right),
\ee
where
\be
\label{eq-fRS-def}
\fRS\left(\BS\right)
\equiv
+\left[\,
\frac{ \left(\PS-\PR\right) }{ \rhoiR }\,
\left(1-\frac{\BR}{\BS}\right)
\right]^{1/2}.
\ee

The matching conditions for rarefactions are governed by the Riemann invariants.
Carrying out steps analogous to the derivation of eq.~(\ref{eq-RiemRz}), we find that
\be
\frac{d\rhoi}{\rhoi}
=
\frac{d\left(\rhoi\vi\right)}{\rhoi\left(\vi\pm\viA\right)}
=
\frac{dB}{B},
\label{eq-riemrare}
\ee
where the upper sign corresponds to \lambdap.
Equating the first and third terms simply gives flux freezing in the charged fluid.
This implies that the magnetic field and hence \viA\ may be viewed as
functions of the ion density alone. Knowing this, we equate the first two terms in
eq.~(\ref{eq-riemrare}) to obtain a differential equation for \vi:
\be
d\vi = \pm\viA\left(\rhoi\right)\,\frac{d\rhoi}{\rhoi}.
\label{eq-ode-vel}
\ee
Integrating eq.~(\ref{eq-ode-vel}) across \lambdam\ gives the matching condition for a 
left rarefaction:
\be
\viS = \viL + \fLR\left(\BS\right),
\ee
where
\be
\fLR\left(\BS\right)
\equiv
+2\viAL\left[1-
\left(\frac{\BS}{\BL}\right)^{1/2}
\right]
\label{eq-fLR-def}
\ee
and $\viAL$ is the ion {\Alf} speed in the left region. For a right rarefaction one finds
similarly that
\be
\viS = \viR + \fRR\left(\BS\right),
\ee
where
\be
\fRR\left(\BS\right)
\equiv
-2\viAR\left[1-
\left(\frac{\BS}{\BR}\right)^{1/2}
\right].
\label{eq-fRR-def}
\ee

\subsection{Flow Configuration and Solution}
\label{sec-flowconfig}

Let RR, SR, RS, and SS denote the four possible flow configurations, where ``RR'' is
a flow with two rarefactions, ``SR'' a left shock plus a right rarefaction, and so on.
If the configuration was known {\it a priori}, one could write the matching conditions 
across the \lambdam\ wave,
\be
\viS = \viL + \fL\left(\BS\right),
\label{eq-match-L}
\ee
and the \lambdap\ wave,
\be
\viS = \viR + \fR\left(\BS\right),
\label{eq-match-R}
\ee
by choosing \fL\ and \fR\ appropriately from the functions \fLS, \fLR, etc.
Subtracting equation (\ref{eq-match-R}) from (\ref{eq-match-L}) gives
\be
\viL-\viR + \fL\left(\BS\right) - \fR\left(\BS\right) = 0,
\label{eq-bstar}
\ee
which is a nonlinear equation for \BS. Once \BS\ has been determined by solving
equation (\ref{eq-bstar}), the velocity in the star region follows from equation (\ref{eq-match-L})
or equation (\ref{eq-match-R}). Finally, the matching conditions for shocks and rarefactions
both require flux freezing in the plasma, and this determines the density in both 
halves of the star region:
\be
\label{rhoiSLeq}
\rhoiSL = \rhoiL\left(\frac{\BS}{\BL}\right)
\ee
and
\be
\label{rhoiSReq}
\rhoiSR = \rhoiR\left(\frac{\BS}{\BR}\right).
\ee

To implement the steps above it remains only to identify the unique flow configuration 
which satisfies the matching conditions across all three characteristics, an exercise in
the process of elimination. For example, suppose that the flow is of type RR, which
requires that $\BS<\BL$ and $\BS<\BR$, for the case $\viL-\viR>0$ (the flows collide).
Equation (\ref{eq-bstar}) has a solution only if $\fL-\fR<0$; but examination of 
equations (\ref{eq-fLR-def}) and (\ref{eq-fRR-def}) shows that $\fLR-\fRR$ is strictly 
positive if $\BS<\BL$ and $\BS<\BR$. We conclude then that the RR configuration 
{\em never}\/ occurs when $\viL - \viR > 0$. Similarly, consider a flow of type SS,
which has $\BS > \BL$ and $\BS > \BR$, for the situation $\viL - \viR < 0$ (the flows 
diverge). For that situation, according to equation (\ref{eq-bstar}) there can only be 
a solution if $\fL - \fR >0$. Equations (\ref{eq-fLS-def}) and (\ref{eq-fRS-def}) reveal
that, for $\BS > \BL$ and $\BS > \BR$, $\fL - \fR$ is always $< 0$. We conclude then that
the SS configuration {\em cannot} occur when $\viL - \viR < 0$. Straightforward but lengthy
analysis of the other possibilities shows that the configuration depends only on three 
dimensionless parameters,
\bml
\be
\label{eq-xidef}
\xi \equiv \frac{\viL-\viR}{\viAL},
\ee
\be
\theta \equiv \BR/\BL,
\ee
and
\be
\delta \equiv \sqrt{\rhoiR/\rhoiL}.
\ee
\eml
The flow classification is given in Table~\ref{table-flowclass}, where
\bml
\bea
\label{eq-Gammadef}
\Gamma
&\equiv&
\frac{1}{\delta\sqrt{2}}\,\left[
\left(1-\theta^2\right)\left(1-\theta\right)
\right]^{1/2}-\xi~~, \\
\label{eq-Psidef}
\Psi
&\equiv&
\frac{1}{\sqrt{2}}\,\left[
\left(\theta^2-1\right) \left(1-\theta^{-1}\right)
\right]^{1/2}
-\xi~~, \\
\label{Phidef}
\Phi &\equiv& \frac{2}{\delta}(\theta - \sqrt{\theta}) + \xi~~, \\
\label{Upsilondef}
\Upsilon &\equiv&  2 (1 - \sqrt{\theta}) + \xi~.
\eea
\eml

Once the flow type is known, which of the expressions
(\ref{eq-fLS-def}), (\ref{eq-fRS-def}), (\ref{eq-fLR-def}), and (\ref{eq-fRR-def}) 
should be used as the functions $\fL(\BS)$ and $\fR(\BS)$ in equation
(\ref{eq-bstar}) is then also known. The value of $\BS$ can then be determined
to any desired level of accuracy from equation (\ref{eq-bstar}) using
an iterative numerical technique such as Newton's method (Aktkinson 1989;
Press et al.\ 1996). We note that for flows with both left and right rarefaction waves 
(type RR), the ion mass densities in the star region $\rhoiSL$ and 
$\rhoiSR$ must be $> 0$; it follows from equations (\ref{rhoiSLeq}) and (\ref{rhoiSReq}) 
that the magnetic field in the star region $\BS$ must also be $> 0$. 
This non-vacuum (or, positivity) condition imposes a lower limit on the value of 
$\xi$, the dimensionless velocity difference between the left and right states
of the initial discontinuity (\ref{eq-xidef}). Using equations (\ref{eq-fLR-def}), (\ref{eq-fRR-def}), and 
(\ref{eq-bstar}), one finds that this condition is satisfied so long as 
$\xi > \xi_{\rm vac}$, where
\be
\label{vaccond}
\xi_{\rm vac} \equiv -2\left(1 +\frac{\theta}{\delta}\right)~.
\ee 

\begin{figure}
\plotone{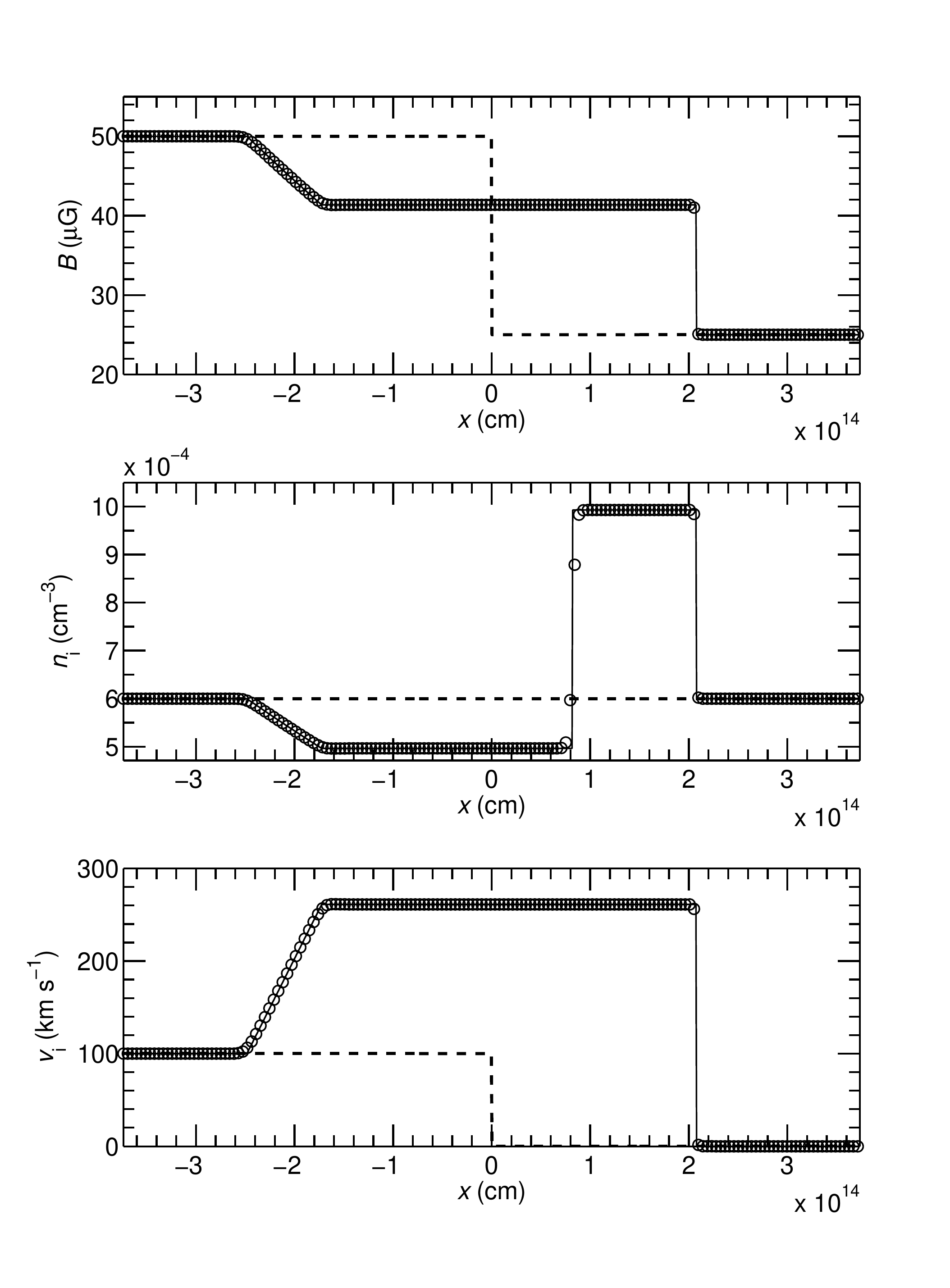}
\caption{First MHD Riemann demonstration problem. Dashed lines indicate the initial state of the 
charged fluid and magnetic field. The solid line is the exact MHD Riemann solution at 
$t=0.1~\yr$. Circles are the solution calculated using the approximate MHD Riemann solver 
(some data points have been omitted for clarity). Top panel: magnetic field. Middle: ion 
number density. Bottom: ion velocity.
\label{fig-Riemanntestone}
}
\end{figure}

As demonstrative examples, we consider two representative MHD Riemann test problems. In 
the first problem, the initial state of the plasma and magnetic field has 
$\BL = 50~\mu{\rm G}$, $\BR = 25~\mu{\rm G}$, $\viL = 200~\kms$, $\viR = 0$, and 
$\rhoiL=\rhoiR=2.51 \times 10^{-26}~{\rm g}~\cc$ (for a cloud with $\mi = 25\mprot$, 
this corresponds to a number density $\nni = 6 \times 10^{-4}~\cc$). For these parameters 
$\BL > \BR$, and $\Gamma = 0.321 > 0$. Examination of Table \ref{table-flowclass} indicates
that the flow configuration for this first test will be of type RS, a left rarefaction with 
a right shock. That this is so can be seen in Figure \ref{fig-Riemanntestone} which shows 
the results for this problem. The initial state of the plasma and magnetic field are displayed
as dashed lines, and the solid curves are the MHD Riemann solution at the time $t=0.1~\yr$. 
The right shock is located at $x_{\rm shk} = +2.07 \times 10^{14}~\cm$, and the head of the 
left rarefaction wave is at $x_{\rm rw} = -2.49 \times 10^{14}~\cm$. A contact discontinuity 
is located at $x_{\rm c} = +8.23 \times 10^{13}~\cm$; that is the position the initial 
discontinuity in the charged fluid's magnetic flux-to-mass ratio $B/\rhoi$ has moved to by 
the time shown.
\begin{center}
\begin{table}
\caption{\mbox{\hspace{6.5em} Classification of Solutions for the Charged Fluid}
\label{table-flowclass}
}
\begin{tabular}{llcl} \hline \hline \\
 \mbox{\hspace{7em}} & \mbox{Initial Conditions \hspace{1em}} & \mbox{\hspace{1em} Flow Type \hspace{1em}} & 
\mbox{\hspace{3em} Note \hspace{7em}} \\ \\
\hline
\\
$\viL - \viR > 0$~: & & & \\
& $B^{L}<B^{R}$, $\Psi<0$ & SS & left shock + right shock       \\ \\
& $B^{L}<B^{R}$, $\Psi>0$ & SR & left shock + right rarefaction \\ \\
& $B^{L}>B^{R}$, $\Gamma<0$  & SS & left shock + right shock       \\ \\
& $B^{L}>B^{R}$, $\Gamma>0$  & RS & left rarefaction + right shock \\ \\
\hline \\
$\viL - \viR = 0$~: & & & \\
& $B^{L}<B^{R}$, $\Psi>0$ & SR & left shock + right rarefaction \\ \\
& $B^{L}>B^{R}$, $\Gamma>0$  & RS & left rarefaction + right shock \\ \\
\hline \\
$\viL - \viR < 0$~: & & & \\
& $B^{L}<B^{R}$, $\Phi<0$  & RR\tbnm{\dagger} & left rarefaction + right rarefaction \\ \\
& $B^{L}<B^{R}$, $\Phi>0$  & SR & left shock + right rarefaction \\ \\
& $B^{L}>B^{R}$, $\Upsilon<0$  & RR\tbnm{\dagger} & left rarefaction + right rarefaction \\ \\
& $B^{L}>B^{R}$, $\Upsilon>0$  & RS & left rarefaction + right shock \\ \\
\hline
\end{tabular}
\\ \\
${}^{\dagger}$The RR-state can exist only if 
$-2(1+\theta/\delta)\viAL < \viL - \viR < 0$ (see eq. [\ref{vaccond}]).
\end{table}
\end{center}

\begin{figure}
\plotone{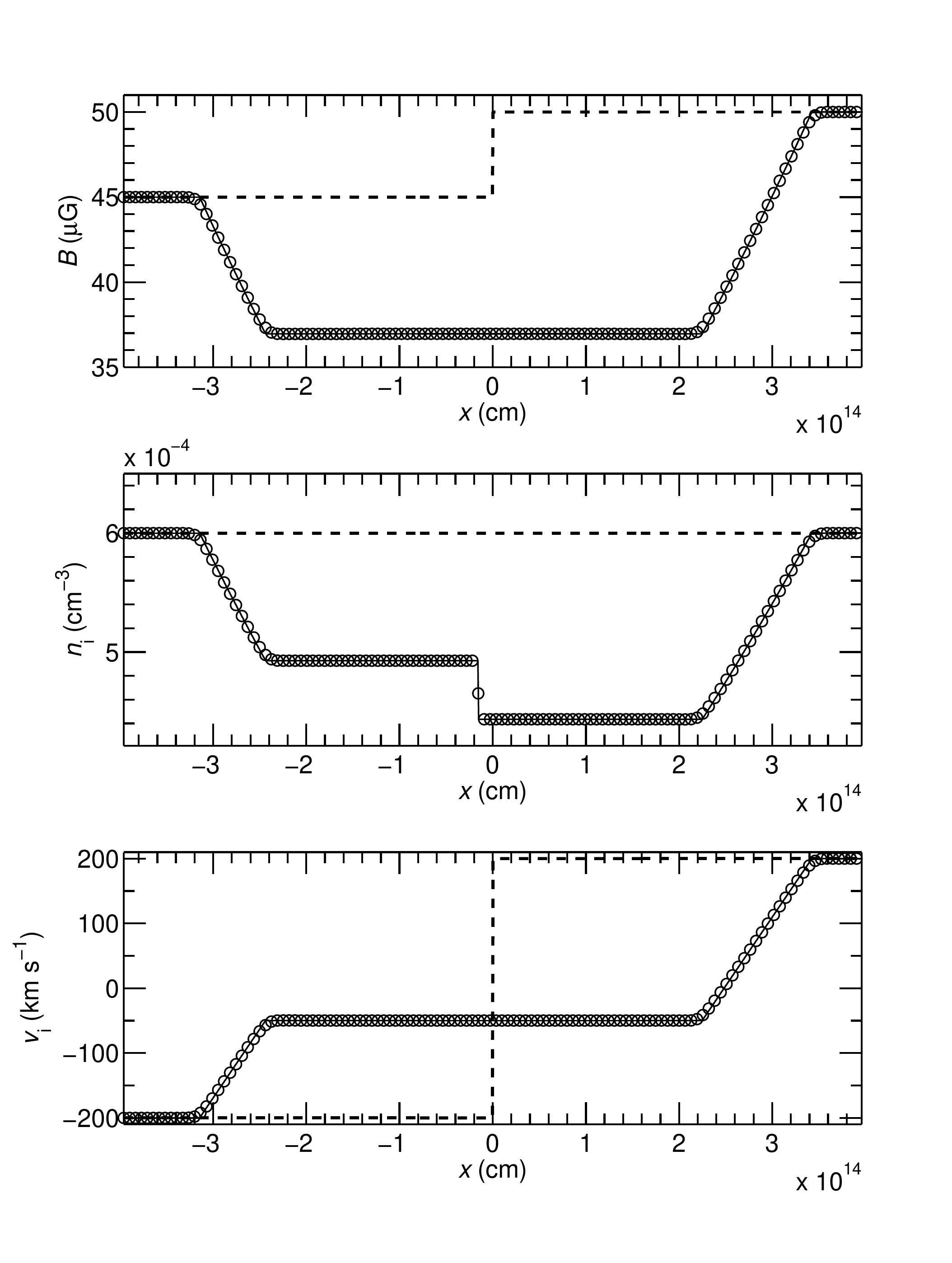}
\caption{Same as in Fig. \ref{fig-Riemanntestone}, but for the second MHD Riemann 
demonstration problem. Data shown at $t = 0.1~\yr$.
\label{fig-Riemanntesttwo}
}
\end{figure}

In the second MHD Riemann test problem we have an initial state with $\BL = 45~\mu{\rm G}$,
$\BR = 50~\mu{\rm G}$, $\viL = -200~\kms$, $\viR =+200~\kms$, and the same constant
ion density as in the first test. This state has $\BL < \BR$, $\viL < \viR$, and
$\Phi = -0.385 < 0$. According to Table \ref{table-flowclass} this should be an RR flow.
This is indeed the case, as can be seen in Figure \ref{fig-Riemanntesttwo} which
displays the exact MHD Riemann solution for this model at $t = 0.1~\yr$. At the time shown, 
the head of the left rarefaction wave is at 
$x_{\rm L,rw} = -3.14 \times 10^{14}~\cm$, and the head of the right rarefaction wave
is at $x_{\rm R,rw} = + 3.42 \times 10^{14}~\cm$. There is also a charged fluid contact
discontinuity at $x_{\rm c} = -1.70 \times 10^{13}~\cm$.

\section{An Approximate MHD Riemann Solver}
\label{sec-approxmhdriemann}

\subsection{The Star Region}
\label{sec-starvalues}

The Riemann solution found in Appendix \ref{sec-exactmhdriemann} can also be used to 
numerically solve the MHD flow problem of equation (\ref{mhdRGprob}) using Godunov's method
(Godunov 1959; Toro 2009). However, doing so can be computationally expensive, because it 
uses an iterative method to solve for $\BS$ from equation (\ref{eq-bstar}). 

It is desirable then, to instead use a more efficient approximate solver to calculate the
variables in the star region ($\BS$, $\viS$, $\rhoiSR$, and $\rhoiSL$) between the 
characteristics $\lambdam$ and $\lambdap$ (see Fig. \ref{fig-characteristics}). This can be 
done by assuming that the flow motion 
is approximately linear, that is, a wave, and then using the characteristic 
relations derived from the Riemann invariants (eq. [\ref{eq-riemrare}]) to connect the 
variables in the star region to those in the adjacent L or R region of the Riemann problem
(see Fig. \ref{fig-characteristics}). But this is just what was done in deriving 
equations (\ref{eq-fLR-def}) and (\ref{eq-fRR-def}). Solving those two equations for $\BS$
and $\viS$ in terms of the L and R region variables yields 
\bea
\label{Bstarapprox}
\BS &=& \left[\frac{\viAL + \viAR + \frac{1}{2}\left(\viL - \viR\right)}{\CiL + \CiR}\right]^{2}~, \\
\label{vistarapprox}
\viS &=& \frac{\CiL\viR + \CiR\viL + 2 \CiL\CiR\left(\sqrt{\BL} - \sqrt{\BR}\right)}
{\CiL + \CiR}~,
\eea
where
\be
\label{Cidef}
\CiL \equiv \frac{\viAL}{\sqrt{\BL}}~~, ~~\CiR \equiv \frac{\viAR}{\sqrt{\BR}}~. 
\ee
With these quantities now known, $\rhoiSL$ and $\rhoiSR$ can be calculated directly from 
equations (\ref{rhoiSLeq}) and (\ref{rhoiSReq}), respectively. 

\subsection{Calculation of Fluxes at the Discontinuity}
\label{sec-fluxes}

To solve the MHD Riemann problem for a state having the discontinuous initial conditions 
(\ref{riemannconditions}) using the method of Godunov (1959), we need to calculate the 
the flux vector $\Fveci(0)$ at the interface separating the two states, $x=0$ (or, 
equivalently, along the ray with the similarity variable $x/t = 0$, which corresponds to 
the $t$-axis in Fig. \ref{fig-characteristics}). To calculate $\Fveci(0)$ from equation 
(\ref{Fvecidef}), we need the array of variables $\Uveci(0)$ (eq. [\ref{Uvecidef}]).

A very thorough discussion of how to obtain the state variables at the interface $x=0$
for the one-dimensional gas dynamic Riemann problem using Godunov's method is presented in
Ch.\ 6 of Toro (2009); the fundamental ideas described there --- in which the 
Riemann-Godunov flux is determined by examining the characteristic wave structure at the 
discontinuity --- remain valid and carry over directly to the MHD Riemann problem that we 
are concerned with here. The first step in the assignment of $\Uveci(0)$ is made by finding
the values of the velocity $\viS$ ($=\lambdaz$) of the contact discontinuity and the magnetic
field of the star region $\BS$ from equations (\ref{Bstarapprox}) and (\ref{vistarapprox}).

For $\viS > 0$, the contact discontinuity is moving to the right. If $\BS > \BL$ the left 
wave is a shock wave with velocity
\be
\label{sLdef}
\sL = \viL - \QL/\rhoiL~ 
\ee
in the stationary frame. This relation was obtained by applying the RH conditions in a 
frame moving with the shock front. The quantity
\be
\label{QLdef}
\QL = \frac{(\BL)^{2}/8\pi - (\BS)^{2}/8\pi}{\viS - \viL}
\ee
is the ion mass flux across the front. If $\sL > 0$, the shock is moving to the
right and $\Uveci(0) = \ULveci$. If $\sL < 0$, the shock is instead heading to the left,
and $\Uveci(0) = \USLveci \equiv \left[\rhoiSL~,~\rhoiSL \viS~,\BS \right]^{\rm T}$

If $\BS < \BL$, the left wave will instead be a rarefaction. As is widely-known (e.g., 
Courant \& Friedrichs 1948; Landau \& Lifshitz 1959; Zeldovich \& Raizer 1966; Toro 2009) 
a rarefaction emanating from a discontinuity has a fan-like phase-space structure (a 
centered simple wave) with a head and a tail. The velocities of the head ($\hL$) and tail 
($\tL$) of the left rarefaction fan are
\be
\label{hLtLdef}
\hL = \viL - \viAL~, \hspace{1em} \tL =\viS - \viASL~,
\ee
respectively, where
\be
\label{viASLdef}
\viASL \equiv \frac{\BS}{\sqrt{4 \pi \rhoiSL}}~.
\ee
For $\hL > 0$, $\tL > 0$, the entire rarefaction fan is 
heading to the right and $\Uveci(0) = \ULveci$, while for $\hL < 0$, $\tL < 0$ the entire
fan is traveling left and $\Uveci(0) = \USLveci$. If $\hL < 0$ and $\tL > 0$ values
inside the rarefaction fan are needed (e.g., see Ch.\ 4 of Toro 2009). In that situation, 
$\Uveci(0) = \ULfanveci \equiv \left[\rhoiLfan~,~\rhoiLfan\viLfan~,~\BLfan \right]^{\rm T}$~.
Using the Riemann invariant relation (\ref{eq-riemrare}) to integrate across the $\lambdam$
characteristic to connect quantities in the left rarefaction fan to those in the left 
state, along with the relations (\ref{ieigenvals}), (\ref{eq-characcurves}), and 
(\ref{rhoiSLeq}), and setting $x/t = 0$ in the resulting expressions yields 
\be
\label{viLfan}
\viLfan = \frac{2}{3}\left(\viAL + \frac{1}{2}\viL\right)~,\hspace{1em}
\BLfan  =  \BL \left(\frac{2}{3} + \frac{\viL}{3 \viAL}\right)^{2} ~, \hspace{1em}
\rhoiLfan  =  \frac{\rhoiL \BLfan}{\BL}~.
\ee

If $\viS < 0$, the contact discontinuity is traveling to the left. For $\BS> \BR$
the right wave is a shock wave. The velocity of the shock in the stationary
frame is
\be
\label{sRdef}
\sR = \viR + \QR/\rhoiR~,
\ee
where
\be
\QR \equiv \frac{(\BS)^{2}/8\pi - (\BR)^{2}/8\pi}{\viS - \viR} 
\ee
is the ion mass flux across the shock front. If $\sR > 0$ the right shock is
traveling to the right and 
$\Uveci(0) =\USRveci \equiv \left[\rhoiSR~,~\rhoiSR\viS~,~\BS\right]^{\rm T}$;
if $\sR < 0$ the shock is heading left, and $\Uveci(0)=\URveci$ instead.

For $\BS < \BR$ the right wave will be a rarefaction. The right rarefaction
fan has head ($\hR$) and tail ($\tR$) velocities
\be
\label{hRdef}
\hR = \viR + \viAR~,\hspace{1em}
\tR = \viS + \viASR~,
\ee
where
\be
\label{viASRdef}
\viASR \equiv \frac{\BS}{\sqrt{4\pi \rhoiSR}}~.
\ee
For $\hR > 0$, $\tR > 0$, the entire fan is rightward traveling and $\Uveci(0)=\USRveci$,
while, for $\hR < 0$, $\tR < 0$ the complete fan is heading to the left, and therefore
$\Uveci(0)=\URveci$. If $\hR < 0$ and $\tR > 0$ values from inside the right rarefaction
fan are required and
$\Uveci(0)= \URfanveci \equiv \left[\rhoiRfan~,~\rhoiRfan\viRfan~,~\BRfan\right]^{\rm T}$. 
Integrating the Riemann invariant (\ref{eq-riemrare}) across $\lambdap$ to
link quantities in the tail of the right rarefaction fan to quantities in
the right state, also using equations (\ref{ieigenvals}), (\ref{eq-characcurves}),
and (\ref{rhoiSReq}), and taking $x/t=0$ in the resulting relations, gives  
\be
\label{viRfan}
\viRfan = \frac{2}{3}\left(-\viAR + \frac{1}{2}\viR \right)~,\hspace{1em}
\BRfan = \BR \left(\frac{2}{3} - \frac{\viR}{3\viAR}\right)^{2}~,\hspace{1em}
\rhoiRfan = \frac{\rhoiR \BRfan}{\BR}~.
\ee

As we have stated earlier, approximate Riemann solvers for the one-dimensional neutral gas
dynamic Riemann-Godunov problem have a very similar structure to that of the MHD solver we
present here. In several instances the algorithm for the MHD and neutral gas Riemann 
solvers are basically the same when a suitable substitution is made. For example, the 
relations we derived for quantities in the left and right rarefaction fans (eqs. 
[\ref{viLfan}] and [\ref{viRfan}]) are identical to the left and right fan relations at 
$x/t = 0$ for the density, velocity and thermal pressure of an ideal non-magnetic gas given
by equations [4.56] and [4.63] of Toro (2009) if one makes the substitution 
$P = B^{2}/8\pi$, replaces the speeds of sound in the left and right states with $\viAL$ 
and $\viAR$, respectively, and uses the fact that the adiabatic index for a flux-frozen 
plasma 
$\gamma \equiv \partial\ln P/\partial\ln\rhoi=\partial\ln(B^{2}/8\pi)/\partial\ln\rhoi=2$
in Toro's expressions. 

Although a linear approximation was made in its derivation, we find the MHD Riemann
solver presented here to be quite accurate and robust, even when dealing with shocks. It 
matches the exact Riemann solution for a variety of test problems, including conditions 
likely to be encountered when studying high-velocity shocks and flows in interstellar 
clouds. Figures \ref{fig-Riemanntestone} and \ref{fig-Riemanntesttwo} show the results of 
RG calculations (circles) for the two example MHD Riemann problems described in 
\S~\ref{sec-flowconfig}. The simulations using the approximate MHD solver are seen to be in
very good agreement with the exact Riemann solutions for those model tests.

\section{Analytic Benchmark Solutions}
\label{sec-benchmarks}

We present here a derivation of the analytical small-amplitude wave solutions used to test 
the split-operator models discussed in \S\S~\ref{sec-firstwavemodel} -
\ref{sec-thirdwavemodel}. For the governing fluid equations we linearize the system in the 
standard way, assuming for any physical quantity $f$ that
\be
\label{perturbeq}
f(x,t) = f_{0} + \delta f(x,t);
\ee
the zero-order value $f_{0}$ is taken to be constant and uniform, and the first-order 
perturbation $|\delta f(x,t)| \ll f_{0}$. Inserting these perturbations into equations 
(\ref{neutmassconteq}) - (\ref{inducteq}) and retaining terms up to first order give the 
linearized system of equations:
\bea
\label{rhonperturb}
\frac{\partial}{\partial t}\left(\delrhon\right) & = & -\rhono\frac{\partial}{\partial x}\left(\delvn\right)~, \\
\label{vnperturb}
\rhono\frac{\partial}{\partial t}\left(\delvn\right) & = & 
-\frac{\partial}{\partial x}\left(\delta \Pn\right) - \frac{\rhono}{\tni}\delvn + \frac{\rhono}{\tni}\delvi~,\\
\label{enperturb}
\frac{1}{\Tn_{0}}\frac{\partial}{\partial t}\left(\delta\Tn\right) &=& (\gamma - 1)\frac{1}{\rhono}
\frac{\partial}{\partial t}\left(\delta\rhon\right)~, \\
\label{rhoiperturb}
\frac{\partial}{\partial t}\left(\delrhoi\right) & = & -\rhoio\frac{\partial}{\partial x}\left(\delvi\right)~, \\
\label{viperturb}
\rhoio\frac{\partial}{\partial t}\left(\delvi\right) & = & -\frac{B_{0}}{4 \pi} 
\frac{\partial}{\partial x}\left(\delB\right) + \frac{\rhoio}{\tin}\delvn - \frac{\rhoio}{\tin}\delvi~,\\
\label{Bperturbb}
\frac{\partial}{\partial t}\left(\delB\right) & = & -B_{0}\frac{\partial}{\partial x}\left(\delvi\right)~,
\eea
where we have used the fact that the charged and neutral fluids are at rest in the zero-order
reference state ($\vn_{0}=\vi_{0}=0$), and also that $\Fnin=\Sn = \Si = \Gn = \Ln = 0$ 
for all of the model tests in \S~\ref{sec-linwaves}. Equation (\ref{Fneq})
was also used to substitute for $\Fn$ in the perturbed force equations 
(\ref{vnperturb}) and (\ref{viperturb}). 

The internal energy for an ideal gas (\ref{eninteq}) having a ratio of specific heats 
$\gamma$ was used in the energy equation of the neutral gas (\ref{neutenergyeq}) to derive 
(\ref{enperturb}); integrating that latter equation and inserting the result into the
linearized ideal gas law 
(\ref{idealgaslaw}), 
\be
\label{linidealgas}
\frac{\delta \Pn}{\Pn_{0}} = \frac{\delrhon}{\rhono} + \frac{\delta \Tn}{\Tn_{0}}~,
\ee
yields the adiabatic relation between density and pressure perturbations in the neutral 
fluid,
\be
\label{linP}
\delta \Pn = \frac{\gamma \Pn_{0}}{\rhono} \delrhon = \Cs^{2} \delrhon~,
\ee
where $\Cs$ is the adiabatic speed of sound (\ref{Csdef}).

Combining equations (\ref{rhoiperturb}) and (\ref{Bperturbb}) and integrating
we immediately obtain the result
\be
\label{fluxfreezeq}
\delrhoi/\delB = \rhoio/B_{0}~,
\ee
which simply expresses freezing of magnetic flux in the charged fluid. Inserting this
result in equation (\ref{perturbeq}) and using $\nni = \rhoi/\mi$, equations
(\ref{iwave}), (\ref{idiff}), and (\ref{nninmseq}) follow directly.

\subsection{Solution for $\lgauss \ll \Lims$}
\label{sec-imssolns}

In this limit the neutral fluid is unable to respond to the very short time and 
length scales of the perturbations; therefore, for this situation the neutrals are
effectively a fixed stationary background with $\delvn = \delrhon = 0$, and the
only equations relevant to these particular modes are (\ref{viperturb}) and (\ref{Bperturbb}).

The Fourier transform $\fcarat$ of any function $f$ is given by
\be
\label{FTeq}
\fcarat(k,t) = \frac{1}{\sqrt{2 \pi}} \int^{\infty}_{-\infty} f(x,t) e^{-ikx} dx~,
\ee
where $k$ is the wavenumber. The inverse transform is
\be
\label{invFTeq}
f(x,t) = \frac{1}{\sqrt{2 \pi}} \int^{\infty}_{-\infty}\fcarat(k,t) e^{ikx} dk
\ee

Fourier transforming equations (\ref{viperturb}) and (\ref{Bperturbb})
gives a system of linear ordinary differential equations with constant
coefficients:
\be
\label{iFTsystem}
\frac{\partial \ycarati}{\partial t} = \matMi \ycarati~,
\ee
where the vector of transformed variables 
$\ycarati(k,t) \equiv [\delvicarat(k,t),\delBcarat(k,t)]^{\rm T}$
and
\be
\label{matMidef}
\matMi
\equiv
\left[
\begin{array}{cc}
       -1/\tin     &  -i B_{0} k/4 \pi \rhoio \\
     -i B_{0} k     &  0 \\
\end{array}
\right].
\ee
The ODE system (\ref{iFTsystem}) has the solution
\bml
\be
\label{ODEsoln}
\ycarati = \aplusi \Eplusi e^{\gplusi t} + \aminusi \Eminusi e^{\gminusi t}
\ee
(e.g., see ch. 3 of Braun 1983),
where the eigenvalues and eigenvectors of $\matMi$ are, respectively,
\be
\label{gpmeq}
\gpmi = -\frac{1}{\tin} \pm i \vims k \left[1 - \left(\frac{\kims}{k}\right)^{2}\right]^{1/2}
\ee
and 
\be
\label{Epmeq}
\Epmi \equiv \left[1 , -i \frac{B_{0} k}{\gpmi} \right]^{\rm T}~.
\ee
\eml
In these expressions we have used equation (\ref{viAdef}) and the fact that $\vims = \viA$
for all conditions of interest in our model interstellar clouds. We also 
introduced the ion magnetosound cutoff wave number
\be
\label{kimsdef}
\kims \equiv \frac{1}{2\vims \tin} = \frac{2 \pi}{\Lims}~;
\ee
modes with $k > \kims$ propagate as ion magnetosound waves. 

The constants $\aplusi$ and $\aminusi$ in equation (\ref{ODEsoln}) are determined
by applying the initial conditions $\delvicarat(k,0) = 0$ (ions initially at rest)
and the Fourier transform of the initial Gaussian pulse in the magnetic
field (see eq. [\ref{Bperturb}]),
\be
\label{Bpertinit}
\delBcarat(k,0) = \frac{B_{0}\ampl \lgauss}{\sqrt{2}} \exp\left(-\frac{\lgauss^{2}k^{2}}{4}\right)~.
\ee
Doing that yields 
\bea
\label{Bcaratsoln}
\delBcarat(k,t) & = &\frac{B_{0} \ampl \lgauss}{2 \sqrt{2}} \exp\left(-\frac{t}{2\tin}\right) \nonumber \\
& & \times \left\{\left[1 - \frac{i \kims/k}{[1 - (\kims/k)^{2}]^{1/2}}\right]
\exp\left(-\frac{\lgauss^{2}k^{2}}{4} + i \vims k t \left[1 - \left(\frac{\kims}{k}\right)^{2}\right]^{1/2} \right) \right. \nonumber \\ 
& & \left. \hspace{1em} +
\left[1 + \frac{i \kims/k}{[1 - (\kims/k)^{2}]^{1/2}}\right]
\exp\left(-\frac{\lgauss^{2}k^{2}}{4} - i \vims k t 
\left[1 - \left(\frac{\kims}{k}\right)^{2}\right]^{1/2}  \right)
\right\} 
\hspace{3em}
\eea
and
\bea
\label{vicaratsoln}
\hspace{-4em} \delvicarat(k,t) &=& \frac{-\vims \ampl \lgauss}{2\sqrt{2}\left[1 - (\kims/k)^{2}\right]^{1/2}}
\exp\left(-\frac{t}{2\tin}\right) 
\left\{\exp\left(-\frac{\lgauss^{2}k^{2}}{4} + i \vims kt\left[1 - \left(\frac{\kims}{k}\right)^{2}\right]^{1/2} \right)
\right. \nonumber \\
& & \left. 
\hspace{16em}- \exp\left(-\frac{\lgauss^{2}k^{2}}{4} - i \vims kt\left[1 - \left(\frac{\kims}{k}\right)^{2}\right]^{1/2} \right) 
\right\}~. \hspace{3em}
\eea

$\delta B(x,t)$ and $\delta \vi(x,t)$ are obtained by inverse transforming (\ref{Bcaratsoln})
and (\ref{vicaratsoln}). Performing the integration in the inverse transforms is greatly 
aided by noting that because the initial perturbation in the magnetic field excites 
wavelengths primarily in the region $\sim \lgauss$, the condition $\lgauss \ll \Lims$ implies
that the contributions in the Gaussian packet are principally from waves with $k \gg \kims$. 
Taking the limit $(\kims/k)^{2} \rightarrow 0$ will therefore introduce only a negligible
error 
when inverting the transforms. Applying this limit gives the final solutions
\bea
\label{Bwavesoln}
\delB(x,t) &=& B_{0}\frac{\ampl}{2}\exp\left(-\frac{t}{2\tin}\right)
\left[\exp\left(\frac{-(x-\vims t)^{2}}{\lgauss^{2}}\right)
+\exp\left(\frac{-(x+\vims t)^{2}}{\lgauss^{2}}\right) \right] \nonumber \\
& & \hspace{1.5em} + B_{0} \frac{\ampl \sqrt{\pi}\lgauss}{8 \vims \tin}
\exp\left(-\frac{t}{2\tin}\right)
\left[{\rm erf}\left(\frac{x+\vims t}{\lgauss}\right)
- {\rm erf}\left(\frac{x-\vims t}{\lgauss}\right)\right]~, \hspace{3em}\\
\label{viwavesoln}
\delvi(x,t) &=& \vims \frac{\ampl}{2} \exp\left(-\frac{t}{2\tin}\right)
\left[\exp\left(\frac{-(x-\vims t)^{2}}{\lgauss^{2}}\right)
-\exp\left(\frac{-(x+\vims t)^{2}}{\lgauss^{2}}\right) \right]~.\hspace{3em}
\eea
The solution corresponds to left- and right-traveling wave pulses propagating at the ion 
magnetosound speed $\vims$ through a background of stationary neutrals. The decay time 
scale for the pulses is $2 \tin$, a value that is consistent with the decay of individual 
Fourier wave modes at wavelengths $\ll \Lims$ (e.g., see \S~3.1.1 of Ciolek et al.\  2004,
or \S~3.2.1 of Mouschovias et al.\ 2011). Equations (\ref{Bwave}) and (\ref{viwave})
follow immediately from inserting (\ref{Bwavesoln}) and (\ref{viwavesoln}) into 
(\ref{perturbeq}).

\subsection{Solution for $\Lims \ll \lgauss \ll \Lnms$}
\label{sec-diffsoln}

In this limit the neutrals still remain motionless. Also, the collisional
drag of the neutrals on the ions essentially balances the driving magnetic pressure gradient
in the charged fluid force equation. The solution in this wavelength regime can be derived 
from the Fourier-transformed magnetic field and velocity relations (\ref{Bcaratsoln}) and 
(\ref{vicaratsoln}), by rewriting them as
\bea
\label{Bdiffcaratsoln}
\delBcarat(k,t) & = &\frac{B_{0} \ampl \lgauss}{2 \sqrt{2}} \exp\left(-\frac{t}{2\tin}\right) \nonumber \\
& & \times \left\{\left[1 - \frac{1}{[1 - (k/\kims)^{2}]^{1/2}}\right]
\exp\left(-\frac{\lgauss^{2}k^{2}}{4} -  \vims \kims t \left[1 - \left(\frac{k}{\kims}\right)^{2}\right]^{1/2} \right) \right. \nonumber \\
& & \left. \hspace{1em} +
\left[1 + \frac{1}{[1 - (k/\kims)^{2}]^{1/2}}\right]
\exp\left(-\frac{\lgauss^{2}k^{2}}{4} + \vims \kims t 
\left[1 - \left(\frac{k}{\kims}\right)^{2}\right]^{1/2}  \right)
\right\}~, 
\hspace{3em}
\eea
and
\bea
\label{vidiffcaratsoln}
\hspace{-4em} \delvicarat(k,t) &=& \frac{-i \vims \ampl \lgauss k}{2\sqrt{2}\kims\left[1 - (k/\kims)^{2}\right]^{1/2}}
\exp\left(-\frac{t}{2\tin}\right) 
\left\{\exp\left(-\frac{\lgauss^{2}k^{2}}{4} -  \vims \kims t\left[1 - \left(\frac{k}{\kims}\right)^{2}\right]^{1/2} \right)
\right. \nonumber \\
& & \left.
\hspace{16em}- \exp\left(-\frac{\lgauss^{2}k^{2}}{4} + \vims \kims t\left[1 - \left(\frac{k}{\kims}\right)^{2}\right]^{1/2} \right) 
\right\}~. \hspace{3em}
\eea

The condition $\lgauss \gg \Lims$ for the modes considered here is equivalent to taking the 
limit $k/\kims \rightarrow 0$. Expanding (\ref{Bdiffcaratsoln}) and (\ref{vidiffcaratsoln})
in the ratio $k/\kims$ and keeping terms only up to second-order will then suffice to give
the principal solution when inverting the transforms. Doing this, and using the relations 
(\ref{kimsdef}) and (\ref{Dicoef}) to replace the cutoff wavenumber $\kims$ with the 
diffusion coefficient $\Di$, gives
\bea
\label{Bdiffsoln}
\delB(x,t) &=& \frac{B_{0}\ampl}{\left(1 + 4 \Di t/\lgauss^{2}\right)^{1/2}}
\exp\left(\frac{-x^{2}/\lgauss^2}{1 + 4 \Di t/\lgauss^{2}}\right) \nonumber \\
& & \hspace{-2em} + \frac{2 B_{0} \ampl \Di \tin/\lgauss^{2}}{\left(1+4\Di \tin/\lgauss^{2}\right)^{3/2}}
\left(1 - \frac{2 x^{2}/\lgauss^{2}}{1 + 4 \Di t/\lgauss^{2}}\right)
\exp\left(\frac{-x^{2}/\lgauss^{2}}{1 + 4 \Di t/\lgauss^{2}}\right) \nonumber \\
& & \hspace{-2em} - \frac{2 B_{0} \ampl \Di \tin/\lgauss^{2}}{\left[1-4\Di \tin/\lgauss^{2}\right]^{3/2}}
\left(1 - \frac{2 x^{2}/\lgauss^{2}}{1 - 4 \Di t/\lgauss^{2}}\right)
\exp\left(\frac{-t}{\tin}\right)
\exp\left(\frac{-x^{2}/\lgauss^{2}}{1 - 4 \Di t/\lgauss^{2}}\right)~, \hspace{3em} \\
\label{vidiffsoln}
\delvi(x,t) &=& \frac{2 \Di \ampl x}{\lgauss^{2}\left(1 + 4\Di t/\lgauss^{2}\right)^{3/2}}
\exp\left(\frac{-x^{2}/\lgauss^{2}}{1 + 4\Di t/\lgauss^{2}}\right) \nonumber 
\\
& & - \frac{2 \Di \ampl x}{\lgauss^{2}\left(1- 4 \Di t/\lgauss^{2}\right)^{3/2}}
\exp\left(\frac{-t}{\tin}\right)
\exp\left(\frac{-x^{2}/\lgauss^{2}}{1 - 4 \Di t/\lgauss^{2}}\right) ~.
\eea
The solution here is one of ambipolar diffusion, in which the charged fluid and magnetic 
field diffuse through a sea of fixed neutrals, as reflected by the first terms 
after the equalities in eqs. (\ref{Bdiffsoln}) and (\ref{vidiffsoln}). 
The diffusion coefficient 
$\Di$ reflects the balance of the magnetic field pressure (since $\vims^{2}\propto B^{2}$) 
and collisional forces (represented by $\tin$). From expressions (\ref{Bdiffsoln})
and (\ref{vidiffsoln}) the characteristic decay time scale $\tau_{\rm dec}$
for this solution is found to be the ambipolar diffusion time 
$\tau_{\rm ad} = \lgauss^{2}/4\Di$.

\subsection{Solution for $\lgauss \gg \Lnms$}
\label{sec-nmssolns}

In this limit, which has length scales $\gg \Lims$ (since $\Lnms \gg \Lims$) for our
clouds, the charged fluid continues to move in a force-free manner, 
For this situation we can make the approximation 
$\rhoio \partial(\delvi)/\partial t \simeq 0$ in the linearized ion momentum equation 
(\ref{viperturb}). Hence, the ions will be traveling at the terminal drift velocity
\be
\label{viterm}
\delvi = -\frac{B_{0}\tin}{4 \pi \rhoio} \frac{\partial}{\partial x}(\delB)~.
\ee 
Inserting the expression 
(\ref{viterm}) into the linearized force equation for the neutral fluid (\ref{vnperturb}) 
and also the linearized magnetic induction equation (\ref{Bperturbb}), then Fourier 
transforming those equations along with the perturbed mass continuity equation 
(\ref{rhonperturb}) yields the approximate system of ODEs for the wave modes in this 
wavelength region,
\be
\label{nFTsystem}
\frac{\partial \ycaratn}{\partial t} = \matMn \ycaratn~,
\ee
where 
$\ycaratn \equiv \left[\delrhoncarat(k,t),~\delvncarat(k,t)~,~\delBcarat(k,t)\right]^{\rm T}$
is the vector of transformed variables, and the $3 \times 3$ array of constant 
coefficients
\be
\label{matMndef}
\matMn
\equiv
\left[
\begin{array}{ccc}
       0     & -i \rhono k & 0 \\
       -i k \Cs^{2}/\rhono & 0 & -i k B_{0}/4\pi \rhono \\
       0    & -i B_{0} k  &  -\Di k^{2} \\
\end{array}
\right].
\ee

In typical clouds with $(\Cs/\vnms)^{2} \ll 1$ (see eqs. [\ref{vnmsdef}]-[\ref{vnAdef}]), the system 
(\ref{nFTsystem}) has the solution
\bml
\be
\label{ycaratnsoln}
\ycaratn = \aplusn \Eplusn e^{\gplusn t} + \aminusn \Eminusn e^{\gminusn t}
+ \azeron \Ezeron e^{\gzeron t}~,
\ee
where
\bea
\label{gpmndef}
\gpmn &=& - \frac{1}{2}\Di k^{2} \pm i \vnms k \left[1 - \left(\frac{k}{\knms}\right)^{2}\right]^{1/2}~, \\
\label{gzerondef}
\gzeron &=& -\left(\frac{\Cs}{\vnms}\right)^{2} \Di k^{2} = -\Dth k^{2}~,
\eea
are the eigenvalues of $\matMn$, and
\bea
\label{Epmndef}
\Epmn &=& \left[1~,~i\frac{\gpmn}{\rhono k}~,\frac{\gpmn B_{0}}{(\gpmn+\Di k^{2})\rhono}\right]^{\rm T} 
\simeq \left[1~,\mp\frac{\vnms}{\rhono}~,~\frac{B_{0}}{\rhono}\right]^{\rm T}~, \\
\label{Ezerondef}
\Ezeron &=& \left[1~,~i\frac{\gzeron}{\rhono k}~,~\frac{\gzeron B_{0}}{(\gzeron + \Di k^{2})\rhono}\right]^{\rm T} 
\simeq \left[1~,-i \frac{\Dth k}{\rhono},~-\left(\frac{\Cs}{\vnms}\right)^{2}
\frac{B_{0}}{\rhono}\right]^{\rm T}
\eea
\eml
are their associated eigenvectors. In the above, 
\be
\label{knmsdef}
\knms \equiv \frac{2 \pi}{\Lnms} = \frac{2 \vnms}{\vnA^{2} \tni} = \frac{2 \vnms}{\Di}~
\ee
is the neutral magnetosound cutoff wave number (eqs. [\ref{Dicoef}], [\ref{viAdef}], 
[\ref{vnAdef}] and [\ref{tnieq}] were used to derive the latter equalities). For 
$k < \knms$ the modes with eigenvalues $\gpmn$ are neutral magnetosound waves; these waves
decay by ambipolar diffusion of the charged fluid and magnetic field with respect to the 
neutrals. The $\gzeron$ mode is a neutral pressure-driven diffusion mode (e.g., see Figs. 
1b and 1d of Ciolek et al.\ 2004; or \S~3.2.1 of Mouschovias et al.\ 2011), in which the 
thermal-pressure gradient force of the neutral fluid drives the neutrals through a 
stationary background of the charged fluid and magnetic field. The balance of 
thermal-pressure and neutral-ion collisional drag in this mode has the neutrals moving at
a terminal drift speed, resulting in neutral diffusion with the coefficient $\Dth$ defined
in equation (\ref{Dthcoef}).
Simplifications resulting in the latter equalities of (\ref{Epmndef}) and 
(\ref{Ezerondef}) were made by noting that for $(\Cs/\vnms)^{2} \ll 1$, 
$|\gzeron| \ll |\gpmn|$. The limit $k \ll \knms$ was also used.

The coefficients $\aplusn$, $\aminusn$, and $\azeron$ in (\ref{ycaratnsoln}) are 
found by applying the initial conditions $\delrhoncarat(k,0)=\delvncarat(k,0)=0$, 
and $\delBcarat(k,0)$ given by (\ref{Bpertinit}). Doing this, we have the solutions
\bea
\label{Bncaratsoln}
\hspace{-1em}\delBcarat(k,t) &=& \frac{B_{0}\ampl\lgauss}{2\sqrt{2}[1 + (\Cs/\vnms)^{2}]}
\nonumber \\
& & \times \left\{\left[1 + i 2\left(\frac{\Cs}{\vnms}\right)^{2}\frac{k}{\knms}\right]
\exp\left(-\left[\frac{\lgauss^{2}}{4}+ \frac{\Di t}{2}\right]k^{2} + i\vnms kt \left[1-\left(\frac{k}{\knms}\right)^{2}\right]^{1/2}\right) \right. \nonumber \\ 
& &
\hspace{1.5em} + \left[1 - i2\left(\frac{\Cs}{\vnms}\right)^{2}\frac{k}{\knms}\right]
\exp\left(-\left[\frac{\lgauss^{2}}{4}+ \frac{\Di t}{2}\right]k^{2} 
- i\vnms kt \left[1-\left(\frac{k}{\knms}\right)^{2}\right]^{1/2}\right) \nonumber \\ 
& &
\hspace{1.5em} + \left. 2 \left(\frac{\Cs}{\vnms}\right)^{2}
\exp\left(-\left[\frac{\lgauss^{2}}{4} + \Dth t\right] k^{2}\right) \right\}~, \\
\label{vncaratsoln}
\hspace{-1em}\delvncarat(k,t) &=& \frac{\ampl \vnms\lgauss}{2 \sqrt{2} [1 + (\Cs/\vnms)^{2}]} \nonumber \\
& & \times \left\{-\left[1 + i2 \left(\frac{\Cs}{\vnms}\right)^{2}\frac{k}{\knms} \right]
\exp\left(-\left[\frac{\lgauss^{2}}{4}+ \frac{\Di t}{2}\right]k^{2} + i\vnms kt \left[1-\left(\frac{k}{\knms}\right)^{2}\right]^{1/2}\right) \right. \nonumber \\ 
& &
\hspace{1.5em} + \left[1 - i2\left(\frac{\Cs}{\vnms}\right)^{2}\frac{k}{\knms}\right]
\exp\left(-\left[\frac{\lgauss^{2}}{4}+ \frac{\Di t}{2}\right]k^{2} 
- i\vnms kt \left[1-\left(\frac{k}{\knms}\right)^{2}\right]^{1/2}\right) \nonumber \\ 
& &
\hspace{1.5em}
\left.
+ i 4 \left(\frac{\Cs}{\vnms}\right)^{2}\frac{k}{\knms}
\exp\left(-\left[\frac{\lgauss^{2}}{4} + \Dth t\right] k^{2}\right) \right\}~, \\
\label{rhoncaratsoln}
\hspace{-1em} \delrhoncarat(k,t) &=& \frac{\ampl \rhono \lgauss}{2\sqrt{2}[1+(\Cs/\vnms)^{2}]} \nonumber \\
& &
\times \left\{\left[1 + i 2 \left(\frac{\Cs}{\vnms}\right)^{2}\frac{k}{\knms}\right]
\exp\left(-\left[\frac{\lgauss^{2}}{4}+ \frac{\Di t}{2}\right]k^{2} + i\vnms kt \left[1-\left(\frac{k}{\knms}\right)^{2}\right]^{1/2}\right) \right. \nonumber \\ 
& &
\hspace{1.5em} + \left[1 - i2\left(\frac{\Cs}{\vnms}\right)^{2}\frac{k}{\knms}\right]
\exp\left(-\left[\frac{\lgauss^{2}}{4}+ \frac{\Di t}{2}\right]k^{2} 
- i\vnms kt \left[1-\left(\frac{k}{\knms}\right)^{2}\right]^{1/2}\right) \nonumber \\ 
& &
\hspace{1.5em}
\left.
- 2 \exp\left(-\left[\frac{\lgauss^{2}}{4} + \Dth t\right] k^{2}\right) \right\}~.
\eea

Inverting the transforms give the solutions for $\delrhon(x,t)$, $\delvn(x,)$,
and $\delB(x,t)$. The inversion is aided by the fact that for the initial perturbation we
consider here, with $\lgauss \gg \Lnms$, the overwhelming bulk of the wave modes contributing
to the wave packet have $k \ll \knms$. Hence, taking the limit $(k/\knms)^{2} \rightarrow 0$ 
will introduce only a minor error when inverting the Fourier transforms (\ref{Bncaratsoln}),
(\ref{vncaratsoln}), and (\ref{rhoncaratsoln}). Doing this yields:
\bea
\label{delBnsoln}
\delB(x,t)&=& 
\frac{\ampl B_{0}\Gonefunc}{2}
\left[\left\{1 - \frac{2(x+\vnms t)\Dth/\lgauss^{2}}{\vnms[1+(2\Di t/\lgauss^{2})]}\right\}
\exp\left(\frac{-(x + \vnms t)^{2}}{\lgauss^{2}[1+(2\Di t/\lgauss^{2})]}\right) \right. \nonumber \\
& &
\hspace{5em} \left. + \left\{1 + \frac{2(x-\vnms t)\Dth/\lgauss^{2}}{\vnms[1+(2\Di t/\lgauss^{2})]}\right\}
\exp\left(\frac{-(x - \vnms t)^{2}}{\lgauss^{2}[1+(2\Di t/\lgauss^{2})]}\right) \right] \nonumber \\
& &
+ \ampl B_{0}\Gtwofunc\left(\frac{\Cs}{\vnms}\right)^{2}\exp\left(\frac{-x^{2}}{\lgauss^{2}[1+(4\Dth t/\lgauss^{2})]}\right)~, \\
\label{delvnsoln}
\delvn(x,t)&=& \frac{\ampl \vnms\Gonefunc}{2}
\left[\left\{\frac{2(x+\vnms t)\Dth/\lgauss^{2}}{\vnms[1+(2\Di t/\lgauss^{2})]}- 1\right\} 
\exp\left(\frac{-(x + \vnms t)^{2}}{\lgauss^{2}[1+(2\Di t/\lgauss^{2})]}\right)
\right. \nonumber \\
& & \hspace{6em} + \left. 
\left\{1 + \frac{2(x - \vnms t)\Dth/\lgauss^{2}}{\vnms[1+(2\Dth t/\lgauss^{2})]}\right\}
\exp\left(\frac{-(x-\vnms t)^{2}}{\lgauss^{2}[1+(2\Di t/\lgauss^{2})]}\right)
\right] \nonumber \\
& & - \frac{2 \ampl \Gtwofunc \Dth x}{\lgauss^{2}[1+(4\Dth t/\lgauss^{2})]}
\exp\left(\frac{-x^{2}}{\lgauss^{2}[1+(4\Dth t/\lgauss^{2})]}\right)~, \\
\label{delrhonsoln}
\delrhon(x,t) &=& \frac{\ampl \rhono\Gonefunc}{2}
\left[\left\{1 - \frac{2(x+\vnms t)\Dth/\lgauss^{2}}{\vnms[1+(2\Di t/\lgauss^{2})]}\right\}
\exp\left(\frac{-(x+\vnms t)^{2}}{\lgauss^{2}[1+(2\Di t/\lgauss^{2})]}\right) \right. \nonumber \\
& & \hspace{5em}
+\left. \left\{1 + \frac{2(x-\vnms t)\Dth/\lgauss^{2}}{\vnms[1+(2\Di t/\lgauss^{2})]}\right\}
\exp\left(\frac{-(x-\vnms t)^{2}}{\lgauss^{2}[1+ (2\Di t/\lgauss^{2})]}\right)\right]
\nonumber \\
& &
- \ampl \rhono \Gtwofunc\exp\left(\frac{-x^{2}}{\lgauss^{2}[1+(4 \Dth t/\lgauss^{2})]}\right)~,
\eea
where the two functions $\Gonefunc$ and $\Gtwofunc$ are respectively defined by equations
(\ref{Gonefuncdef}) and (\ref{Gtwofuncdef}). The solutions clearly show that
for initial perturbation with $\lgauss \gg \Lnms$ left- and right-propagating neutral 
pulses traveling at the magnetosound speed $\vnms$. There is some damping iin the pulses
because of ambipolar diffusion, occurring on a time scale 
$\tau_{\rm dec} = 2 \tau_{\rm ad} = \lgauss^{2}/2\Di$. There is also a part
to the above solutions that indicate an effect due to pressure-driven diffusion of the 
neutrals, as can be seen by the terms in the above equations involving $\Dth$.
Those terms show that pressure and density gradients are created in the neutrals 
(by collisional drag from the ions) that can also affect the motion of the
neutral fluid. However, these effects are small compared to that which 
results from the magnetically-driven collisional drag of the ions on the 
neutrals. 

Finally, equations (\ref{Bnmseq})-(\ref{nnnmseq}) are obtained by
combining (\ref{perturbeq}), (\ref{delBnsoln}) - (\ref{delrhonsoln}), 
and (\ref{fluxfreezeq}), and using the relations $\nn = \rhon/\mn$ and 
$\nni = \rhoi/\mi$. Equation (\ref{vinmseq}) follows from taking the partial derivative
$\partial/\partial x$ of (\ref{delBnsoln}) and then inserting the result into (\ref{viterm}). 
\section{Numerical Convergence}
\label{sec-convergence}
Here we provide quantitative data verifying the scaling and convergence of the numerical 
algorithm described in this paper. We do this by modeling the evolution of small-amplitude
disturbances and comparing the numerical model results against exact solutions of the
linearized system of equations (\ref{rhonperturb}) - (\ref{Bperturbb}). Inserting 
(\ref{linidealgas}) and (\ref{linP}) into those equations, and then Fourier transforming
them in space (eq. [\ref{FTeq}]) gives 
\be
\label{FTtotsys}
\frac{\partial \ycarattot}{\partial t} = \matMtot \ycarattot~,
\ee
where 
$\ycarattot \equiv [\delrhoncarat(k,t),~\delvncarat(k,t),~\delrhoicarat(k,t),
~\delvicarat(k,t),~\delBcarat(k,t)]^{\rm T}$
is the vector of transformed variables for the full system, and
\be
\label{matMtotdef}
\matMtot
\equiv
\left[
\begin{array}{ccccc}
       0 & -i\rhoio k  & 0  & 0 & 0\\
       0 &  -1/\tin  & -iB_{0}k/4\pi\rhoio & 0 & 1/\tin\\
       0 & -ikB_{0}  & 0  & 0 & 0\\
       0 & 0  & 0  & 0 & -i\rhono k\\
       0 & 1/\tni  & 0  & -i\Cs^{2}k/\rhono & -1/\tni \\
\end{array}
\right]. 
\ee
This particular ODE system has the solution
\be
\label{genFTeigsoln}
\ycarattot(k,t) = \sum_{m=1}^{5} a_{m}\Evec_{m}e^{g_{m}t}~~
\ee
(Braun 1983, Ch. 3), where $g_{m}$ is the $m$th eigenvalue of $\matMtot$ and $\Evec_{m}$ is 
its corresponding eigenvector. The constants $a_{m}$ are determined by the initial 
conditions. Making the subsitution 
\bml
\be
\label{gm-omegam-relation}
g_{m} = -i \omega_{m}, 
\ee
where
\be
\label{compomega}
\omega_{m} = \omega_{m,R} + i \omega_{m,I}
\ee
\eml
is a complex frequency (with real and imaginary parts $\omega_{m,R}$ and $\omega_{m,I}$,
respectively) and inverting the transform (eq. [\ref{invFTeq}]), yields
\be
\label{geneigsoln}
\yvec_{\rm tot}(x,t) = \frac{1}{\sqrt{2\pi}}\sum_{m=1}^{5}\int_{-\infty}^{+\infty} a_{m} \Evec_{m} e^{i(kx-\omega_{m} t)} dk~;
\ee
$\yvec_{\rm tot} = [\delrhon(x,t),~\delvn(x,t),~\delrhoi(x,t)~, ~\delvi(x,t),~\delB(x,t)]^{\rm T}$
is the total vector of perturbed variables. 

A single mode with $m=l$ at wavelength $\lambdaeig$ and wavenumber
$\keig = 2\pi/\lambdaeig$ is excited from the initial complex perturbation
\be
\label{initialeig}
\yvec_{\rm tot}(x,0) = \beig \Evec_{l} e^{i (k_{\rm p} x + \phaseeig)}~.
\ee
The constant $\beig$ is taken to be real, and $\phaseeig$ is the phase constant
of the initial state. Setting $t=0$ in (\ref{geneigsoln}) and equating 
it to (\ref{initialeig}) reveals that for the former relation to describe the
evolution of the latter, we must have 
$a_{m} = \sqrt{2\pi} \beig \delta_{lm} \delta(k-\keig) e^{i\phaseeig}$.
Hence,
\be
\label{compeigsoln}
\yvec_{\rm tot}(x,t) = \beig \Evec_{l} e^{i(k_{\rm p} x - \omega_{l} t +\phaseeig)}~.
\ee
$\Evec_{l}$ is also complex, and can be written as
\be
\label{compeigvec}
\Evec_{l} = \Evec_{l,R} + i \Evec_{l,I}
\ee 
where $\Evec_{l,R}$ and $\Evec_{l,I}$ are both real vectors. Inserting (\ref{compomega})
and (\ref{compeigvec}) into (\ref{compeigsoln}), it follows that the real part of the
singly-excited eigenmode of the system is given by
\be
\label{realeigsoln}
\Re[\yvec_{\rm tot}(x,t)] = \beig \left[\Evec_{l,R}\cos\left(\keig x - \omega_{l,R} t + \phaseeig \right)
- \Evec_{l,I}\sin\left(\keig x - \omega_{l,R} t + \phaseeig \right)\right] e^{\omega_{l,I}t}~.
\ee

To test convergence we excite a single mode of the multifluid system and compare the
resulting evolution of our numerical code to the exact solution (\ref{realeigsoln}). A 
$\lambdaeig$ is selected and the eigenvalues and eigenvectors of the array $\matMtot$ at 
that wavelength are calculated using widely available numerical routines (e.g., EISPACK, 
LAPACK) or software packages (e.g., MATLAB, Maple, IDL). To ensure that an excited 
mode always remains linear, the value of the numerical constant $\beig$ of the initial 
state (the real part of eq. [\ref{initialeig}], or equivalently, eq. [\ref{realeigsoln}] at
$t=0$) is chosen so that its magnetic field perturbation has a relative amplitude that is 
$10^{-6}$ that of the background magnetic field. For our convergence test runs we assume
the same background reference state and conditions for all of the physical variables 
($\nno$, $\nnio$, $B_{0}$, $\Tn$, etc.) as in the models described in \S~\ref{sec-linwaves}. 
All of the speeds ($\vims$, $\vnms$, and $\Cs$), collision times ($\tin$ and $\tni$),
and characteristic length scales ($\Lims$ and $\Lnms$) are therefore also the same as
for those models. Additionally, we set the eigenmode wavelengths $\lambdaeig$ equal to the
widths $\lgauss$ of the Gaussian test packets of \S~\ref{sec-linwaves}; hence, the
underlying physics of the waves discussed for those models also applies to the models 
presented here.

\begin{figure}
\epsscale{0.83}
\plotone{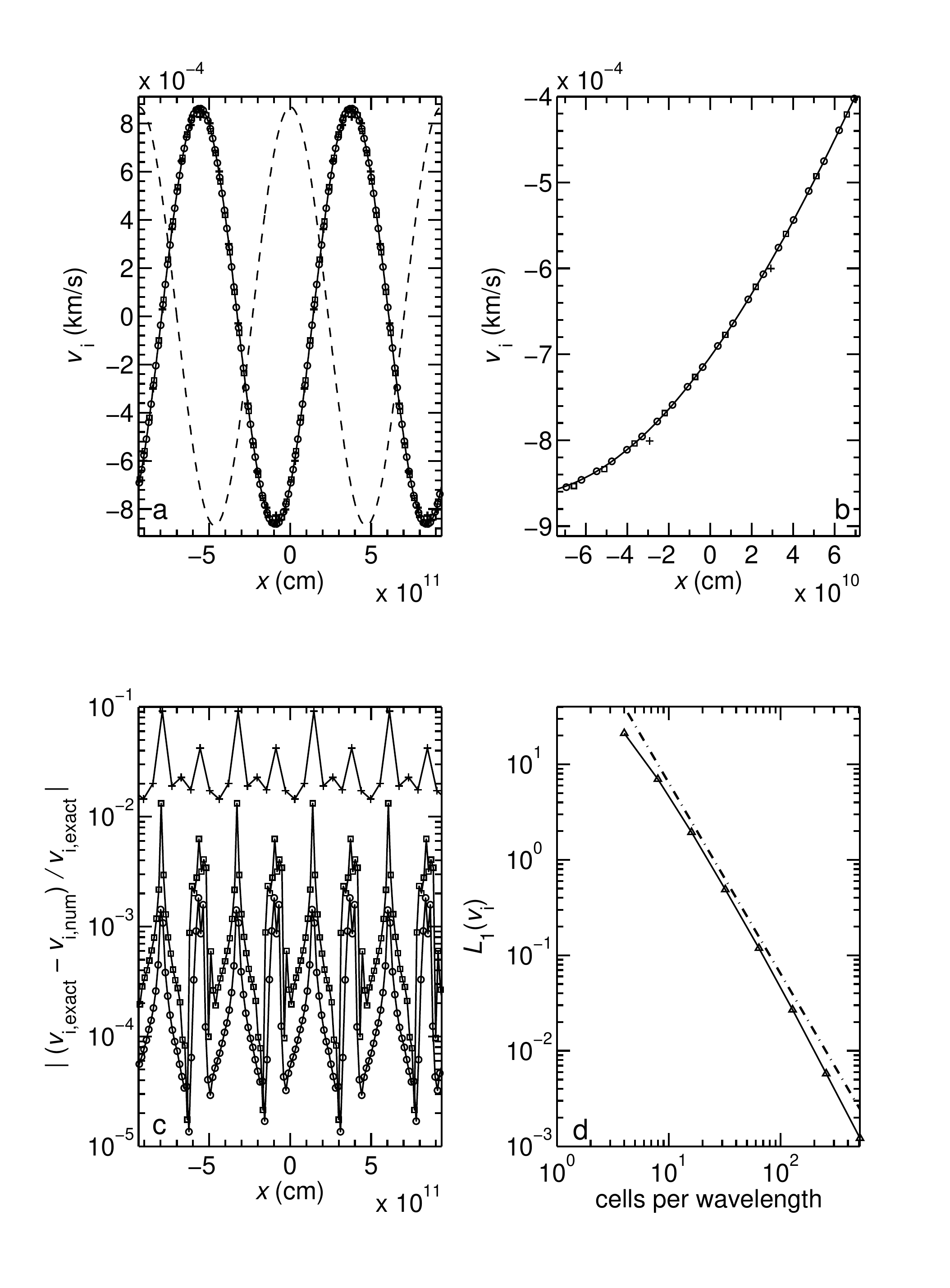}
\caption{Rightward-traveling ion magnetosound wave with 
$\lambdaeig=9.35\times 10^{11}~\cm < \Lims$ at time $t=1.37 \times 10^{-4}~\yr$. 
({\it a}) Ion velocity. Solid line is the exact eigensolution (eq. [\ref{realeigsoln}])at 
that time, dashed-line is the initial state. Also shown are numerical code results for 
three models
having different numbers of mesh cells per wavelength of the initial perturbation: 16 cells
(plus signs), 64 cells (squares), and 128 cells (circles). ({\it b}) Zoom-in of the ion 
velocity data about the origin. ({\it c}) Relative error in the ion velocity for the same 
three numerical models. ({\it d}) Numerically-determined error measure of the ion velocity 
$\Lone$ as a function of the number of cells per wavelength in a model (curve with 
triangles). The dash-dot line is a reference curve with a logarithmic slope of -2.    
\label{fig-imswave}
}
\end{figure}

Our first excited eigenmode model has $\lambdaeig = 9.35 \times 10^{11}~\cm$, which
is less than the upper ion magnetosound wave cutoff length $\Lims$. The selected 
mode is an ion magnetosound wave traveling in the $+x$-direction with velocity 
$+\vims$. At this wavelength the mode has $\omega_{l,R} = 5.84 \times 10^{-4}~{\rm s}^{-1}$
and $\omega_{l,I} = -1.26 \times 10^{-6} {\rm s}^{-1}$, the latter value corresponding to a
decay rate $\tau_{\rm dec} = 0.0251~\yr = 2 \tin$. Figure \ref{fig-imswave}$a$ shows the 
exact solution for the ion velocity (solid curve) at $t=1.37 \times 10^{-4}~ \yr$, by which 
time the wave has advanced by more than {\onefourth} of a wavelength from its initial state 
(dashed curve). Also shown are the results of three models having different spatial 
resolution: the first model (displayed as plus signs) has 16 mesh cells within each 
wavelength of the initial perturbation, the second (squares) 64 cells per wavelength, and the
third (circles) 128 cells per wavelength. For this perturbation the numerical time step is 
determined by the CFL time step (\ref{deltCFLeq}); the CFL number $\nu = 0.8$ in each of 
these models. Figure \ref{fig-imswave}$b$ displays a zoom-in of the same data about $x=0$. 
The local relative error in the ion velocity 
$\left|\left[v_{\rm i,exact}(x,t)-v_{\rm i,num}(x,t)\right]/v_{\rm  i,exact}(x,t)\right|$
is presented in Figure \ref{fig-imswave}$c$, showing how the numerical accuracy increases
with finer mesh spacing. Finally, Figure \ref{fig-imswave}$d$ shows the error measure in 
the ion velocity across the entire compuational domain of N cells
\be
\label{Lonedefeq}
\Lone
\equiv \frac{1}{N} \sum_{j=1}^{N} \left|v_{\rm i,exact}(\xj,t)-v_{\rm i,num}(\xj,t)\right|
\ee 
(curve with triangles) for several different model runs, all with $\nu =0.8$, as a 
function of each model's numerical resolution. The range is from 4  to 512 cells per 
wavelength of the perturbation. Also shown (dash-dot curve) is a reference curve 
with a logarithmic slope $=-2$, the value expected for a code that is second-order 
accurate in space and time. Our numerical results are seen to be consistent with 
second-order accuracy. 

\begin{figure}
\epsscale{0.83}
\plotone{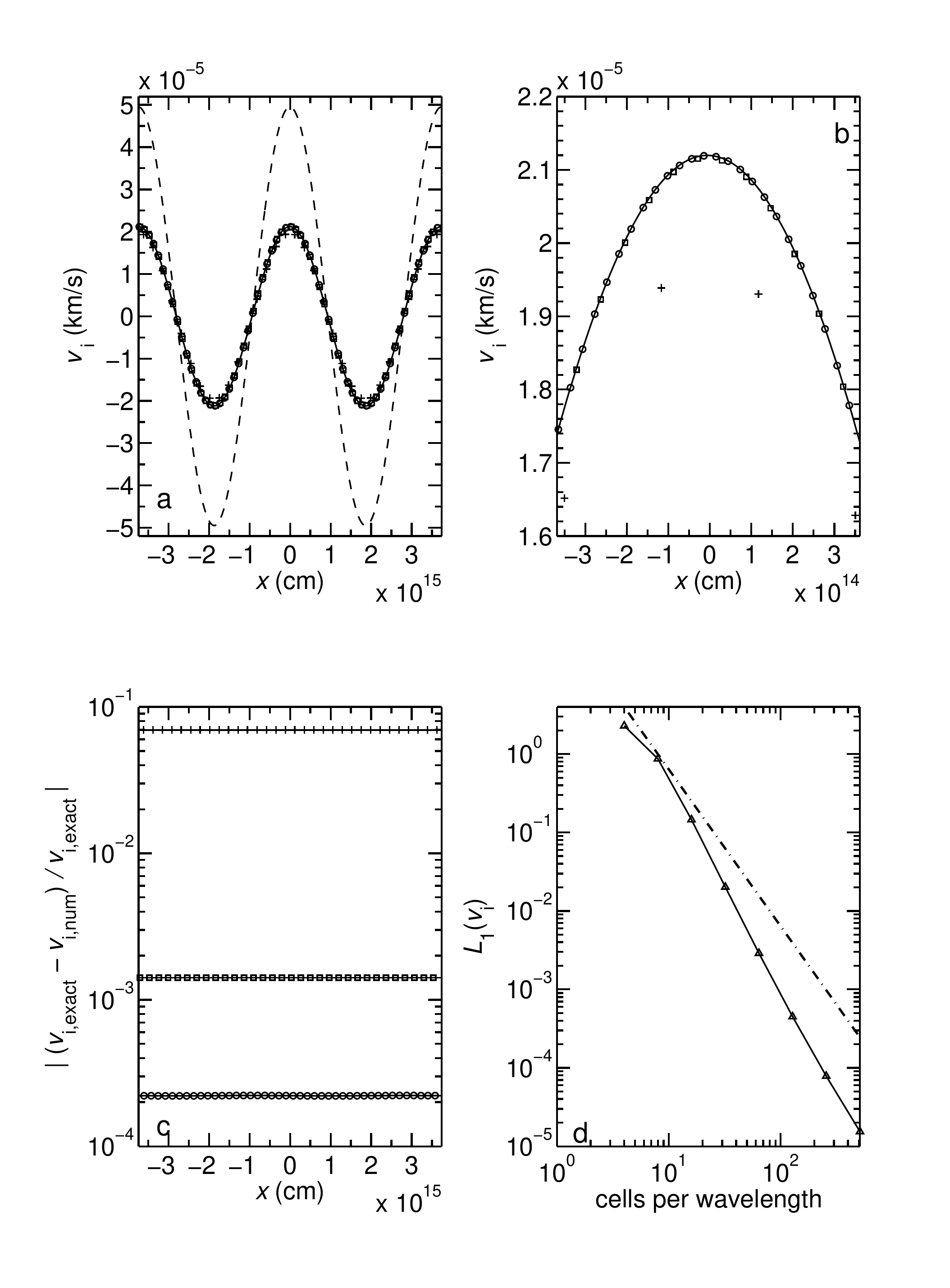}
\caption{Ambipolar diffusion mode with $\lambdaeig =  3.74 \times 10^{15}~\cm$ 
($\Lims < \lambdaeig < \Lnms$) at $t=3.17~\yr$. All symbols and curves have the same meaning 
as in Fig. \ref{fig-imswave}.
\label{fig-idiff}
}
\end{figure}

The second excited eigenmode is at $\lambdaeig = 3.74 \times 10^{15}~\cm$, which
is in the range $\Lims < \lambdaeig < \Lnms$. The ambipolar diffusion mode that exists
within this range is chosen for this test: at this wavelength that particular mode has
$\omega_{l,R} = 1.30 \times 10^{-26}~{\rm s}^{-1}$ and 
$\omega_{l,I} = -8.47 \times 10^{-9}~{\rm s}^{-1}$. Its decay time is therefore
$\tau_{\rm dec} = 3.74~\yr$. Figure \ref{fig-idiff}$a$ displays the initial state and
the exact solution for the ion velocity at $t=3.17~\yr$. Also shown are the results of 
numerical runs having resolutions of 16, 64, and 128 cells per wavelength of the 
perturbation. For these model runs, the upper limit to the maximum stable numerical time is 
limited by the time scales occurring in the source terms (here, the ion-neutral collision
time $\tin$) instead of that which would be calculated with a traditional CFL time step
with $\nu \sim 0.1 - 1$ (see \S~\ref{sec-algorithm}). Stability of the time integration 
for these models thus requires that their time step $\delt = \deltsource \leq \tin$.
In analogy to convergence tests in which the CFL number $\nu \propto \delt/\delx$ is
kept fixed, the models for this test have $\delt/\delx = {\rm constant}$, such that
\be
\label{convsourcetimestep}
\delt = \frac{(4/\lambdaeig)}{(N_{\rm p}/\lambdaeig)} \tin~,
\ee
where $N_{\rm p}$ ($\geq 4$) is the number of cells within one wavelength of the
perturbation. A close-up showing the results for this eigenmode about the origin is
presented in Figure \ref{fig-idiff}$b$, and the local relative error is displayed in
Figure \ref{fig-idiff}$c$. The $\Lone$ error measure of this mode can be seen in Figure 
\ref{fig-idiff}$d$. Examination of that panel indicates that our code tends to scale 
slightly better than second-order accuracy in this instance. 

\begin{figure}
\epsscale{0.83}
\plotone{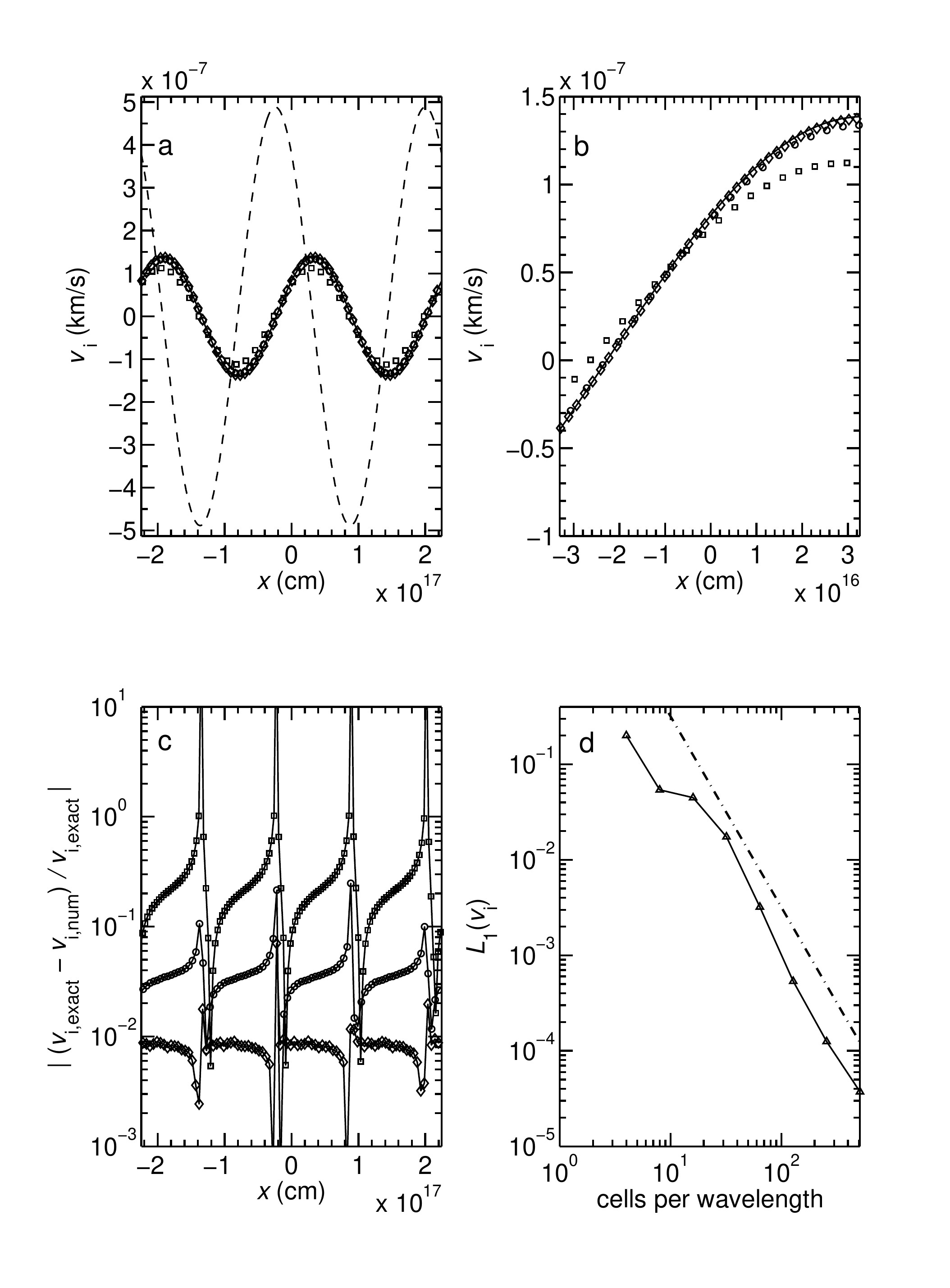}
\caption{Right-traveling neutral magnetosound wave with $\lambdaeig = 2.24 \times 10^{17}~\cm > \Lnms$ 
at $t= 5.03 \times 10^{4}~\yr$. ({\it a}) Ion velocity: exact eigensolution
(solid curve) and initial state (dashed). Also displayed are results for models having 
different
numerical resolutions: 64 cells per initial perturbation wavelength (squares), 128 cells 
per wavelength (circles), and 256 cells per wavelength (diamonds). ({\it b}) Close-up of
the ion velocity data about the origin. ({\it c}) Relative error in the ion velocity
for the three numerical models. ({\it d}) Error measure $\Lone$ as a function of
model resolution. The dash-dot reference curve has a logarithmic slope $=-2$. 
\label{fig-nmswave}
}
\end{figure}

The final excited eigenmode has $\lambdaeig = 2.24 \times 10^{17}~\cm > \Lnms$. For this
test we chose the mode that corresponds to a neutral magnetosound wave propagating in
the $+x$-direction. At this wavelength $\omega_{l,R} = 1.02 \times 10^{-12}~{\rm s}^{-1}$
and $\omega_{l,I} = -7.93 \times 10^{-13}~{\rm s}^{-1}$; hence, the decay time by ambipolar
diffusion for this mode is $\tau_{\rm dec} = 3.99 \times 10^{4}~\yr$. Figure 
\ref{fig-nmswave}$a$ shows the eigenmode solution for the ion velocity at 
$t=5.03 \times 10^{4}~\yr$, by which time the wave has moved about $\onefourth$ of a 
wavelength from its initial state. Three model runs with different numerical 
resolution are also shown: one with 64 cells per wavelength (squares), another 
with 128 cells per wavelength (circles), and the last with 256 cells per 
wavelength (diamonds). For these eigenmode runs it is again the case that the maximum
stable numerical time step cannot exceed $\tin$, so the relation (\ref{convsourcetimestep})
was used to fix the ratio $\delt/\delx$ to the same value in each model to
test convergence in the usual way. An expanded view of the ion 
velocity data near $x=0$ is shown in Figure \ref{fig-nmswave}$b$, and the local 
relative error in $\vi$ as a function of position 
for the three models is found in Figure \ref{fig-nmswave}$c$. The error measure $\Lone$ as
a function of numerical resolution is provided in Figure \ref{fig-nmswave}$d$; the data
there show for the large part that our code again exhibits second-order accuracy.

\end{document}